\documentclass{ws-wsarpp}
\usepackage[super,sort&compress,comma]{natbib}
\usepackage{hyperref}
\usepackage{doi}

\usepackage{graphicx}

\usepackage{units}
\usepackage{enumerate}
\usepackage{pdflscape}
\usepackage{afterpage}
\usepackage{lineno}
\usepackage{xspace}
\usepackage{heppennames2}
\usepackage{cernunits}
\usepackage{epigraph}
\usepackage{xcolor}
\renewcommand{\wsdraftnote}{}

\begin{document}

%
\catchline{}{}{}{}{}
%


\let\Xspace\xspace

\DeclareRobustCommand{\Pepem}{\HepParticle{\Pe}{}{+}\HepParticle{\Pe}{}{-}\Xspace} 
\DeclareRobustCommand{\PGmpGmm}{\HepParticle{\PGm}{}{+}\HepParticle{\PGm}{}{-}\Xspace} 
\newcommand{\pT}{\ensuremath{p\sb{\scriptstyle\mathrm{T}}}\Xspace}

\newcommand{\sqrts}{\ensuremath{\sqrt{s}}\Xspace}
\newcommand{\sNNraw}{\ensuremath{s_{\mbox{\tiny NN}}}}
\newcommand{\sqrtsNN}{\ensuremath{\sqrt{\sNNraw}}\Xspace}
\newcommand{\sNN}{\sqrtsNN}
\newcommand{\Npart}{\ensuremath{N_{\rm part}}\Xspace}
\newcommand{\Ncoll}{\ensuremath{N_{\rm coll}}\Xspace}
\newcommand{\RpPb}{\ensuremath{R_{\rm pPb}}\Xspace}
\newcommand{\RAA}{\ensuremath{R_{\rm AA}}\Xspace}
\newcommand{\RpA}{\ensuremath{R_{\rm pA}}\Xspace}
\newcommand{\TAA}{\ensuremath{T_{\rm AA}}\Xspace}
\newcommand{\RCP}{\ensuremath{R_{\rm CP}}\Xspace}
\newcommand{\vtwo}{\ensuremath{v_{\rm 2}}\Xspace}
\newcommand{\vone}{\ensuremath{v_{\rm 1}}\Xspace}
\newcommand{\vthree}{\ensuremath{v_{\rm 3}}\Xspace}
\newcommand{\vfour}{\ensuremath{v_{\rm 4}}\Xspace}
\newcommand{\vfive}{\ensuremath{v_{\rm 5}}\Xspace}
\newcommand{\vsix}{\ensuremath{v_{\rm 6}}\Xspace}
\newcommand{\vseven}{\ensuremath{v_{\rm 7}}\Xspace}
\newcommand{\veight}{\ensuremath{v_{\rm 8}}\Xspace}
\newcommand{\vnine}{\ensuremath{v_{\rm 9}}\Xspace}
\newcommand{\vn}{\ensuremath{v_{\rm n}}\Xspace}

\newcommand{\nch}         {\ensuremath{N}\Xspace}
\newcommand{\mch}         {\ensuremath{M_{\mathrm {ch}}}\Xspace}
\newcommand{\meannch}     {\ensuremath{\langle N_{\mathrm {ch}} \rangle}\Xspace}
\newcommand{\meanpT}      {\ensuremath{\langle\pT\rangle}}
\newcommand{\dNdeta}      {\mathrm{d}N_\mathrm{ch}/\mathrm{d}\eta}
\newcommand{\dNdy}        {\mathrm{d}N_\mathrm{ch}/\mathrm{d}y}
\newcommand{\kT}          {\ensuremath{k\sb{\scriptstyle\mathrm{T}}}}
\newcommand{\ptt}         {\ensuremath{p_{\mathrm{T, trig}}}}
\newcommand{\pta}         {\ensuremath{p_{\mathrm{T, assoc}}}}

\newcommand{\ee}          {\ensuremath{\mathrm{e^+e^-}}\Xspace}
\newcommand{\pp}          {$\mathrm{pp}$\Xspace}
\newcommand{\pPb}         {$\mathrm{p}$--$\mathrm{Pb}$\Xspace}
\newcommand{\Pbp}         {$\mathrm{Pb}$--$\mathrm{p}$\Xspace}
\newcommand{\pO}          {$\mathrm{p}$--$\mathrm{O}$\Xspace}
\newcommand{\Op}          {$\mathrm{O}$--$\mathrm{p}$\Xspace}
\newcommand{\OO}          {$\mathrm{O}$--$\mathrm{O}$\Xspace}
\newcommand{\pA}          {$\mathrm{p}$--$\mathrm{A}$\Xspace}
\newcommand{\dA}          {$\mathrm{d}$--$\mathrm{A}$\Xspace}
\newcommand{\AOnA}        {$\mathrm{A}$--$\mathrm{A}$\Xspace}
\newcommand{\PbPb}        {$\mathrm{Pb}$--$\mathrm{Pb}$\Xspace}
\newcommand{\ArAr}        {$\mathrm{Ar}$--$\mathrm{Ar}$\Xspace}
\newcommand{\XeXe}        {$\mathrm{Xe}$--$\mathrm{Xe}$\Xspace}
\newcommand{\KrKr}        {$\mathrm{Kr}$--$\mathrm{Kr}$\Xspace}
\newcommand{\AuAu}        {$\mathrm{Au}$--$\mathrm{Au}$\Xspace}
\newcommand{\CuCu}        {$\mathrm{Cu}$--$\mathrm{Cu}$\Xspace}
\newcommand{\pAu}         {$\mathrm{p}$--$\mathrm{Au}$\Xspace}
\newcommand{\dAu}         {$\mathrm{d}$--$\mathrm{Au}$\Xspace}
\newcommand{\HeAu}         {$\mathrm{^{3}He}$--$\mathrm{Au}$\Xspace}

\newcommand{\sigmaBFPP}   {\ensuremath{\sigma_\mathrm{BFPP}}\Xspace}

\newcommand{\lsim}        {\,{\buildrel < \over {_\sim}}\,}
\newcommand{\gsim}        {\,{\buildrel > \over {_\sim}}\,}
\newcommand{\co}[1]       {\relax}
\newcommand{\nl}          {\newline}
\newcommand{\el}          {\\\hline\\[-0.4cm]}

\newcommand{\gmom}{\ensuremath{\mathrm{GeV}\kern-0.05em/\kern-0.02em c}}
\newcommand{\antip}{\ensuremath{\overline{\mathrm{p}}}}
\newcommand{\antid}{\ensuremath{\overline{\mathrm{d}}}}
\newcommand{\tritium}{\ensuremath{{}^{3}\mathrm{H}}}
\newcommand{\antitritium}{\ensuremath{{}^{3}\overline{\mathrm{\mathrm{He}}}}}
\newcommand{\hethree}{\ensuremath{{}^{3}\mathrm{He}}}
\newcommand{\hefour}{\ensuremath{{}^{4}\mathrm{He}}}
\newcommand{\antihethree}{\ensuremath{{}^{3}\overline{\mathrm{He}}}}
\newcommand{\antihefour}{\ensuremath{{}^{4}\overline{\mathrm{He}}}}
\newcommand{\hyp}        {\ensuremath{^{3}_{\Lambda}\mathrm{H}}}
\newcommand{\antihyp}{\ensuremath{^{3}_{\bar{\Lambda}}\overline{\mathrm{H}}}}
\newcommand{\hypfour}    {\ensuremath{^{4}_{\Lambda}\mathrm{H}}}
\newcommand{\antihypfour}{\ensuremath{^{4}_{\bar{\Lambda}}\overline{\mathrm{H}}}}
\newcommand{\hyphefour}    {\ensuremath{^{4}_{\Lambda}\mathrm{He}}}
\newcommand{\antihehypfour}{\ensuremath{^{4}_{\bar{\Lambda}}\overline{\mathrm{He}}}}
\newcommand{\sigmahyp}     {\ensuremath{^{3}_{\Sigma^{0}}\mathrm{H}}}
\newcommand{\antisigmahyp} {\ensuremath{^{3}_{\bar{\Sigma}^{0}}\overline{\mathrm{H}}}}

\newcommand{\sla}{\slash \hspace{-0.2cm}}
\newcommand{\slam}{\slash \hspace{-0.25cm}}
\newcommand{\no}{\nonumber}
\def\lsim{\mathrel{\rlap{\lower4pt\hbox{\hskip1pt$\sim$}}
    \raise1pt\hbox{$<$}}}         
\def\gsim{\mathrel{\rlap{\lower4pt\hbox{\hskip1pt$\sim$}}
    \raise1pt\hbox{$>$}}}         

\newcommand{\Anucl}{$\mathrm{A}$\Xspace}
\newcommand{\nbInv}{$\mathrm{nb}^{-1}$\Xspace}
\newcommand{\pbInv}{$\mathrm{pb}^{-1}$\Xspace}
\newcommand{\isospin}{$I$\Xspace}
\newcommand{\spinJ}{$J$\Xspace}
\newcommand{\BA}{$B_{\mathrm{A}}$\Xspace}
\newcommand{\significance}{$\frac{\mathrm{S}}{\sqrt{\mathrm{S}+\mathrm{B}}}$\Xspace}
\newcommand{\Tchem}{\ensuremath{T_{\mathrm{chem}}}\Xspace}

\def\Bs{{\overline{B}}_s}
\def\R{\mathcal{R}}
\newcommand{\gev}{\mathrm{GeV}}
\newcommand{\tev}{\mathrm{TeV}}
\newcommand{\mev}{\mathrm{MeV}}
\newcommand{\e}{\epsilon}
\newcommand{\tce}{\frac{t_{\rm cool}(\e)}{t_{\rm esc}(\e)}}
\newcommand{\tcer}{\frac{t_{\rm c}(\R)}{t_{\rm esc}(\R)}}
\def\Xe{X_{\rm esc}}
\def\X{X_{\rm esc}}
\def\te{t_{\rm esc}}
\def\tc{t_{\rm cool}}
\def\nb{n_{\rm B}}
\def\nc{n_{\rm C}}
\def\ni{n_{i}}
\def\rism{\rho_{\rm ISM}}
\def\nism{n_{\rm ISM}}
\def\x{(\R,\vec r,t)}
\def\xo{(\R,\vec r_\odot,t_\odot)}
\def\ap{\overline{\rm p}}
\def\ad{\overline{\rm d}}
\def\ep{e^+}
\def\Qep{Q_{e^+}}
\def\epm{$e^\pm$\ }
\def\ah{\overline{\rm ^3He}}
\def\at{\overline{\rm t}}
\def\s{$(*)$}
\newcommand{\dd}{\text{d}}
\newcommand{\gaga}{\gamma\gamma}
\newcommand{\Rp}{\mathcal{R}^\prime}
\newcommand{\Lp}{L^{\prime}}

\newcommand{\Dsc}{\ensuremath{D_{\rm s}}\Xspace}
\newcommand{\twopiTDsc}{\ensuremath{2 \pi T D_{\rm s}}\Xspace}
\newcommand{\ToverTc}{\ensuremath{T/T_{\rm c}}\Xspace}
\newcommand{\Tc}{\ensuremath{T_{\rm c}}\Xspace}
\newcommand{\chisquared}{\ensuremath{\chi^{\rm 2}}\Xspace}

\newcommand{\RunsThreeFour}{Runs~3 \& 4\Xspace}

\newcommand{\ttbar}{\ensuremath{t\overline{t}}\Xspace}



\newcommand{\qty}[2]{\ensuremath{#1\,\mathrm{#2}}}  
\newcommand{\enum}[2]{\ensuremath{#1\times10^{#2}}} 
\newcommand{\NQTY}[2]{\mbox{$[#1/{\rm #2}]$}}     
\newcommand{\UQTY}[2]{\ensuremath{#1/\mathrm{#2}}}  
\newcommand{\eqty}[3]{\qty{\enum{#1}{#2}}{#3}}  
\newcommand{\invnb}{\mathrm{nb}^{-1}}
\newcommand{\invpb}{\mathrm{pb}^{-1}}

\newcommand{\elumi}[2]{\qty{\enum{#1}{#2}}{cm^{-2}s^{-1}}}
\newcommand{\murad}[1]{\qty{#1}{\mu rad}}
\newcommand{\intlumimub}[1]{\qty{#1}{\mu b^{-1}}}

\newcommand{\yNN}{\ensuremath{y_{\mbox{\tiny NN}}}}
\newcommand{\bstar}{\ensuremath{\beta^{*}}}
\newcommand{\emittn}{\ensuremath{\varepsilon_n}}
\newcommand{\LAA}{\ensuremath{L_\text{AA}}}
\newcommand{\LpA}{\ensuremath{L_{pA}}}
\newcommand{\Lpp}{\ensuremath{L_{pp}}}
\newcommand{\Lpeak}{\ensuremath{\hat{L}}}
\newcommand{\LNN}{\ensuremath{ L_{\text{NN}}}}

\newcommand{\isotope}[3]{\ensuremath{^{#1}\mathrm{#2}^{#3}}}

\newcommand{\speciesheader}{ &
\isotope{16}{O}{8+}&
\isotope{40}{Ar}{18+}&
\isotope{40}{Ca}{20+}&
\isotope{78}{Kr}{36+}&
\isotope{129}{Xe}{54+}&
\isotope{208}{Pb}{82+}
}

\newcommand{\bfunc}{$\beta$-function}
\newcommand{\bstarval}[1]{$\bstar = #1\,\mbox{m}$}
\newcommand{\betarel}{\ensuremath{\beta_\text{rel}}}
\newcommand{\emittnx}{\ensuremath{\epsilon_{n,x}}}
\newcommand{\emittny}{\ensuremath{\epsilon_{n,y}}}
\newcommand{\emittnxy}{\ensuremath{\epsilon_{n,xy}}}
\newcommand{\emitts}{\ensuremath{\epsilon_s}}
\newcommand{\sigs}{\ensuremath{\sigma_s}}
\newcommand{\sigp}{\ensuremath{\sigma_p}}
\newcommand{\kb}{\ensuremath{k_b}}
\newcommand{\frev}{\ensuremath{f_0}}
\newcommand{\Nb}{\ensuremath{N_b}}
\newcommand{\Eb}{\ensuremath{E_b}}
\newcommand{\emittval}[1]{\ensuremath{\emittn=\qty{#1}{\mu m\,rad}}}
\newcommand{\Nbval}[2]{\ensuremath{\Nb=\enum{#1}{#2}}}
\newcommand{\taul}{\ensuremath{\tau_l}}
\newcommand{\taulval}[1]{\ensuremath{\taul=\qty{#1}{ns}}}
\newcommand{\sigzval}[1]{\ensuremath{\sigz=\qty{#1}{cm}}}
\newcommand{\etev}[1]{\ensuremath{\Eb=\qty{#1}{TeV}}}
\newcommand{\VRF}{\ensuremath{V_{\mathrm{RF}}}}
\newcommand{\lumival}[2]{\ensuremath{L=\qty{#1\times 10^{#2}}{cm^{-2} s^{-1}}}}

\newcommand{\aibsx}{\ensuremath{\alpha_{\mathrm{IBS},x}}}
\newcommand{\aibsy}{\ensuremath{\alpha_{\mathrm{IBS},y}}}
\newcommand{\aibsxy}{\ensuremath{\alpha_{\mathrm{IBS},x,y}}}
\newcommand{\aradd}{\ensuremath{\alpha_{\mathrm{rad}}}}
\newcommand{\aradds}{\ensuremath{\alpha_{\mathrm{rad},s}}}
\newcommand{\araddx}{\ensuremath{\alpha_{\mathrm{rad},x}}}
\newcommand{\araddy}{\ensuremath{\alpha_{\mathrm{rad},y}}}
\newcommand{\araddxy}{\ensuremath{\alpha_{\mathrm{rad},x,y}}}
\newcommand{\Z}{\ensuremath{Z_\text{ion}}}
\newcommand{\A}{\ensuremath{A_\text{ion}}}
\newcommand{\Circ}{\ensuremath{C_\text{ring}}}
\newcommand{\lumi}{\ensuremath{\mathcal{L}}}
\newcommand{\Lb}{\ensuremath{\mathcal{L}_b}}
\newcommand{\Lint}{\ensuremath{L_{\text{int}}}}
\newcommand{\Lbint}{\ensuremath{L_{b,\text{int}}}}
\newcommand{\Lbpeak}{\ensuremath{\mathcal{L}_{b,\text{peak}}}}


\newcommand{\mlna}{\langle \ln\!A \rangle}
\newcommand{\nmu}{N_\mu}
\newcommand{\lnnmu}{\ln\!\nmu}
\newcommand{\xmax}{X_\text{max}}
\newcommand{\nmult}{N_\text{mult}}
\newcommand{\tocite}{{\bf REF}}
\newcommand{\si}[1]{\ensuremath{\text{#1}}}
\newcommand{\SI}[2]{\ensuremath{#1\,\si{#2}}}

\newcommand{\Ntrig}        {\ensuremath{N_{\mathrm{trig}}}}
\newcommand{\Nassoc}       {\ensuremath{N_{\mathrm{assoc}}}}
\newcommand{\dNassoc}      {\ensuremath{\frac{\dd^2N_{\mathrm{assoc}}}{\dd\Deta\dd\Dphi}}}
\newcommand{\Dphi}         {\ensuremath{\Delta\varphi}}
\newcommand{\Deta}         {\ensuremath{\Delta\eta}}

\title{A Decade of Collectivity in Small Systems}

\author{JAN FIETE GROSSE-OETRINGHAUS}

\address{CERN, 1211 Geneva, Switzerland\\
jgrosseo@cern.ch}

\author{URS ACHIM WIEDEMANN}

\address{CERN, 1211 Geneva, Switzerland\\
urs.wiedemann@cern.ch}

\maketitle


\begin{abstract}
Signatures of collectivity, including azimuthally anisotropic and radial flow as well as characteristic hadrochemical dependencies, have been observed since long in (ultra)relativistic nucleus--nucleus collisions. They underpin the interpretation of these collision systems in terms of QGP formation and close-to-perfect fluidity. Remarkably, however, essentially all these signatures of collectivity have been identified within the last decade in collision systems as small as \pp and \pPb, where collective phenomena had been assumed to be absent traditionally. Precursor phenomena may have been found even in ep and \ee collisions. This article provides a complete review of all data on small system collectivity. It reviews model simulations of these data where available. However, in the absence of a phenomenologically fully satisfactory description of collectivity across all system sizes, we focus in particular on the theoretical basis of all dynamical frameworks of collectivity invoked in heavy ion collisions, and their expected scaling with system size. Our discussion clarifies to what extent all dynamical explanations are challenged by the available data. 
\end{abstract}

\clearpage
\setcounter{tocdepth}{2}
\tableofcontents

\clearpage

\markboth{J.F. Grosse-Oetringhaus, U.A. Wiedemann}{A Decade of Collectivity in Small Systems}

\epigraph{In high-energy physics, we have concentrated on experiments in which we distribute a higher and higher amount of energy into a region with smaller and smaller dimensions. But, in order to study the question of ‘vacuum’, we must turn to a different direction; we should investigate bulk phenomena by distributing high energy over a relatively large volume.}{T.D. Lee, 1975 }

\section{Introduction}

The LHC discovery of collectivity in small systems relates measurements in proton--proton, proton--nucleus and nucleus--nucleus collisions in a remarkably smooth way without any  step-wise discontinuity.
In addition to an intense experimental effort at the LHC, it has motivated the study of dedicated small collision systems at RHIC, the re-analysis and re-interpretation of data in \ee and ep collisions, and extensive theoretical work.
As this review will emphasize, this discovery impacts both: our understanding of the dynamics of nucleus--nucleus collisions (in which signatures of collectivity have been searched for traditionally) and our understanding of the dynamics of proton--proton collisions (in which signatures of collectivity had been assumed to be absent). To put these LHC measurements into a broader scientific context, we recall the following:

With the discovery of QCD in 1973\cite{Politzer:1973fx,Gross:1973id} and with first insights into the QCD phase structure of the strong interactions at finite temperature and density~\cite{Collins:1974ky,Cabibbo:1975ig,Shuryak:1978ij}, confinement and the chiral condensate were understood to be characteristics of the QCD vacuum. The strong interactions at finite temperature became calculable, see e.g. Ref.~\citenum{Karsch:1987kz}. Their physical origin became testable by creating the conditions for the QCD high-temperature phase --- the quark--gluon plasma (QGP) --- in laboratory-based experiments. As captured succinctly in the above quote of T.D. Lee~\cite{Lee:1974kn}, this led since the 1980s to research in heavy-ion physics (HIP) that took a different direction from the main efforts of high-energy physics (HEP).  

HEP focused on “hard” high-momentum transfer processes that are calculable perturbatively with controlled precision~\cite{Ellis:1996mzs}. Aimed initially at establishing QCD as the theory of the strong interactions, this precision frontier is advanced since the 1990s to control as precisely as possible the QCD background on top of which new physics is searched for in collider experiments. Most relevant for this program are processes in which the systematically calculable “hard” short-distance physics factorizes from the “soft” long-distance contributions which are process-independent and thus independent of the partonic environment in which they may occur. Models of soft multi-particle production in HEP multi-purpose event generators were initially based on extrapolating this picture of collinearly factorized QCD beyond its regime of guaranteed validity. 
Modern model implementations in PYTHIA\cite{Sjostrand:2014zea}, HERWIG\cite{Bellm:2015jjp} and SHERPA\cite{Gleisberg:2008ta}
go beyond a purely incoherent superposition of multiple parton interactions (e.g. by implementing color reconnections in hadronization~\cite{Christiansen:2015yqa}) but HEP multi-purpose event generators are not partonic transport codes: the partonic fragmentation processes and essential parts of the hadronization process are modeled as in vacuum, i.e., independent of the phase-space density within which they occur.

HIP, in contrast, focuses on testing QCD at finite phase-space density by searching for collective phenomena in soft multi-particle production, and by characterizing the modification of hard processes embedded in dense QCD matter. Throughout the 1980s and 90s, fixed target experiments with ultra-relativistic nuclear beams at the BNL AGS and at CERN SPS searched for signals of the QGP phase transition in terms of step-like structures that would occur in multiple classes of measurements at a common critical energy density~\cite{Harris:1996zx,Harris:2024aov}. These experiments stated circumstantial evidence for QGP formation based on their discovery of strong sensitivities of soft particle production to the nuclear environment in the collision region~\cite{Heinz:2000bk}. However, the original simple search strategy (that was akin of the methodology used to characterize phase transitions in non-evolving solid state physics) had failed. No step-like changes were discovered. 

In the early 2000s, due to major theoretical advances~\cite{Policastro:2001yc,Teaney:2003kp,Baier:2007ix} and in comparison with RHIC data~\cite{BRAHMS:2004adc,PHENIX:2004vcz,PHOBOS:2004zne,STAR:2005gfr}, it became clear that dissipative fluid dynamics provides a successful phenomenology of soft multi-particle production in heavy-ion collisions~\cite{Romatschke:2007mq,Song:2010mg}. 
This resulted in a radically different approach towards testing hot and dense QCD. Since dissipative fluid dynamics is formulated in terms of thermodynamic quantities that are calculable from first principles in QCD, the rapid (hydro)dynamic evolution from extreme initial conditions to a diluted final hadronic state was not viewed any more as a confounding factor in testing equilibrium QCD but it became the very tool via which QCD equilibrium properties would reveal themselves. 
The resulting phenomenology supports the formation of a QCD fluid of close to minimal dissipative properties in which the mean free path of quasi-particles is so short that the very concept of a medium composed of propagating quasi-particles becomes questionable~\cite{Casalderrey-Solana:2011dxg}. It cannot be overemphasized that this HIP perfect fluid picture of nucleus--nucleus collisions is the extreme opposite of the HEP picture for proton--proton collisions. In the HIP case, no degree of freedom propagates through a sizable fraction of the collision region without interaction with the surrounding medium; in the HEP case, all degrees of freedom free-stream and fragment through the entire collision region without interacting with any medium.

These maximally different default pictures for soft multi-particle production in proton--proton (HEP) and nucleus--nucleus (HIP) collisions raise a set of common sense questions, including: Can both pictures be correct? Is there a critical size or density at which a QCD system produced in hadronic collisions transits from free-streaming to fluid-dynamic behavior? Can precursors of fluid dynamics be identified and could they inform us about the QCD non-equilibrium dynamics that leads to fast hydrodynamization and equilibration?
All these questions point to the importance of understanding how collectivity and medium-modifications arise as a function of systems size\footnote{Varying the spatial extension of a system is not the only way of changing conditions of collectivity. For instance, 
in simple conformal models, the degree to which a collision system develops collective phenomena is known to depend on one dimensionless product of energy density and transverse extension, the so-called opacity\cite{Kurkela:2018qeb,Kurkela:2019kip,Ambrus:2022qya,Ambrus:2022koq,Arslandok:2023utm}. In practice, however, a change of beam particle from Pb to p changes the transverse spatial extension by a factor 10 -- to change energy density by a comparable factor is more difficult because of its approximately logarithmic $\sqrt{s}$-dependence.}
It is this point that is informed by the LHC discovery of collectivity in small systems. These measurements marked a paradigm shift as, historically, the experimental programs with nuclear beams focused on central collisions of the heaviest nuclei to distribute high energy over the largest achievable volume and to create conditions most favorable for the formation of a QGP. 
In this context, the study of small collision systems like \pA or pp was seen as a reference for large systems. The aim was to constrain the nuclear dependence of parton distribution functions and to study so-called cold nuclear matter effects~\cite{Armesto:2018ljh}, i.e. the particle production expected in the absence of QGP formation.
This logic prevailed throughout the CERN SPS and RHIC experimental programs and it was the original motivation for supplementing the LHC heavy-ion program with \pA collisions~\cite{Salgado:2011wc}. Heavy-ion experiments therefore always had an important small systems program but without the expectation of gaining significant additional understanding about QCD physics at extreme temperature and density. However, with the LHC discovery of long-range correlation structures in \pp collisions in 2010\cite{Khachatryan:2010gv} and in \pPb collisions in 2012\cite{Aad:2012gla,Abelev:2012ola,CMS:2012qk}, for the first time, effects well-known from the study of the QGP, and not accounted for by standard multi-purpose HEP event generators~\cite{Fischer:2016zzs}, occurred in high-multiplicity \pp and \pPb collisions. 
That the near-side ridge in \pp collisions was only the first observation of a completely generic phenomenon, namely small system collectivity, became gradually clear with the observation of similar effects in \pPb collisions and with the equally unexpected discovery of a second ridge structure on the away side. 
A new research direction emerged. This manuscript refers to about 200 experimental publications related to small systems, and it reviews the theoretical research motivated by these data. 
For earlier reviews, see Refs.~\citenum{Dusling:2015gta,Song:2017wtw,Nagle:2018nvi,Noronha:2024dtq}.

This review separates the discussion of definitions of measurements in chapter~\ref{sec2}, from the discussion of experimental data in chapter~\ref{sec3} and from the discussion of theoretical concepts informed by these data in chapter~\ref{sec4}. To the extent possible, the chapters are self-contained. In particular, chapter~\ref{sec2} provides a detailed and necessarily technical discussion of how measures of collectivity are defined. But one may want to skip chapter~\ref{sec2} in a first reading if one is mainly interested in the state of art of experiment or theory.

\section{Measures of collectivity and medium-modifications}
\label{sec2}
In the broader physics debate, it has been emphasized repeatedly that ``more is different”\cite{Anderson:1972pca} in the sense that macroscopic volumes of matter manifest qualitatively new phenomena that are distinct from those that can be discerned in interactions among few elementary constituents. In a broad sense, collectivity refers to such phenomena that emerge with increasing size and complexity of a physical system. Emergent collective phenomena pose novel challenges for physics, as they often require distinct experimental techniques for their characterization, distinct theoretical methods for their description and as there is the fundamental challenge of understanding how collectivity in many-body systems emerges from the fundamental degrees of freedom. 

In the phenomenological practice of collider physics, a measure of “collectivity” is more than a measurement. It is a measurement supplemented by a baseline that is free of collective phenomena and on top of which “collectivity” can be established. 
Baselines are chosen to separate microscopic physics (that is thought to arise from the incoherent superposition of elementary interactions) from those qualitatively novel ``collective" phenomena 
that can arise only in sufficiently macroscopic systems. However, the simple idea of separating microscopic from macroscopic physics faces obvious challenges if applied to small mesoscopic systems that interpolate smoothly between the limiting cases of the small and the big. 

\subsection{Measures of anisotropic flow}
\label{sect:methods}
Hadronic wavefunctions have a finite transverse extent. When they collide, energy is deposited in a finite region of the plane transverse to the beam direction. The shape of this initial {\it spatial} deposition depends on the kind of hadron or nucleus that is collided, on the impact parameter of the collision and on event-by-event fluctuations. In an azimuthal Fourier decomposition, this spatial information is typically characterized in terms of {\it spatial} eccentricities $\epsilon_n$. However, what is measured are certain correlations of final azimuthal asymmetries $v_n$ in {\it momentum} space. If the quanta produced in different positions in the transverse plane would not interact with each other, the measured $v_n$ would not be sensitive to the $\epsilon_n$'s. One important aspect discussing collectivity is to ask: What is the non-trivial collective dynamics that turns spatial gradients and asymmetries $\epsilon_n$ into momentum-anisotropies $v_n$,
\begin{equation}
    \underbrace{\lbrace \epsilon_1, \epsilon_2, \epsilon_3,\dots \rbrace}_{\rm spatial\, asymmetries}   \qquad
    \xrightarrow[\rm dynamics?]{\rm which\, collective}
    \qquad
    \underbrace{\lbrace v_1, v_2, v_3,\dots \rbrace}_{\rm momentum\, asymmetries}
    \label{trans}
\end{equation}
Experimentally, the initial conditions $\lbrace \epsilon_1, \epsilon_2, \epsilon_3,\dots \rbrace$ in \eqref{trans} can be changed by choosing different nuclear or hadron beams, by varying the centrality of the collision, or by selecting particular event classes via event shape engineering~\cite{Schukraft:2012ah}. As reviewed in the following section~\ref{sec3}, a wealth of data maps out the dynamical response \eqref{trans} as a function of system size.
Here, we start by reviewing different experimental definitions of the anisotropic flow coefficients $v_n$.

If $P(\vec{p}_1,\vec{p}_2)$ is the probability to produce a pair of particles with momenta $\vec{p}_1$ and $\vec{p}_2$ in a collision, and if $P(\vec{p}_i)$ is the corresponding probability to produce a single particle with momentum $\vec{p}_i$, then $P(\vec{p}_1,\vec{p}_2) \not= P(\vec{p}_1) P(\vec{p}_2)$ signals correlated particle production. However, not all correlated production is ``collective". Measures of anisotropic flow aim at separating few-particle correlations (that originate from microscopic sources like jets or resonance decays, also referred to as non-flow) from collective phenomena (that involve the correlation amongst essentially all particles in an event). Both sources of correlations scale differently with charged event multiplicity\footnote{For reasons of readability, we denote charged-particle multiplicity with $N$ and without the usual subscript ``ch''.} $\nch$. If the correlation function contains a collective phenomenon that affects with signal strength $S_{\rm collective}$ {\it each} particle pair, then it is multiplied by the number of pairs $ \tfrac{\nch(\nch-1)}{2}$. This collective contribution dominates parametrically by $O(\nch)$ compared to the microscopic production, since
\begin{equation}
     P(1,2) = \frac{2}{\nch (\nch-1)}
        \left( \frac{\nch(\nch-1)}{2} S_{\rm collective}(\nch) 
            + \frac{\nch}{2} S_{\rm microscopic} \right)\, .
\end{equation}
Here, $S_{\rm collective}$ is a function of \nch
since it depends also on system size.  In contrast, the signal strength $S_{\rm microscopic}$ of microscopic correlations is by definition independent of the environment it is embedded in and it is thus independent of $\nch$. 

As a next step we discuss {\bf what one would like to measure and what one would like to remove from the measurement:}
Ideally, one would like to characterize the azimuthal distribution of particles into Fourier coefficients 
\begin{equation}
\vn \equiv \langle\langle e^{i n \varphi} \rangle\rangle = \langle \int \dd\varphi e^{i n \varphi} \tfrac{d\nch}{\dd\varphi} \rangle \, ,
\label{vn_theory}
\end{equation}
 where the two brackets denote averages over all particles in an event and over event samples. Then, one would like to remove from the measurement of $\vn$ all physics of microscopic origin, so that it becomes a measure of collectivity. The problem here is two-fold. First, without reference to a physically preferred azimuthal orientation (such as the orientation of the reaction plane), the azimuthal integral will average to zero. Eq.~\ref{vn_theory} does not correspond to an experimental measurement. 
To bypass this problem, one constructs azimuthal multi-particle correlations that are independent of  overall azimuthal orientation. Second, any subtraction of microscopic physics depends on a assumption of what this microscopic physics is and it is therefore, strictly speaking, a choice. The notion of {\it pseudorapidity gap} plays in this context a central role. Namely, the two main few-particle correlation phenomena, jets and resonance decays, occur at small angular differences on the near side, i.e. at small $\Deta$ and $\Dphi$, as we will see later in the left panel of Fig.~\ref{fig:ridge1}. Those can therefore be separated from long-range contributions by considering only particle pairs with $|\Deta| > \eta_{\rm gap}$ ($\eta$-gap method). 
    However, a jet has a back-to-back component due to momentum conservation\footnote{In pp collisions, contrary to elementary \ee collisions, the parton--parton scattering frame is not the lab frame and therefore the back-to-back side of the jet is found at $\Dphi = \pi$ but at different $\Deta$.} leading to a broad away-side structure. This structure cannot be separated through an $\eta$-gap and in consequence, a considerable amount of the jet correlation remains.

{\bf What one measures and what one removes from the measurement.}
In the following, we summarize different procedures that are in use to quantify azimuthal correlations and that are used to reduce few-particle correlations in these measurements. 

\begin{enumerate} 
\item {\it Near-side ridge yields}\\ 
    Measurements of near-side ridge yields are arguably the simplest way of bypassing the above-mentioned problem that jet-like correlations lead to a broad away-side structure. As basic \ee or pp inspired calculations and Monte Carlo simulations do not have a mechanism which produces near-side long-range correlation, the yield measurement carries a strong physics message. Some uncertainties remain from  the definition of the baseline of the combinatorial background, and one may underestimate this yield at low $p_T$ where the away-side peak is wide in azimuth. Also, from the ridge yield alone one cannot extract $\vn$ coefficients while the converse is possible.

\item {\it Event-plane method}\\
    In this method~\cite{Voloshin:1994mz}, two different detectors are used. Typically, a forward rapidity detector determines a preferred azimuthal orientation called event plane $\Psi_{\rm EP}$ which in large collision systems is a proxy for the orientation of the reaction plane. This angle is then used to extract Fourier coefficients similar to Eq.~\ref{vn_theory} from particles measured in a mid-rapidity detector:
    \begin{equation}
        \vn \equiv \langle\langle e^{i n (\varphi - \Psi_{\rm EP})} \rangle\rangle = \langle \int \dd\varphi e^{i n (\varphi-\Psi_{\rm EP})} \tfrac{dN}{\dd\varphi} \rangle \, .
    \label{eq:vn_eventplane}
    \end{equation}
    The rapidity separation of the two detectors allows to reduce non-flow contributions. In small systems such as \pPb collisions, this method has limitations, since there may not be a clear definition of the reaction plane. For instance, if the system is not rapidity-invariant or if it has small multiplicity, $\Psi_{\rm EP}$ can depend on rapidity. The precision of the event plane depends on the number of considered particles and its resolution is proportional to $\sqrt{\nch}$. If the event plane is misestimated, the extracted signal~\eqref{eq:vn_eventplane} is then reduced as it is averaged out. In practice, if the two used detector systems have a fixed rapidity gap between them by the experiment's design, an experimental study of the signal as function of the rapidity separation is not possible and the associated uncertainties cannot be experimentally assessed. 

\item {\it Decomposing associated per-trigger yields into Fourier coefficients}\\
    Measuring the distance in azimuth ($\Delta\varphi$) and pseudorapidity ($\Deta$) of   
    a so-called \textit{trigger particle} from a second particle (commonly referred to as \textit{associated particle}), one constructs the normalized per-trigger yield $Y_N$ for events of multiplicity $N$
    \begin{equation}
        \frac{1}{\Ntrig} \frac{\dd \Nassoc}{\dd\Dphi\dd\Deta} \equiv Y_N \equiv J_N(\Dphi,\Deta)
            +  N \left[ 1 + \sum_n 2 \vn^2 \cos n\Dphi \right]\, . \label{eq:vn}
    \end{equation}
    This quantity is normalized as particle yield, i.e. an integral over a certain phase space region expresses the number of associated particles to each trigger particle.
    This method does not require to determine an event plane, as pointed out e.g. in Ref.~\citenum{Wang:1991qh}.
    To characterize a collective component in terms of Fourier coefficients $v_n$ and to distinguish it from  a jet-like component $J_N(\Dphi,\Deta)$ due to microscopic interactions, one then exploits  that both contributions should scale differently with event multiplicity. To this end, one chooses
    a factor $a$ such that the jet-like near-side contribution cancels in the difference between two different multiplicities $N$ and $M$
    \begin{equation}
        \Delta Y(\Dphi) = Y_N - a\, Y_M\, ,
    \end{equation}
    and one decomposes this difference in terms of flow coefficients $v_n$. This subtraction assumes that the away-side structures scale in the same way with $a$ as the near-side structures.
    Two different subtraction techniques have been used regularly: 
    \begin{enumerate}
        \item {\it The low-multiplicity subtraction method\cite{Abelev:2012ola,Aad:2012gla}}\\
        assumes that the collective component vanishes at low multiplicity $M$ and extracts the $v_n$'s from the ansatz   $$\Delta Y(\Dphi) =G' + N \sum_n 2 v_n^2\, \cos(n\Dphi)\, .$$
        \item {\it The template method\cite{Aad:2015gqa}}\\
        assumes that the collective component is significant at low multiplicity and extracts the $v_n$ from the ansatz\footnote{For pedagogical reasons we present here the typical simultaneous fit of $a$, $G$ and \vn in several steps.}
        $$\Delta Y(\Dphi) = G \left(1 +  \sum_n 2 v_n^2\, \cos(n\Dphi)\right).$$
    \end{enumerate}
    To illustrate the subtle differences between both subtraction methods, one may consider the case that $a=1$ and that the values of the flow coefficients $v_{n,N}$ in
    \eqref{eq:vn} vary with multiplicity like $v_{n,M}^2  = \alpha_N v_{n,N}^2$. For the two methods, the measured $v_{n,{\rm method}}^2$ deviates then from the true $v_{n,N}^2$ like
    $$ \frac{v_{\rm n, low\, mult\, subtraction}^2}{v_{ n,N}^2} = \frac{N-\alpha_N M}{N} \qquad  \hbox{versus}\qquad  \frac{v_{\rm n, template}^2}{v_{n,N}^2} = \frac{N-\alpha_N M}{N-M}. $$
    If one of these ratios equals unity then 
    the ``true" signal $v_{n,N}^2$ for the $n$-th harmonic at multiplicity $N$ is returned by the chosen method.
    The low-multiplicity subtraction is correct for $\alpha_N = 0$, and the template method is correct of $\alpha_N = 1$. As illustrated in Fig.~\ref{fig:methods}, the larger a true collective component will be at low multiplicity, the more it will be underestimated by the low-multiplicity subtraction method and the more it will be overestimated by the template method.   
This also becomes apparent in measurements if both methods are applied to the same dataset. For instance this explains the differences seen at low event multiplicity in the $v_2$ measurements displayed later in Fig.~\ref{fig:atlas_methods}. In the limit of high multiplicity, the results of both methods converge. However,
the discussion of what constitutes collectivity at small multiplicity requires a particularly critical assessment of the method employed since even qualitatively different conclusions may be reached at face value. We therefore suggest to make the differences explicit in the nomenclature, for instance by using $v_{\rm n, low\, mult\, subtraction}$ and $v_{\rm n, template}$.
    
    \begin{figure}[ht]
    \centering
    \includegraphics[width=0.7\textwidth]{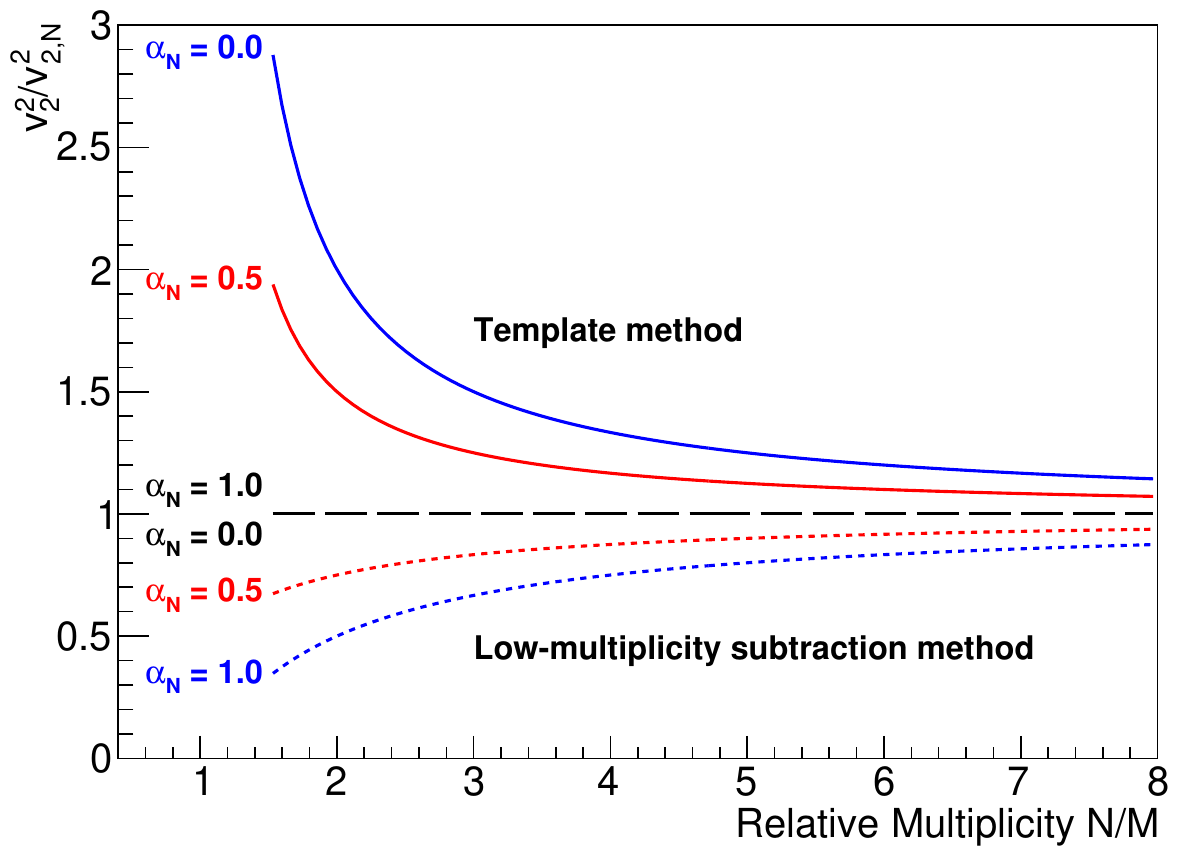}
    \caption{\label{fig:methods} Ratio of the measured $\vtwo^2$ to the true $\vtwo^2$ as a function of relative multiplicity $N/M$ for different cases of $\alpha_N$ ($v_{2,M}^2 = \alpha_N v_{2,N}^2$). The solid lines in the upper part show the template method where the \vtwo is exact or overestimated depending on $\alpha_N$, while the dashed lines in the lower part show the low-multiplicity subtraction method where the \vtwo is exact or underestimated depending on $\alpha_N$. Each of the two methods is exact only for a single $\alpha_N$ and the two values are maximally different.}
    \end{figure}    

\item {\it Two-particle cumulant flow $v_n\lbrace 2\rbrace$ and higher orders $v_n\lbrace 2m\rbrace$}\\
    The azimuthal angles of all pairs $(1,2)$ in an event can be directly used to compute a cumulant\cite{Borghini:2000sa,Borghini:2001vi}:
    \begin{equation}
        \langle\langle e^{i n (\varphi_1 - \varphi_2)} \rangle\rangle  = 
        v_n\lbrace 2\rbrace \, v_n\lbrace 2\rbrace + 
         \underbrace{ \langle e^{i n (\varphi_1 - \varphi_2)} \rangle_{\rm micro} }_{\sim O(1/N_s)}\,,
    \end{equation}
    where the brackets denote averages over all particles in an event and over event samples.
    Non-flow effects are assumed to result only from correlations between pairs of particles that both originate from the same microscopic source. If there are $N_s$ such sources in an event, the second ``non-flow" contribution is parametrically $O(1/N_s)$ smaller. The first collective contribution dominates the measured two-particle correlation if the observed signal satisfies $v_n\lbrace 2\rbrace \gsim O(1/N_s^{1/2})$.

    The cumulant method 
    may be regarded as being theoretically preferred since it has the important property that the sensitivity to ``non-flow" few-particle correlations can be systematically reduced by purely statistical arguments.
    To this end, one studies connected 
$2m$-point correlation functions (``cumulants"). 
For instance, the fourth order cumulant reads 
    \begin{eqnarray}
        \langle\langle e^{i n (\varphi_1 +\varphi_2 - \varphi_3 -\varphi_4)} \rangle\rangle_c  &\equiv& 
        \langle\langle e^{i n (\varphi_1 +\varphi_2 - \varphi_3 -\varphi_4)} \rangle\rangle
        - \langle\langle e^{i n (\varphi_1 -\varphi_3)} \rangle\rangle\,
        \langle\langle e^{i n (\varphi_2 -\varphi_4)} \rangle\rangle
         \nonumber \\
        &&       
         -\langle\langle e^{i n (\varphi_1 -\varphi_4)} \rangle\rangle\,
        \langle\langle e^{i n (\varphi_2 -\varphi_3)} \rangle\rangle
         \nonumber \\       
       &=&  - (v_n\lbrace 4\rbrace)^4 \, + 
        O\left( \frac{1}{N_s^3} \right)\, .
        \label{eq:cumulant4}
    \end{eqnarray}
    In general, cumulants are constructed such that a non-flow correlation that manifests itself only on the level of $(2m-2)$-particle correlations does not manifest itself in the $(2m)$-th cumulant. The signal strength needs to be $v_n\lbrace 2m\rbrace > O\left(1/N_s^{(2m-1)/m}\right)$ for flow contributions to dominate the signal. In the phenomenological practice, if the signal $v_n\lbrace 2m\rbrace$ stabilizes with increasing $m$, this is taken as evidence for collectivity. Mathematically, the limit $m \rightarrow \infty$ of these higher-order cumulants corresponds to the so-called LYZ (Lee-Yang-Zero) method~\cite{Bhalerao:2003xf}. The parametric separation of flow- and non-flow effects via $O\left(1/N_s^{(2m-1)/m}\right)$ power counting is powerful if $N_s$ is large, i.e. if event multiplicities are large. Therefore, the capability of the cumulant method to separate by purely statistical arguments flow from non-flow is limited for small systems.
\item {\it Subevent technique}\\
    Higher-order cumulants are computationally intense and naively scale with $O(\nch^{2m})$. Even for a 4-particle cumulant this is computationally challenging if applied to large event samples. The complexity can be reduced to $O({\nch})$ by the calculation of so-called $Q$-vectors whose expressions instead of considering $2m$ particles sum over single particles\cite{Borghini:2000sa,Bilandzic:2010jr}.
    As mentioned before, jet-like fragmentation patterns and resonance decays lead mainly to correlations between particles that are close in $\Delta\eta \times \Delta \varphi$. In order to exclude those when computing the $Q$-vectors, where an $\eta$-gap between particles cannot be applied (as only single particles are considered in the calculation), the event is split into several subevents\cite{Jia:2017hbm}, each of which is used for the computation of a $Q$-vector. The extracted coefficient is then:
    \begin{equation}
         \left(v_n\lbrace 2, \vert\Delta\eta\vert > \gamma\rbrace \right)^2 =
         \langle\langle e^{i n (\varphi_i - \varphi_j)} \rangle\rangle   \,,
    \end{equation}
    where all particles $i$ are in the range $\eta > \gamma/2$ while all particles $j$ are in the range $\eta < -\gamma/2$ with $\gamma \geq 0$.
    Experimentally, one can then test the dependence on $\gamma$ and if the signal stabilizes with increasing $\gamma$, this is taken as evidence for collectivity. It should be noted that regardless how large $\gamma$, the back-to-back jet structure is not excluded from $v_n\lbrace 2, \vert\Delta\eta\vert > \gamma\rbrace$, but can be suppressed by extending this method into calculating higher-order cumulants from more than two subevents\cite{Jia:2017hbm}.
\end{enumerate}

This enumeration demonstrates that there are numerous experimental measures for \vn coefficients. Those use different definitions with different underlying assumptions, in particular on what is the definition of the physics in absence of collectivity. Only some of the different methods have separate notation, e.g. $v_n\lbrace 2m\rbrace$. Therefore, care has to be taken when comparing different \vn between each other or with theoretical calculations.

\subsection{Hadrochemical measures of collectivity}
\label{sec22}
Hadrochemistry addresses the question with which relative abundance different hadron species are produced in a collision. What is measured are identified particle yields and their ratios, particle-identified transverse momentum spectra and particle-identified flow coefficients. The baseline on top of which collective effects are established is traditionally chosen to be set by minimum bias \pp collisions. There are some technical challenges related to the question of how feed-down contributions from resonance decays are included in or subtracted from particle-identified data\cite{Acharya:2018orn}, and how the experimental procedure is paralleled in theoretical descriptions. On the modelling side, an additional challenge arises from the incomplete knowledge of short-lived higher mass resonance states which feed down inevitably into the measured yields, see e.g. Ref.~\citenum{Andronic:2017pug}.
In general, however, what is measured is easily stated and it seems unnecessary to describe the definition of hadrochemical measures of collectivity in as much detail as we did for measures of anisotropic flow. In the following, we therefore focus almost exclusively on the physics motivations for which hadrochemical measures of collectivity are studied:

\begin{enumerate}
    \item  Hadrochemical equilibration\\
    In QCD at sufficiently high energy density, the gluon fusion process  $g g \leftrightarrow s\bar{s}$ can drive the strangeness content to chemical equilibrium. The resulting expectation of strangeness enhancement in heavy-ion collisions counts amongst the very first proposed signatures of QGP formation~\cite{Rafelski:1982pu,Koch:1986ud}. In nucleus--nucleus collisions, strangeness enhancement (compared to a baseline set by minimum bias \pp collisions) is observed in the phase-space integrated relative abundances of all identified strange and multi-strange hadrons at all relativistic center of mass energies from BNL AGS to the LHC~\cite{Braun-Munzinger:2003pwq}. These observations
    can be accounted for by fitting to a statistical hadronization model that involves two intensive quantities (temperature and a baryo-chemical potential) and one extensive one (volume)~\cite{Braun-Munzinger:2003pwq,Andronic:2017pug}. Also the relative abundances of charm and beauty hadrons can be included in the systematics of this statistical hadronization model~\cite{Andronic:2017pug,Andronic:2021erx}, if their total production rate is set by initial hard (vacuum-like) processes. That means that these rates are not thermally produced but statistically distributed.
    \item  Particle-identified flow\\
    To the extent to which collective flow is built up in the partonic phase of a collision, all hadron species emerge from the same common flow field $u_\mu$. According to the Cooper-Frye freeze-out prescription, all hadrons will decouple with statistical weights $\exp\left[-p^\mu u_\mu/T \right]$ from locally equilibrated patches of a globally expanding system~\cite{Cooper:1974mv}. In analytical blast-wave models\cite{Retiere:2003kf} and in fluid-dynamical simulations, this leads to a characteristic mass hierarchy in the $p_T$-dependent particle-identified flow coefficients $v_n(p_T)$ and in the single inclusive $p_T$ spectra (``radial flow"). These dependencies have been observed in central and semi-peripheral nucleus--nucleus collisions at the CERN SPS, at RHIC and at the LHC for sufficiently small transverse momenta (see Figure~\ref{fig:pid_vn} and related discussion). \\
    One central question is whether the onset of flow seen in the smallest pp and \pPb collision systems has the same collective dynamical origin as the flow in nucleus--nucleus collisions. Testing whether anisotropic flow in large and small collision systems shows the same PID-dependence is of obvious interest in this context. 
    \item  Hadronization mechanisms\\
    The dynamics of hadronization in QCD is not understood. In high-momentum transfer processes
    where QCD collinear factorization applies, high-$p_T$ partons hadronize the same irrespective of the hard partonic process in which they are produced. Remarkably, also the Lund string\cite{Andersson:1997xwk} or cluster hadronization models\cite{Bellm:2015jjp} implemented in multi-purpose \pp event generators are process-independent in this sense. On the other hand, the observation of significant strangeness enhancement (as well as a suite of other measurements) falsifies the assumption of a process-independent hadronization in nucleus--nucleus collisions. Hadronization in a dense nuclear environment leads to a characteristically different hadrochemical composition compared to that in the vacuum.
    One important question is whether a process-independent hadronization mechanism of bulk-particle production can also be falsified in small pp and \pPb collision systems. 
    \\
    There are process-dependent hadronization models that are sensitive to the environment in which hadronization occurs. In particular, 
    in recombination models hadrons result from combining valence-like partons that are produced within the same collision in different partonic interactions\cite{Fries:2003kq,Fries:2003vb}. 
    Signatures of recombination are normally searched for at intermediate transverse momenta ($2 \lsim p_T \lsim \unit[5]{GeV}$). The reason is that this mechanism cannot be dominant for soft bulk particle production as it reduces the number of independent degrees of freedom and thus entropy. Also, fragmentation from a high-$p_T$ power-law tail dominates over recombination from a low-$p_T$ exponential slope\cite{Fries:2003kq,Fries:2003vb}.
    One qualitative prediction of recombination is a characteristic baryon-meson splitting as a consequence of the different number of valence-like quarks that need to be combined.  
    It is of obvious interest to study whether this baryon-meson separation persists across system size. 
\end{enumerate}

\subsection{Hard processes in small systems}
\label{sec23}

In sufficiently central nucleus--nucleus collisions, essentially all high-$p_T$ hadronic final states show numerically large medium-modifications. Our understanding of these jet quenching~\cite{Connors:2017ptx} phenomena is based on mechanisms of radiational and collisional parton energy loss that strongly suppress  high-$p_T$ colored probes in sufficiently large collision systems. Jet quenching arises from interactions of jets with a medium that exhibits collective phenomena. 
In theory, there is therefore a direct relation between jet quenching and flow. If fluid dynamics or kinetic transport is the dynamical origin of flow, then there must be medium interactions with colored degrees of freedom at {\it all} momentum scales, and thus, there must be  jet quenching. 

In this context, it is remarkable that in the smallest collision systems (pp, \pPb), LHC experiments have discovered sizable anisotropic flow without finding evidence for jet quenching. Jet quenching in small systems may have escaped detection so far since effects become smaller with decreasing system size and since systematic uncertainties in measuring jet quenching become more important. If this is so, it calls for studies based on more sensitive observables and supported by more accurate calculations. 
We therefore comment here on problems related to determining the system size dependence of jet quenching in small systems.

In nucleus--nucleus collisions, one of the simplest and most widely used measures of jet quenching is the nuclear modification factor 
\begin{align}
  \left. R^{h,j}_\text{AA}(p_T,y) \right|_\text{cent}= 
 \frac{1}{\langle T_{AA} \rangle} \frac{ \left.
   \frac{1}{N_\text{ev}}\frac{dN^{h,j}_\text{AA}}{dp_T dy} \right|_\text{cent}}{ \frac{d\sigma_{pp}^{h,j}}{dp_T dy}}\, .
\label{RAA}
\end{align}
This factor compares the number $dN^{h,j}_\text{AA}$ of charged hadrons ($h$) or calorimetrically defined jets ($j$) (produced in AA collisions in some range of transverse momentum $p_T$ and rapidity $y$) to an equivalent number of such hadrons or jets produced in pp collisions. 
Here, the equivalent number of collisions $\langle N_{\rm coll}\rangle =
\sigma_{\rm pp}^{\rm inel} \, \langle T_{AA} \rangle$
is the product of the total inelastic pp cross section and the nuclear overlap function 
$\langle T_{AA} \rangle$. $R_{AA}(p_T, y)$ is typically determined as a function of centrality.  
Centrality is defined in terms of the fraction of highest multiplicity events (or highest total transverse energy events) of the total inelastic nucleus--nucleus cross section, and it is related to  $\langle T_{AA} \rangle$  by Glauber-type models~\cite{Miller:2007ri}. 

For centrality to be a good proxy of system size, event-multiplicity should be tightly correlated to the geometry of the collision and it should not be correlated with the hard process. These assumptions are satisfied approximately in sufficiently large systems but they become problematic in small systems. Estimates of the uncertainties associated to $\langle T_{AA} \rangle$ range from 3\% in central to 15\% in  peripheral \PbPb collisions~\cite{CMS:2016xef}. 
In addition, in peripheral \PbPb collisions (centralities larger than $\sim$ 70\%), Eq.~\eqref{RAA} is known to be affected by significant event selection and geometry biases~\cite{Loizides:2017sqq,ALICE:2018ekf}. The centrality dependence of hard processes in \pPb collisions reveals even larger biases. Some of these biases arise from the larger influence of fluctuations in smaller systems, other biases originate in correlations between the hard process and the global event multiplicity\cite{Adam:2014qja,Armesto:2015kwa}. Finally, the nuclear modification factor \eqref{RAA} measures jet quenching  with respect to a baseline set by pp collisions, and it is therefore not a good starting point for analyzing jet quenching in pp collisions.

These findings indicate that the measurement of single inclusive jet or hadron spectra as a function of event multiplicity (centrality) is of limited use for the characterization of jet quenching in the smallest collision systems.  
For small collision systems, it is clearly problematic to include geometrical information such as $\langle T_{AA} \rangle$ in the definition of jet quenching measures, as such information depends on soft physics models that are difficult to constrain for small systems. One way out is the use of inclusive, minimum bias spectra in light nucleus collisions for which a nuclear modification factor can be defined without reference to soft physics modeling, 
$\left. R^{h,j}_\text{AA}(p_T,y) \right|_\text{min bias} = 
 \frac{1}{A^2} \frac{d\sigma^{h,j}_\text{AA}}{dp_T dy} \Big/  \frac{d\sigma_{pp}^{h,j}}{dp_T dy}$.
It has been shown that such inclusive measurements have much smaller uncertainties and that they therefore allow for the unambiguous detection of smaller jet quenching signals in nucleus--nucleus collisions with $A$ as small as oxygen~\cite{Huss:2020dwe,Huss:2020whe}. A second general possibility of detecting signatures of jet quenching in small collision systems is the use of self-normalizing  processes such as dijet asymmetries. We recall that the very first proposal of a jet quenching effect in small p$\bar{\rm p}$ collisions made by Bjorken in 1982 was of that type~\cite{Bjorken:1982tu}. 

\section{Experimental Overview}
\label{sec3}
\subsection{Long-range correlations}
\label{sect:longrangecorr}
\begin{figure}[ht]
\includegraphics[width=0.48\textwidth,trim={10.5cm 1.1cm 0.1cm 10cm},clip]{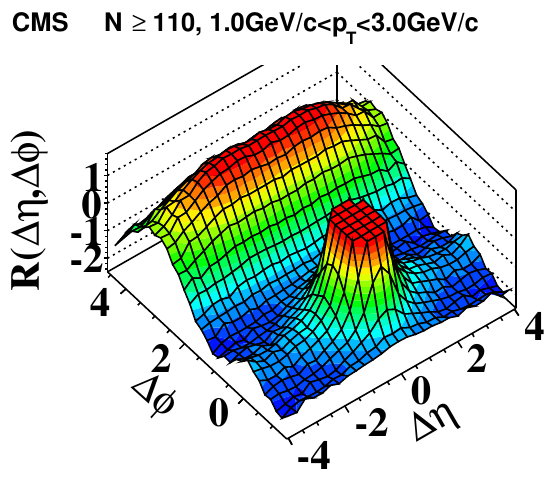}
\hfill
\includegraphics[width=0.48\textwidth]{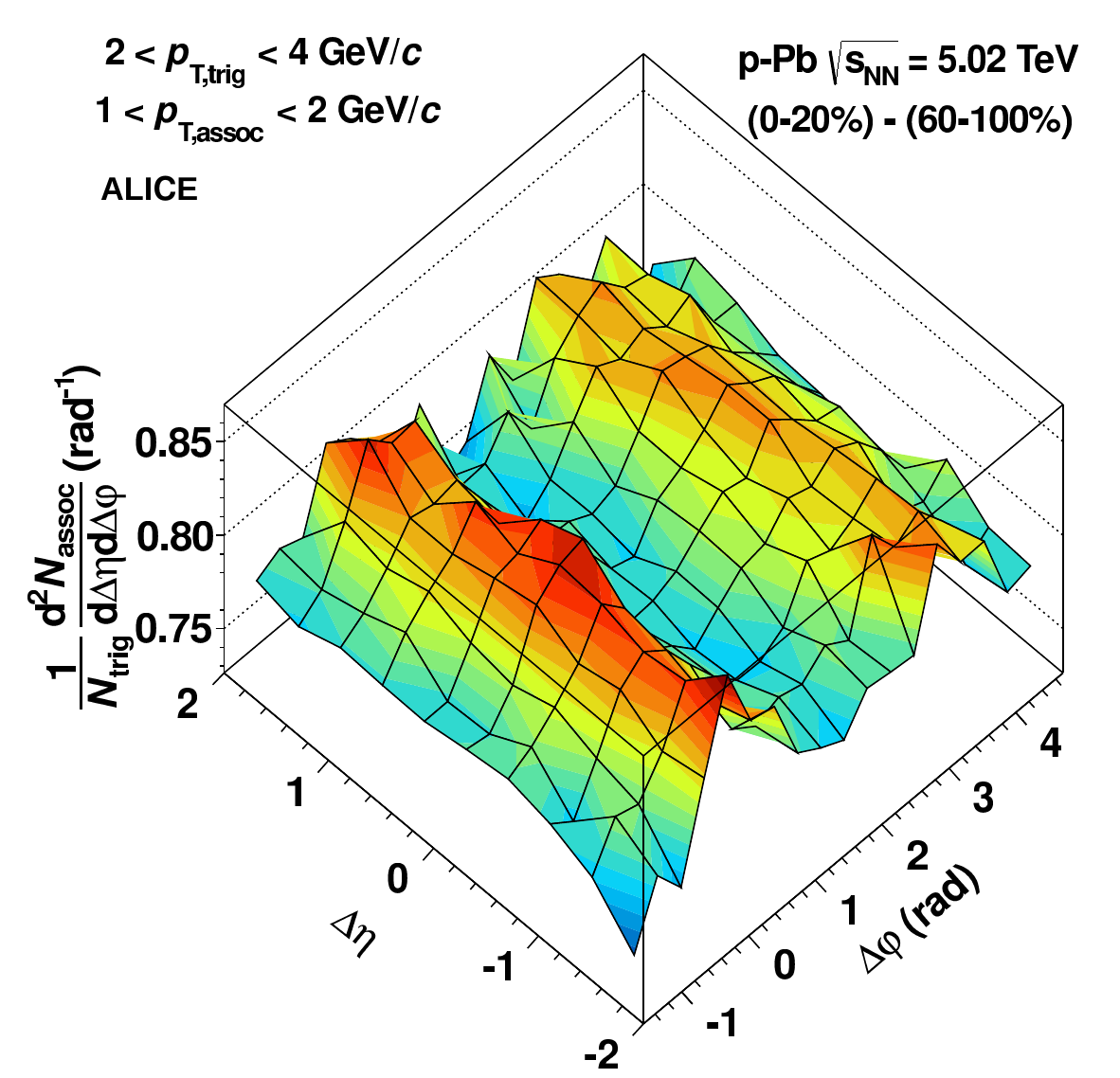}
\caption{\label{fig:ridge1} Two-particle correlation showing the first observation of the long-range correlation on the near side in pp collisions (left panel; Figure from Ref.~\citenum{Khachatryan:2010gv}) and the first observation of the double-ridge structure in \pPb collisions (right panel; Figure from Ref.~\citenum{Abelev:2012ola}).
The left panel shows the two-particle correlation without any subtraction. Hence the near-side jet peak is clearly visible and the the away side is dominated by the jet component. The right panel uses the low-multiplicity subtraction described in section~\ref{sect:methods} which effectively reduces the jet component. This results in an almost complete suppression of the near-side jet peak allowing one to observe two ridges.
The right panel is normalized as per-trigger yield, while the left panel uses a relative normalization, see Ref.~\citenum{Khachatryan:2010gv} for details.
}
\end{figure}

The most prominent and from the point of view of analysis methodology rather simple observation is the one of long-range effects in two-particle correlations~\eqref{eq:vn}. 
 QCD effects implemented in standard multi-purpose event generators of proton--proton collisions lead to distinct patterns in these correlations: At small angular distances ($\Delta\varphi \approx 0$ and $\Delta\eta \approx 0$) a so-called \textit{near-side} peak results from jet-like fragmentation and resonance decays. 
A back-to-back structure from momentum conservation is observed at $\Delta\varphi \approx \pi$, called \textit{away side}. This away-side structure is not concentrated around $\Delta\eta \approx 0$ since
the center-of-mass systems of any two partons in a hadronic collision is distributed widely in rapidity. In addition to these expected patterns, 
Fig.~\ref{fig:ridge1} (left panel) shows also positive long-range rapidity correlations for $|\Delta\eta| \gtrsim 2$ on the near side. This was surprising since the known microscopic dynamics operational in proton--proton collisions does not lead to long-range near-side correlations. Fig.~\ref{fig:ridge1}  displays the first ever observed ``collective effect" in proton--proton collisions. At the time, it was based on the first sizable LHC high-multiplicity sample although from today's perspective this sample is very small.

The second hallmark result was the observation of a double-ridge structure in \pPb collisions\cite{Aad:2012gla,Abelev:2012ola}, displayed on the right panel of Fig.~\ref{fig:ridge1}. For reasons which are detailed in section~\ref{sect:methods}, the observation in small systems (contrary to large systems) required to take the difference between high-multiplicity and low-multiplicity collisions. Fig.~\ref{fig:ridge1} shows clearly an almost rapidity-independent second harmonic $v_2$ imprinted on the azimuthal two-particle correlation. This is the telltale sign of an elliptic flow signal that had been observed in heavy-ion collisions at all fixed-target and collider energies. Elliptic flow was thought to arise from the collective dynamical response of a system to pressure gradients related to initial spatial anisotropies $\varepsilon_2$. However, such a collective effect had not been expected in the small \pPb collision system. At this moment, the scientific community was convinced that something extraordinary was going on and embarked on understanding 
if the observed effects were truly of similar nature as in heavy-ion collisions. To this end, studies focused first on characterizing higher order flow harmonics $v_n$ and their relation to finer details $\varepsilon_n$ of the initial conditions (for motivation, see the discussion of \eqref{trans}). Soon after,  the searches for high-density QCD phenomena in small systems extended to nearly all areas of heavy-ion physics, see Table~\ref{table:smallsystems}.

\afterpage{%
\begin{landscape}
\begin{table}[ht!]
\caption{Summary of observables or effects in \PbPb, \XeXe and \AuAu collisions, as well as in high multiplicity \pPb, a--A and \pp collisions. References to key measurements are given. See text for details. Table adapted from Ref.~\citenum{Citron:2018lsq} and extended by publications of the last 5 years.}
  \small
  { \begin{tabular}{p{5cm}|p{3.4cm}|p{3.1cm}|p{2.5cm}| p{3.5cm} }
    Observable or effect                             & \PbPb, \XeXe, \AuAu                              & \pPb, a--A (high \nch)   & \pp (high \nch)                         & Refs.\\
    \hline
    \hline
    Near-side ridge yields                           & yes                                              & yes                                           & yes   & \mbox{\citenum{Khachatryan:2010gv,Chatrchyan:2011eka,Aad:2012gla,Abelev:2012ola,CMS:2012qk,Chatrchyan:2012wg,Aad:2014lta,Khachatryan:2015lva,Khachatryan:2015lva,ALICE:2023ulm}} \el
    Azimuthal anisotropy                       & \vone--\vnine                                    & \vone--\vfive                                 & \vtwo--\vfour                  & \mbox{\citenum{Aamodt:2011by,Chatrchyan:2011eka,ATLAS:2012at,Aad:2012gla,Abelev:2012ola,CMS:2012qk,Chatrchyan:2012wg,Aad:2014lta,PHENIX:2014fnc,Aad:2015gqa,Khachatryan:2015lva,STAR:2015kak,Adam:2016nfo,Adam:2016ows,Khachatryan:2016txc,Aaboud:2017acw,Acharya:2017ino,Sirunyan:2017uyl,Acharya:2018zuq,Aidala:2018mcw,ALICE:2019zfl,ALICE:2020sup,STAR:2022pfn}} \el
    Weak $\eta$ dependence                           & yes                                              & yes                                           & yes                  & \mbox{\citenum{ATLAS:2011ah,Aad:2014eoa,Aaij:2015qcq,Adam:2015bka,Aaboud:2016jnr,Adam:2016ows,Khachatryan:2016ibd,Sirunyan:2017igb,ATLAS:2023rbh,ALICE:2023gyf}} \el
    Characteristic mass dependence                   & \vtwo--\vfive                                    & \vtwo, \vthree                                & \vtwo                         & \mbox{\citenum{Abelev:2012di,ABELEV:2013wsa,Abelev:2014pua,Khachatryan:2014jra,PHENIX:2014fnc,Adam:2016nfo,Khachatryan:2016txc,PHENIX:2017djs,Acharya:2018zuq,CMS:2018loe,ALICE:2022zks,CMS:2022bmk}} \el
    Higher-order cumulants \nl
    (mainly $v_{\rm 2}\{n\}$, $n\ge4$)               & \mbox{``$4\approx6\approx8\approx$ LYZ''} \mbox{+higher harmonics}&\mbox{``$4\approx6\approx8\approx$ LYZ''} \mbox{+higher harmonics}  & \mbox{``$4\approx6$''}  & \mbox{\citenum{Aamodt:2010pa,ALICE:2011ab,Chatrchyan:2012ta,Aad:2013fja,Chatrchyan:2013kba,Chatrchyan:2013nka,Aad:2014vba,Abelev:2014mda,Khachatryan:2015waa,Adam:2016izf,Khachatryan:2016txc,Aaboud:2017acw,Aaboud:2017blb,Sirunyan:2017igb,Sirunyan:2017pan,ALICE:2019zfl,CMS:2019wiy,ALICE:2022zks}} \\
    \hline \hline
    Symmetric cumulants (SC)                             & up to $(5,3)$                             & only $(4,2)$, $(3,2)$             & only $(4,2)$, $(3,2)$                                   & \mbox{\citenum{Aad:2014fla,Aad:2015lwa,ALICE:2016kpq,Acharya:2017gsw,Sirunyan:2017uyl,Aaboud:2018syf,ALICE:2019zfl,CMS:2019lin,STAR:2022vkx}}  \el
    Non-linear flow modes                            & up to \vseven                                    & not measured                                  & not measured & \mbox{\citenum{Acharya:2017zfg,CMS:2019nct,ALICE:2020sup}} \el
    Factorization breaking                           & $n=2$--$4$, $\{2\},\{4\}$               & $n=2,3$, $\{2\}$                                 & not measured                  & \mbox{\citenum{Aad:2014lta,Khachatryan:2015oea,Aaboud:2017tql,Acharya:2017ino,Sirunyan:2017gyb,ATLAS:2020sgl,ALICE:2022dtx}} \el
    Event-by-event \vn distributions                 & \vtwo--\vfour                                    & not measured                                  & not measured                  & \mbox{\citenum{Aad:2013xma,Sirunyan:2017fts,Acharya:2018lmh}} \el
    Flow--\pT correlation                            & up to \vfour                                     & \vtwo                                         & not measured & \mbox{\citenum{ATLAS:2019pvn,ALICE:2021gxt}} \el
    Directed flow (from spectators)                  & yes                                              & no                                            & no                            & \mbox{\citenum{Abelev:2013cva}} \el
    Charge-dependent correlations                    & yes                                              & yes                                           & yes                            & \mbox{\citenum{Abelev:2013csa,Adam:2015gda,Adam:2015vje,Khachatryan:2016got,Acharya:2017fau,CMS:2017pah,Sirunyan:2017quh}} \\
    \hline \hline
    Low \pT spectra (``radial flow'')                & yes                                              & yes                                           & yes                            & \mbox{\citenum{Abelev:2012wca,Chatrchyan:2012qb,Abelev:2013haa,Abelev:2013vea,Chatrchyan:2013eya,Andrei:2014vaa,Adam:2015vsf,Adam:2016bpr,Acharya:2017dmc,Adam:2017zbf,Acharya:2018orn,ALICE:2019hno}} \el
    Intermediate \pT (``recombination'')             & yes                                              & yes                                           & yes                            & \mbox{\citenum{Abelev:2013haa,Abelev:2013xaa,Abelev:2014uua,Andrei:2014vaa,Adam:2015jca,Adam:2016dau,Khachatryan:2016yru,Adam:2017zbf}} \el
    Particle ratios                                  & GC level                                         & GC level                                      & GC level                       & \mbox{\citenum{ABELEV:2013zaa,Abelev:2013haa,Abelev:2013vea,Adam:2015vsf,Adam:2016bpr,ALICE:2017jyt}} \\
    Statistical model                                & $\gamma^{\rm GC}_s=1$                            & $\gamma^{\rm GC}_s\approx1$                   & $\gamma^{\rm C}_s<1$           & \mbox{\citenum{Acharya:2018orn,ALICE:2019hno,ALICE:2020nkc,ALICE:2021lsv,ALICE:2022wpn}} \el
    HBT radii ($R(\kT)$, $R(\sqrt[3]{\nch})$)        & $R_{\rm{out}}/R_{\rm side}\approx1$              & $R_{\rm out}/R_{\rm side}\lsim1$              & $R_{\rm out}/R_{\rm side}\lsim1$       & \mbox{\citenum{Aamodt:2011kd,Abelev:2014pja,Adam:2015pya,Adam:2015vja,Adam:2015vna,Aaboud:2017xpw,Acharya:2017qtq,CMS:2023jjt,CMS:2023xyd}} \el
    Direct photons at low \pT                        & yes                                              & not measured                                  & not observed                  & \mbox{\citenum{Adam:2015lda,Acharya:2018dqe,ALICE:2023jef}} \\
    \hline \hline
    \vn in events with Z, jets                       & not measured                                     & up to \vthree                                 & \vtwo & \mbox{\citenum{ATLAS:2019vcm,ATLAS:2019wzn,ATLAS:2023bmp}} \el
    Jet constituent \vn                              & \vtwo                                            & \vtwo                                         & \vtwo in jet frame & \mbox{\citenum{ALICE:2022cwa,CMS:2023iam}} \el
    Jet quenching through \RAA                       & yes                                              & not observed                                  & not observed                  & \mbox{\citenum{Aamodt:2010jd,ALICE:2012ab,Abelev:2012hxa,CMS:2012aa,ATLAS:2014cpa,Aad:2014bxa,Aad:2015wga,Adam:2015ewa,Khachatryan:2015xaa,Adam:2016jfp,CMS:2016xef,Sirunyan:2016fcs,ALICE:2018ekf,ALICE:2018hza,ALICE:2018vuu,CMS:2018yyx,ATLAS:2022kqu,ATLAS:2023iad}} \el  
    \hspace{0.2cm}... through dijet asymmetry            & yes                                              & not observed                                  & not observed & \mbox{\citenum{Aad:2010bu,Chatrchyan:2011sx,Chatrchyan:2014hqa,Khachatryan:2016tfj,Sirunyan:2017bsd,Sirunyan:2018jju,CMS:2021nhn,ATLAS:2023xzy}} \el
    \hspace{0.2cm}... through correlations               & yes ($Z$--jet, $\gamma$--jet, h--jet)            & not obs. (h--jet, jet--h)                 & not measured                  & \mbox{\citenum{Adam:2015doa,Adam:2016xbp,Aaboud:2017eww,Acharya:2017okq,CMS:2017ehl,Sirunyan:2017jic,Sirunyan:2018jqr,Sirunyan:2018qec,ATLAS:2022iyq,ATLAS:2022zbu,ATLAS:2023iad}} \el
    \hspace{0.2cm}... through high \pT \vn and jet-\vn   & yes                                              & yes                                           & not measured & \mbox{\citenum{CMS:2012tqw,ATLAS:2019vcm,ATLAS:2021ktw,CMS:2022nsv}} \\
    \hline \hline
    Heavy flavour anisotropy                         & up to \vthree (c), up to \vtwo (b)               & up to \vtwo                                   & up to \vtwo                   & \mbox{\citenum{ALICE:2013xna,Abelev:2013lca,Abelev:2014ipa,Adam:2015pga,ALICE:2016clc,Adam:2016ssk,Khachatryan:2016ypw,Acharya:2017qps,Acharya:2017tfn,Acharya:2017tgv,Sirunyan:2017plt,Acharya:2018dxy,CMS:2018loe,ALICE:2020pvw,ATLAS:2020yxw,CMS:2020qul,CMS:2021qqk,ALICE:2022ruh,CMS:2022vfn,CMS:2023mtk,CMS:2023dse,ALICE:2019pox,ATLAS:2019xqc,CMS:2020efs}} \el
    Quarkonia production                             & suppressed                                       & suppressed                                    & not measured                  & \mbox{\citenum{Abelev:2012rv,CMS:2012bms,Chatrchyan:2012lxa,ALICE:2013snh,Chatrchyan:2013nza,ALICE:2014ict,Abelev:2014zpa,ALICE:2015sru,Adam:2015isa,Adam:2015jsa,Adam:2015rba,Adam:2016ohd,Adam:2016rdg,Khachatryan:2016xxp,Khachatryan:2016ypw,Sirunyan:2016znt,Aaboud:2017cif,Aaij:2017cqq,Adamova:2017uhu,CMS:2017uuv,Sirunyan:2017lzi,Sirunyan:2017mzd,ALICE:2018mml,ALICE:2018szk,ALICE:2018wzm,ATLAS:2018hqe,CMS:2018zza,ALICE:2019qie,STAR:2019fge,ALICE:2020wwx,ALICE:2022jeh,ALICE:2022zig,ATLAS:2022exb,CMS:2022wfi,ALICE:2023gco,ALICE:2023hou,CMS:2023lfu}} \\ 
    \hline
    \hline
  \end{tabular} \label{table:smallsystems} }
\end{table}
\end{landscape}
}

\begin{figure}[ht]
\begin{minipage}[t]{.49\textwidth}
\includegraphics[width=\textwidth]{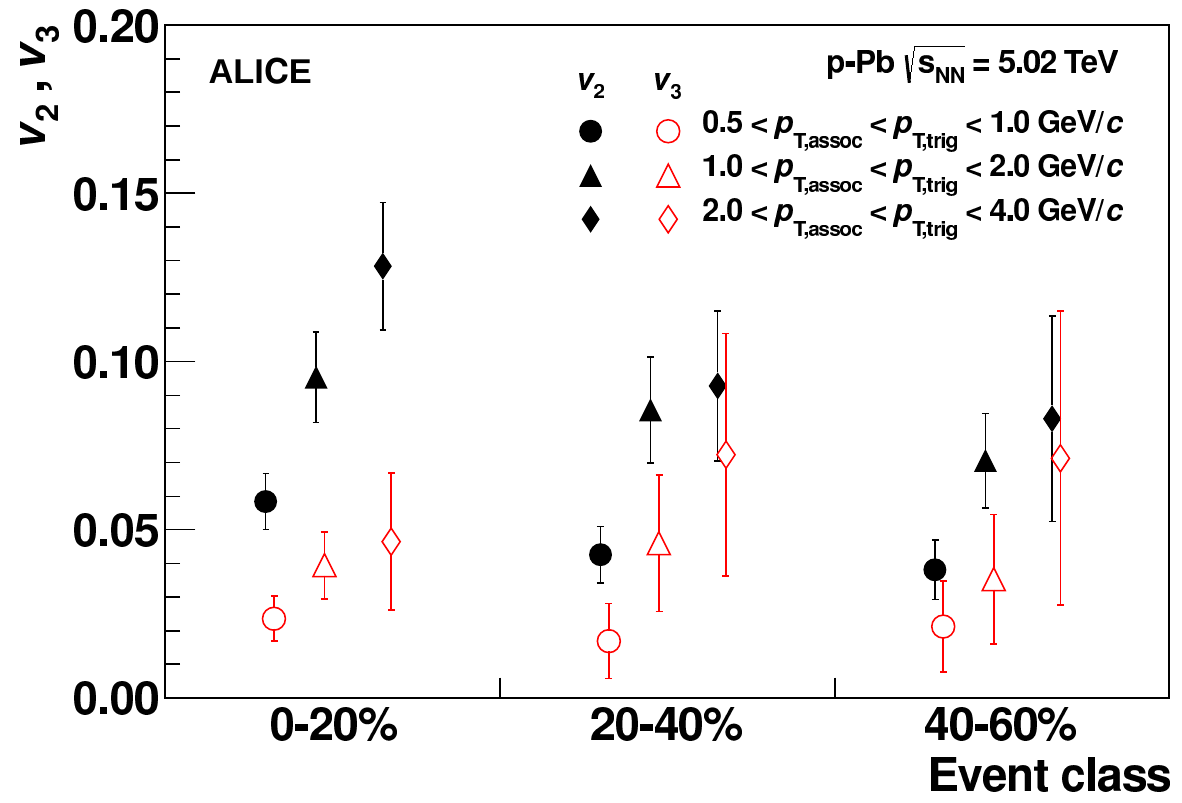}
\caption{\label{alice_vn} \vtwo and \vthree coefficients extracted from low-multiplicity subtracted two-particle correlations in \pPb collisions (Figure from Ref.~\citenum{Abelev:2012ola}). For details on the methodology see Sec.~\ref{sect:methods}.}
\end{minipage}
\hfill
\begin{minipage}[t]{.49\textwidth}
\includegraphics[width=\textwidth]{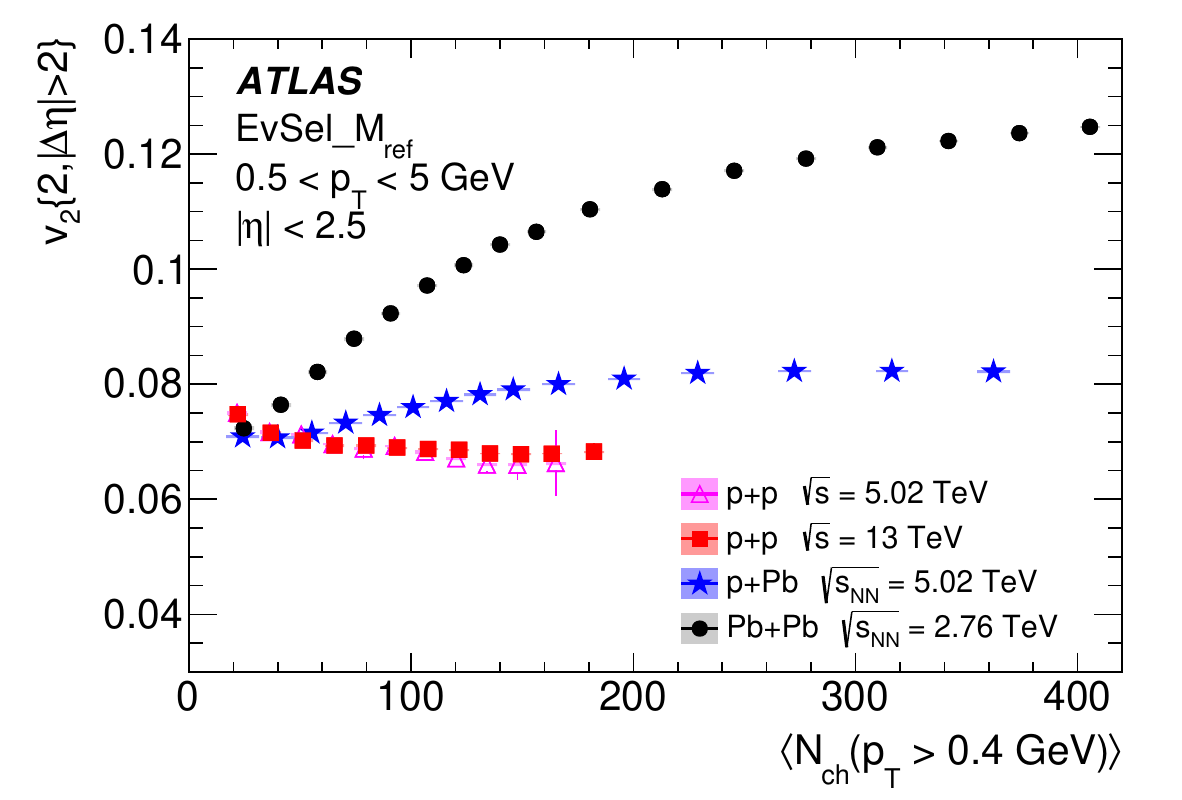}
\caption{\label{atlas_vn} The \vtwo coefficient extracted with the template method in pp, \pPb and \PbPb collisions (Figure from Ref.~\citenum{Aaboud:2017acw}). For details on the methodology see Sec.~\ref{sect:methods}.}
\end{minipage}
\end{figure}

\begin{figure}[ht]
\includegraphics[width=\textwidth]{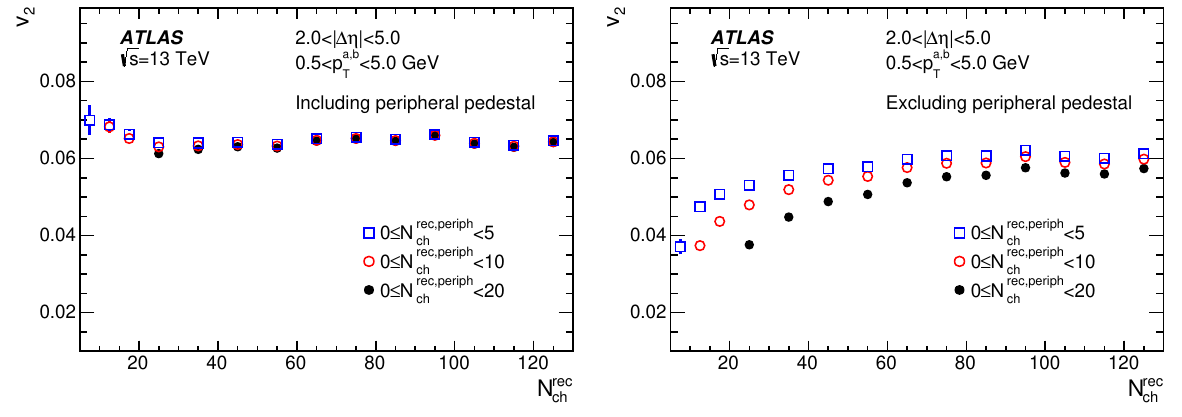}
\caption{\label{fig:atlas_methods} The \vtwo coefficient extracted with the template method (left panel) and the low-multiplicity subtraction method (right panel). The overestimation with the template method and underestimation with the low-multiplicity subtraction method are clearly visible. Figure from Ref.~\cite{Aad:2015gqa}. For details on the methodology see Sec.~\ref{sect:methods}, in particular Fig.~\ref{fig:methods}}
\end{figure}

\textbf{Azimuthal anisotropies}
The observed long-range structures are quantified by decomposing the per-trigger yield into Fourier coefficients $v_n$, see Eq.~\ref{eq:vn}.
The first measurement of non-zero \vtwo and \vthree in \pPb collisions, shown in Fig.~\ref{alice_vn}, provided another indication that the phenomena seen in small systems are heavy-ion like. Since then, a large number of detailed studies have been done on different coefficients of anisotropic flow. 
Today, measurements extend up to \vnine in \PbPb, \vfive in \pPb, \vthree in \dAu, and \vfour in \pp collisions for charged particles. 
Fig.~\ref{atlas_vn} shows the state-of-the art measurement of \vtwo in \pp, \pPb and \PbPb collisions. These coefficients exhibit a weaker multiplicity dependence in \pp and \pPb collisions than in \PbPb collisions where this is closely related to the shape of the overlap region, while \vthree are independent of the collision system as a function of multiplicity\cite{Aamodt:2011by,Chatrchyan:2011eka,ATLAS:2012at,Aad:2012gla,Abelev:2012ola,CMS:2012qk,Chatrchyan:2012wg,Aad:2014lta,PHENIX:2014fnc,Aad:2015gqa,Khachatryan:2015lva,STAR:2015kak,Adam:2016nfo,Adam:2016ows,Khachatryan:2016txc,Aaboud:2017acw,Acharya:2017ino,Sirunyan:2017uyl,Acharya:2018zuq,ALICE:2019zfl,ALICE:2020sup}.
The slight trend of increasing \vtwo for decreasing multiplicity in pp collisions is related to the extraction method used, see section~\ref{sect:methods}. The sensitivity to the method becomes clear when both methods are applied to the same dataset, which is shown in Fig.~\ref{fig:atlas_methods}. The $\eta$ dependence is overall weak and similar in small and large systems~\cite{ATLAS:2011ah,Aad:2014eoa,Aaij:2015qcq,Adam:2015bka,Aaboud:2016jnr,Adam:2016ows,Khachatryan:2016ibd,Sirunyan:2017igb,ATLAS:2023rbh,ALICE:2023gyf}.

\begin{figure}[ht]
\begin{minipage}[t]{.49\textwidth}
\includegraphics[width=\textwidth]{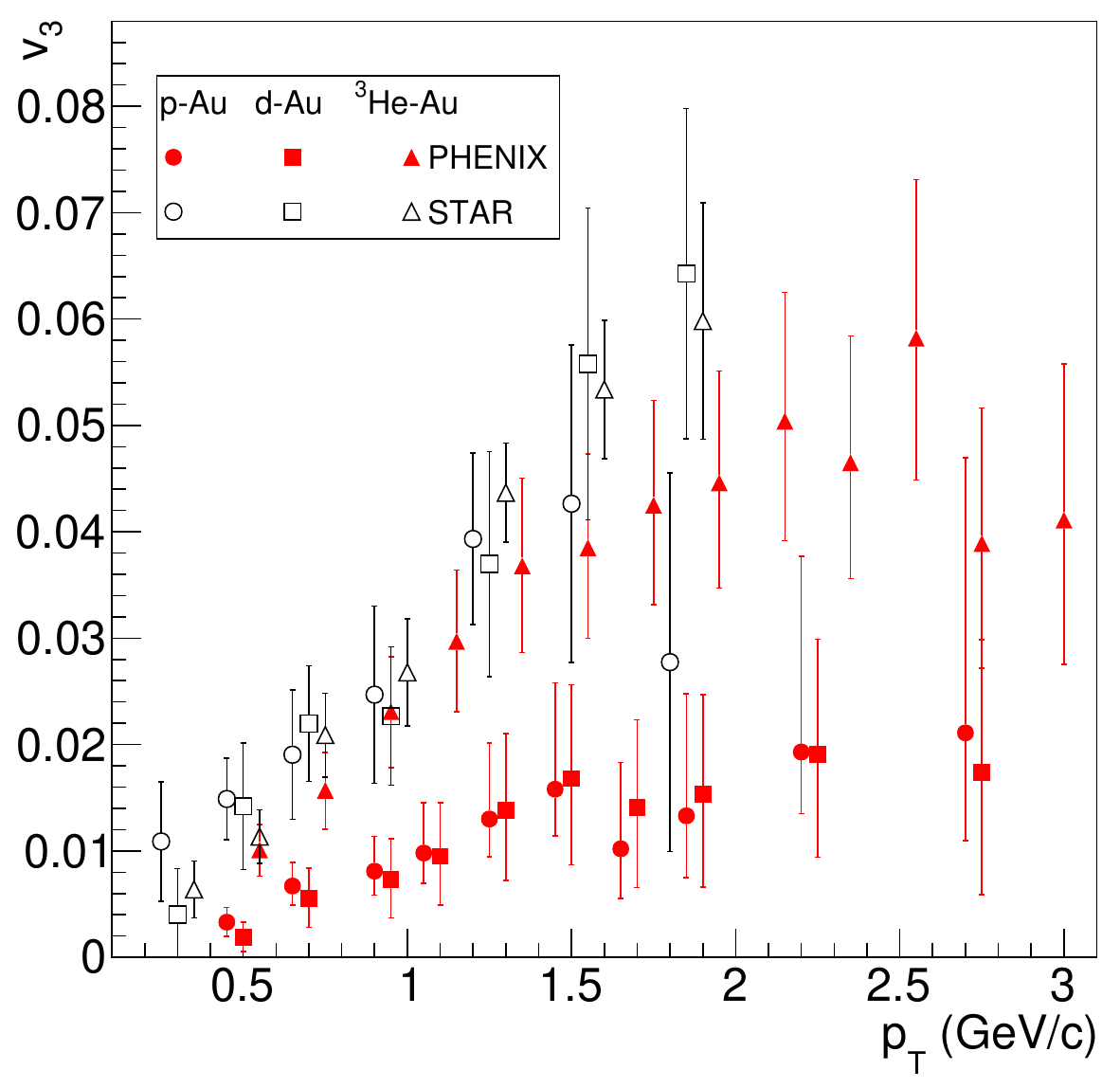}
\caption{\label{fig:rhic} Comparison of the \vthree measurement of PHENIX\cite{Aidala:2018mcw} (red filled symbols) and STAR\cite{STAR:2022pfn} (black open symbols) in three different collision systems which favor round (\pAu), elliptical (\dAu) and triangular (\HeAu) shapes in the initial state.}
\end{minipage}
\hfill
\begin{minipage}[t]{.49\textwidth}
\includegraphics[width=\textwidth,trim={0cm 0cm 12.6cm 0cm},clip]{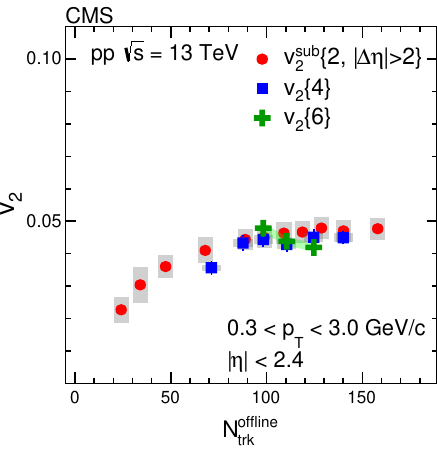}
\caption{\label{fig:vn_cumulants} Higher-order correlations of the \vtwo coefficient measured in pp collisions (Figure adapted from Ref.~\citenum{Khachatryan:2016txc}).}
\end{minipage}
\end{figure}   

At RHIC, the influence of the initial geometry has been studied with special collision configurations that favor round (\pAu), elliptical (\dAu) and triangular (\HeAu) shapes in the initial state. Fig.~\ref{fig:rhic} compares the $v_3$ coefficients measured by the PHENIX\cite{Aidala:2018mcw} and STAR\cite{STAR:2022pfn} collaborations. The larger signal strength in \HeAu as compared to \pAu and \dAu observed by PHENIX\cite{Aidala:2018mcw} supports the conclusion that a larger triangular eccentricity leads to a larger triangular \vthree also in these small systems.
However, this conclusion is not supported by the STAR\cite{STAR:2022pfn} data which differ from those of PHENIX by up to a factor 3 in \pAu and \dAu, while they are consistent for \HeAu. In model studies, half of this difference could be attributed to the different methods and rapidity ranges\footnote{PHENIX uses the event-plane method with detectors at $-3.9 < \eta < -3.1$ and $|\eta| < 0.35$ while STAR uses the template method with both particles within $|\eta| < 0.9$ and an $\eta$ gap of 1 unit.} used in the extraction methods\cite{Zhao:2022ugy} but this does not reconcile the different data. The discrepancy is of a qualitative, not only a quantitative nature. To draw lasting physics conclusions from the unique RHIC opportunity of studying these different small collision systems, a resolution of the apparent inconsistencies between STAR and PHENIX remains important.

\textbf{Multi-particle nature of the observed correlation} In heavy-ion collisions, collective phenomena are a correlation between all particles (through common symmetry planes) and therefore correlations are in principle present at all orders. To test this assumption, higher-order cumulants (see Eq.~\eqref{eq:cumulant4} and following text) have been measured using the Lee-Yang Zeros (LYZ) method and multi-particle correlation techniques with up to 8 particles in both \PbPb and \pPb collisions and up to 6 particles in \pp collisions (shown in Fig.~\ref{fig:vn_cumulants}) for \vtwo\cite{Aamodt:2010pa,ALICE:2011ab,Chatrchyan:2012ta,Aad:2013fja,Chatrchyan:2013kba,Chatrchyan:2013nka,Aad:2014vba,Abelev:2014mda,Khachatryan:2015waa,Adam:2016izf,Khachatryan:2016txc,Aaboud:2017acw,Aaboud:2017blb,Sirunyan:2017igb,Sirunyan:2017pan,ALICE:2019zfl,CMS:2019wiy}. 
The measurements of the cumulants at different orders ($n \geq 4$) are similar within 10\% within each collision system. This is a tell-tale sign of collectivity, but it does not identify unambiguously the dynamical mechanism underlying collectivity. From fluid dynamics (see section~\ref{sec411}) to one-hit kinetic theory (see Section~\ref{sec412}), very different dynamical frameworks can lead to correlations amongst essentially all particles in an event. 

\begin{figure}[ht]
\includegraphics[width=0.49\textwidth]{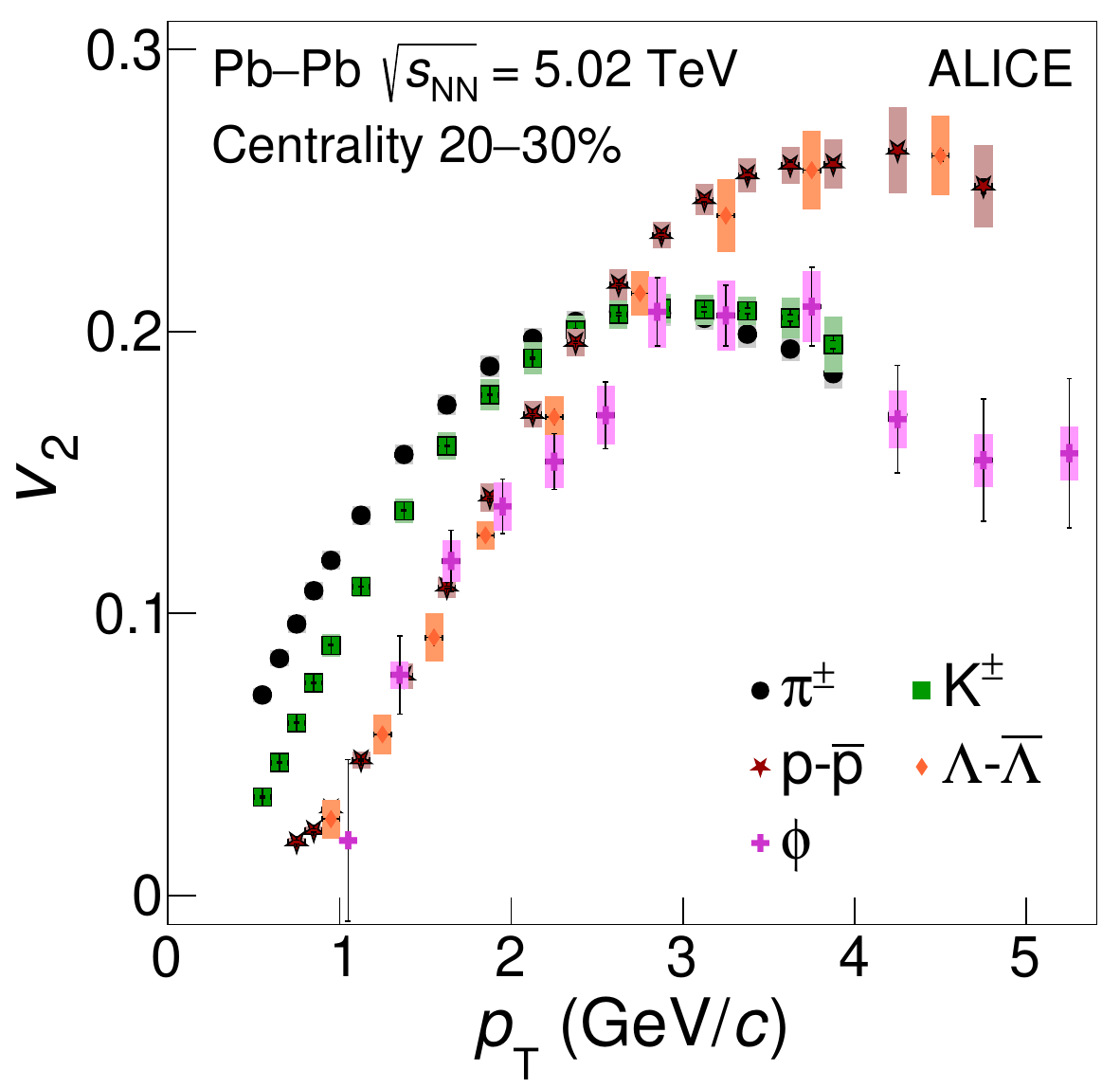}
\hfill
\includegraphics[width=0.49\textwidth]{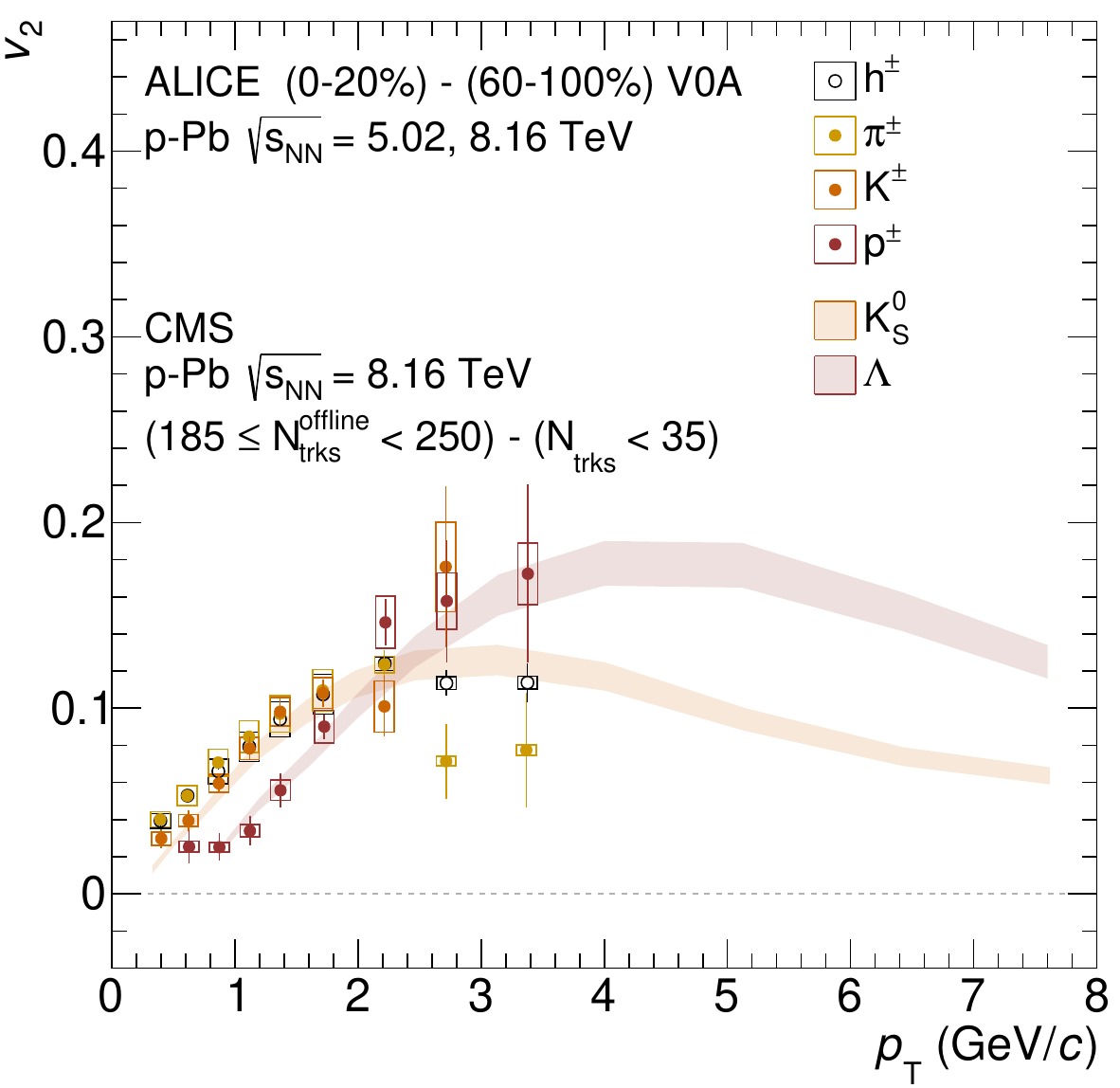}
\caption{\label{fig:pid_vn} Compilation of identified-particle \vtwo in \PbPb collisions (left panel) compared to \pPb collisions (right panel). Figures from Ref.~\citenum{ALICE:2022wpn}).}
\end{figure}

\textbf{Particle-mass dependence}
Studying the \vn coefficients as a function of \pT for different particle species shows a characteristic mass dependent pattern up to \vfive in \PbPb (left panel of Fig.~\ref{fig:pid_vn}), \vthree in \pPb (right panel of Fig.~\ref{fig:pid_vn}) and other small systems, as well as \vtwo in \pp collisions where heavier particles are depleted at low \pT~\cite{Abelev:2012di,ABELEV:2013wsa,Abelev:2014pua,Khachatryan:2014jra,PHENIX:2014fnc,Adam:2016nfo,Khachatryan:2016txc,PHENIX:2017djs,Acharya:2018zuq,CMS:2018loe,CMS:2022bmk}. 
If all hadron species were emitted with thermal weight $\exp\left[-p_\mu u^\mu/T\right]$ from the same common flow field $u^\mu$, then this would lead to a characteristic $m_T = \sqrt{m^2 + p_T^2}$-scaling of radial and anisotropic flow (see section~\ref{sec22}). If, on the other hand, hadron species would result from the recombination of valence-like quarks, then a baryon-meson grouping of observables may be expected. We emphasize that $m_T$-scaling and baryon-meson grouping are distinct dependencies that cannot be realized simultaneously. Existing data show approximate $m_T$-scaling at low and approximate meson-baryon grouping at intermediate $p_T$. Remarkably, essentially the same hadrochemical and $p_T$-dependencies are observed across all systems from \PbPb to \pPb to \pp although both mechanisms may be expected to depend on system size or phase space density.

The extent to which signatures of collectivity show close commonalities between small and large collision systems had not been expected. Beyond the early findings described above, commonalities have been found by now in a large number of refined measurements listed in  Table~\ref{table:smallsystems}. In the following, we summarize the main experimental conclusions as follows:

\textbf{Non-linear response} Correlations between different anisotropic flow harmonics characterize the non-linear coupling between different collective excitations. 
In the modelling of heavy-ion collisions, non-linear mode-mode coupling is a generic feature of fluid dynamics\cite{Teaney:2010vd,Teaney:2012ke,Floerchinger:2013tya,Noronha-Hostler:2015dbi} where it is found to be sensitive to transport properties. However, non-linear mode-mode coupling is also a generic feature of kinetic transport theory where it persists even in the dilute one-hit approximation~\cite{Borghini:2018xum,Kurkela:2018ygx}, see section~\ref{sec412} for details. Mixed harmonics or so-called symmetric cumulants have been measured up to $\rm{SC}(5,3)$ in \PbPb and $\rm{SC}(4,2)$ in \AuAu, \XeXe, \pPb and \pp collisions~\cite{Aad:2014fla,Aad:2015lwa,ALICE:2016kpq,Acharya:2017gsw,Sirunyan:2017uyl,Aaboud:2018syf,ALICE:2019zfl,CMS:2019lin,STAR:2022vkx}. A further class of observables, linear ($\vn^{\rm{L}}$) and non-linear ($\vn^{\rm{NL}}$) flow modes, have been investigated in \PbPb collisions up to the seventh order flow harmonic~\cite{Acharya:2017zfg,CMS:2019nct,ALICE:2020sup} but not yet in \pp and \pPb collisions. In general, the measured non-linear response varies only mildly with system size.

\textbf{More detailed $v_n$-characterizations} 
Numerous other characterizations constrain the phenomenology of $v_n$. The factorization of long-range azimuthal two-particle correlations into a product of single particle anisotropies and the breaking of this factorization due to event-plane angle decorrelations in \pT and $\eta$ has been measured in both \PbPb and \pPb collisions~\cite{Khachatryan:2015oea,Acharya:2017ino,Sirunyan:2017gyb,ATLAS:2020sgl}. Recent measurements show this behavior using four-particle correlators~\cite{ALICE:2022dtx}. With the existing data such measurements are not yet possible in \pp collisions. It would also be interesting to study event-by-event distributions of \vn coefficients in both \pPb and \pp collisions, which have so far been done in \PbPb collisions~\cite{Aad:2013xma,Sirunyan:2017fts,Acharya:2018lmh}.
The interplay of flow harmonics and radial expansion is studied with correlations of \vn coefficients and the mean \pT event-per-event. 
The correlation has been measured through modified Pearson correlation coefficient as a function of event activity up to fourth order in \PbPb, third order in \XeXe collisions and second order in \pPb collisions~\cite{ATLAS:2019pvn,ALICE:2021gxt}. First measurements of higher-order correlations, e.g. \vtwo--\vthree--\pT have been made~\cite{ALICE:2021gxt}. 

\textbf{Directed flow} In \PbPb collisions, directed flow
of charged particles at mid-rapidity was measured relative to the collision symmetry plane defined by the spectator nucleons, and evidence for dipole-like initial-state density fluctuations in the overlap region was found\cite{Abelev:2013cva}. In small systems, the concept of directed flow is less clear, especially in \pp collisions. If there is collectivity in \pp collisions, one could also expect a non-zero directed flow measurement. This is technically challenging since the measurement of the spectator plane is not feasible in small systems and, hence, \vone could only be measured using higher-order ($n\geq 4$) cumulants which has not been achieved to date.

\textbf{Charge-dependent correlations} The balance function probes the charge creation time and the development of collectivity in the produced system. Its width, $\langle\Delta\eta\rangle$ and $\langle\Delta\varphi\rangle$, has been measured for charged particles in \pp, \pPb and \PbPb collisions\cite{Abelev:2013csa,Adam:2015gda}. 
A picture emerges where the system exhibits larger radial flow with increasing multiplicity but also whose charges are created at the later stages of the collision. Charge-dependent azimuthal correlations are measured in both \PbPb and \pPb collisions\cite{Aamodt:2011kd,Abelev:2014pja,Adam:2015pya,Aaboud:2017xpw}. 
These studies assess the chiral magnetic effect (CME) and the chiral magnetic wave (CMW) on the produced particles. Their interpretation is still today influenced by strong background contributions, for example from local charge conservation and possibly radial and anisotropic flow.

\subsection{Bulk properties}
\begin{figure}[ht]
\centering
\includegraphics[width=0.5\textwidth]{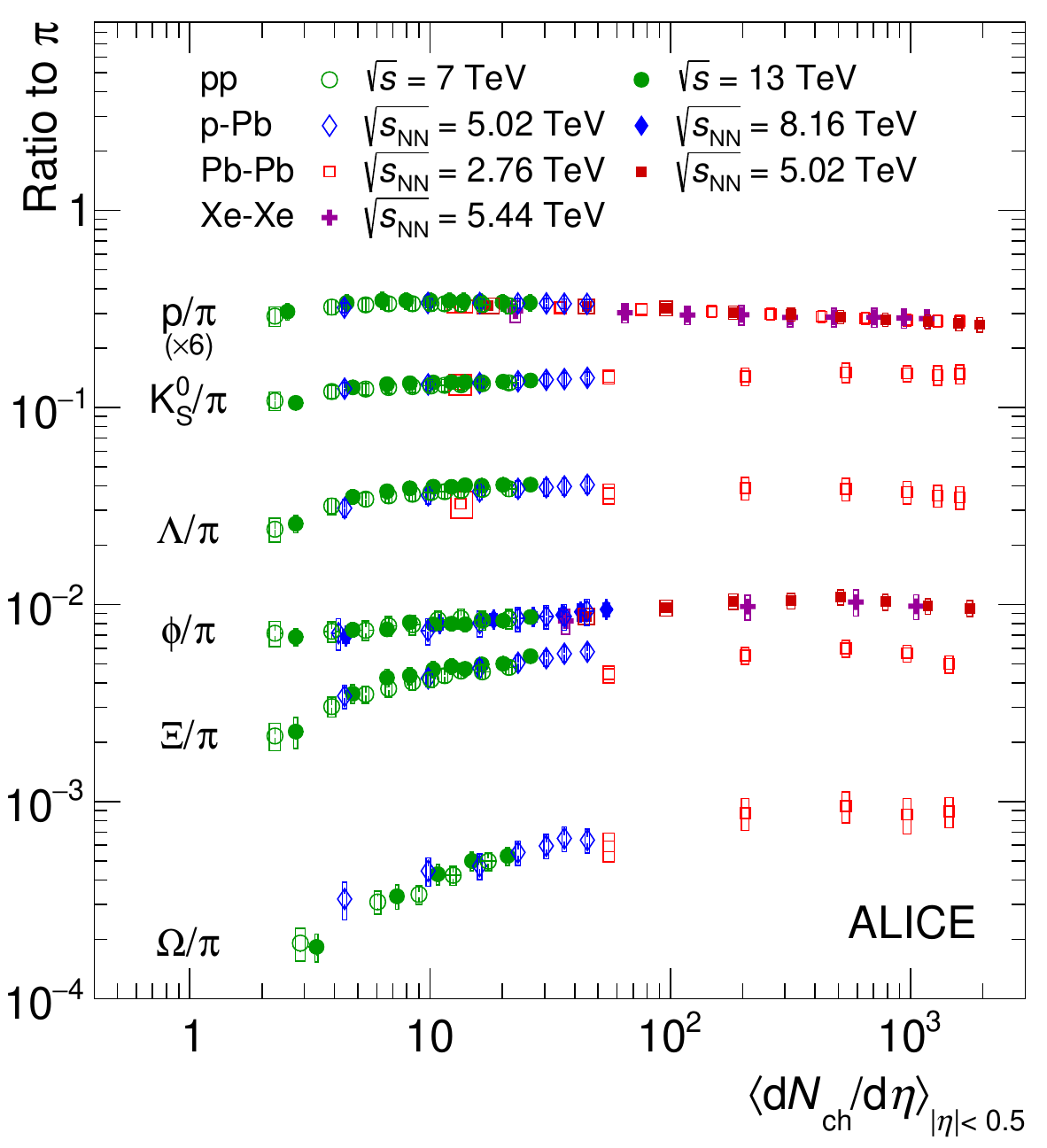}
\caption{\label{fig:strangeness} Particle production yields of strange and non-strange particles normalized to the pion yield as a function of multiplicity in pp, \pPb, \PbPb and \XeXe collisions (Figure from Ref.~\citenum{ALICE:2022wpn}).}
\end{figure}

\textbf{Particle spectra and production}
Particle production yields have been measured for numerous particle species in small and large collision systems\cite{ABELEV:2013zaa,Abelev:2013haa,Abelev:2013vea,Adam:2015vsf,Adam:2016bpr,ALICE:2017jyt,Acharya:2018orn,ALICE:2019hno,ALICE:2020nkc,ALICE:2021lsv}.
Figure~\ref{fig:strangeness} shows the production yields of different identified non-strange, strange and multiple-strange hadrons normalized to the pion yield. As a function of multiplicity, these ratios increase more for strange than for non-strange hadrons, and they increase more for multi-strange hadrons than for strange hadrons~\cite{ALICE:2017jyt,Acharya:2018orn,ALICE:2020nkc}. That strangeness in nucleus--nucleus collisions is enhanced compared to minimum-bias proton--proton collisions has been observed at all relativistic collision energies at the BNL AGS, the CERN SPS, the RHIC\cite{Braun-Munzinger:2003pwq} and now the LHC. Up until today, the factor 25 enhancement of the $\Omega/\pi$ ratio observed by the NA57 Collaboration at the CERN SPS in comparing yields in \PbPb and \pPb collisions is arguably one of the numerically largest medium-modification ever observed in heavy-ion collisions~\cite{WA97:1999uwz}. The qualitatively novel information of Figure~\ref{fig:strangeness} is that this strangeness enhancement sets in smoothly with increasing multiplicity so that even \pp collisions show an enhanced strangeness content for higher multiplicity, and \pPb collisions must not be considered as a baseline in which strangeness enhancement is absent. Moreover, strangeness enhancement shows a universal dependence on event multiplicity irrespective of the collision system. This is in marked contrast to kinetic signatures of collectivity such as elliptic flow (see e.g. Fig.~\ref{atlas_vn}) that depend not only on event multiplicity as they contain also spatial information about the collision systems and their centrality dependence.     

The \pT spectra of identified particles harden with increasing multiplicity for all studied collision systems. The hardening is stronger for heavier 
particles\cite{Abelev:2013haa,Abelev:2013xaa,Abelev:2014uua,Andrei:2014vaa,Adam:2015jca,Adam:2016dau,Khachatryan:2016yru,Adam:2017zbf}. In central nucleus--nucleus collisions, the slopes
of soft charged pion, kaon and (anti-)proton \pT spectra are consistent with models in which all hadron spectra are blue-shifted by the same transverse radial flow~\cite{Abelev:2012wca}.  In this framework, a larger radial flow is observed\cite{Acharya:2018orn} in \pp and \pPb collisions as compared to \PbPb collisions at the same multiplicity.  

\textbf{Source size} The measurement of quantum-statistic correlations between pairs of same-charge particles in particular pions at low-momentum transfer allows to assess the size of the emitting source~\cite{Lisa:2005dd}. These freeze-out radii are measured in three orthogonal directions (``out", ``side", ``long"). While clearly the expectation is that these are vastly different in small and large collision systems, their dependence on event and particle pair variables can be compared. The radii scale with $\sqrt[3]{\nch}$ in all collision systems indicating a constant density at freeze-out and decrease with increasing pair momentum \kT. This is consistent with the expectation that velocity gradients in the collision region should lead to smaller ``homogeneity regions" with increasing \kT~\cite{Lisa:2005dd}. The size along the emission direction is similar to the geometric size of the system ($R_{\rm out}/R_{\rm side} \approx 1$) in \PbPb collisions\cite{Abelev:2014pja,Adam:2015vja,Adam:2015vna,Acharya:2017qtq,Acharya:2017qtq,CMS:2023jjt,CMS:2023xyd} while this ratio is smaller than unity for both \pPb and \pp collisions\cite{Aamodt:2011kd,Abelev:2014pja,Adam:2015pya,Aaboud:2017xpw}.

\textbf{Direct photons}
Direct-photon measurements have been performed in \PbPb and \pp collisions in the \pT region sensitive to thermal photon production. For central \PbPb collisions, an effective temperature (averaged over the evolution) of about \unit[300]{\UMeV} has been extracted from the slope of the photon transverse momentum spectrum~\cite{Adam:2015lda,ALICE:2023jef}.  Fluid models that reproduce these spectra start from an initial QGP temperature of more than \unit[400]{\UMeV}\cite{Adam:2015lda} in central \PbPb collisions. 
In \pp collisions, no significant direct-photon signal has been identified at low momentum\cite{Acharya:2018dqe}. The reported upper bounds lie well above expectations of next-to-leading order QCD. A significant further increase in experimental accuracy would be needed to gain access to a possible thermal component.

\subsection{Hard probes}
\begin{figure}[ht]
\includegraphics[width=0.49\textwidth]{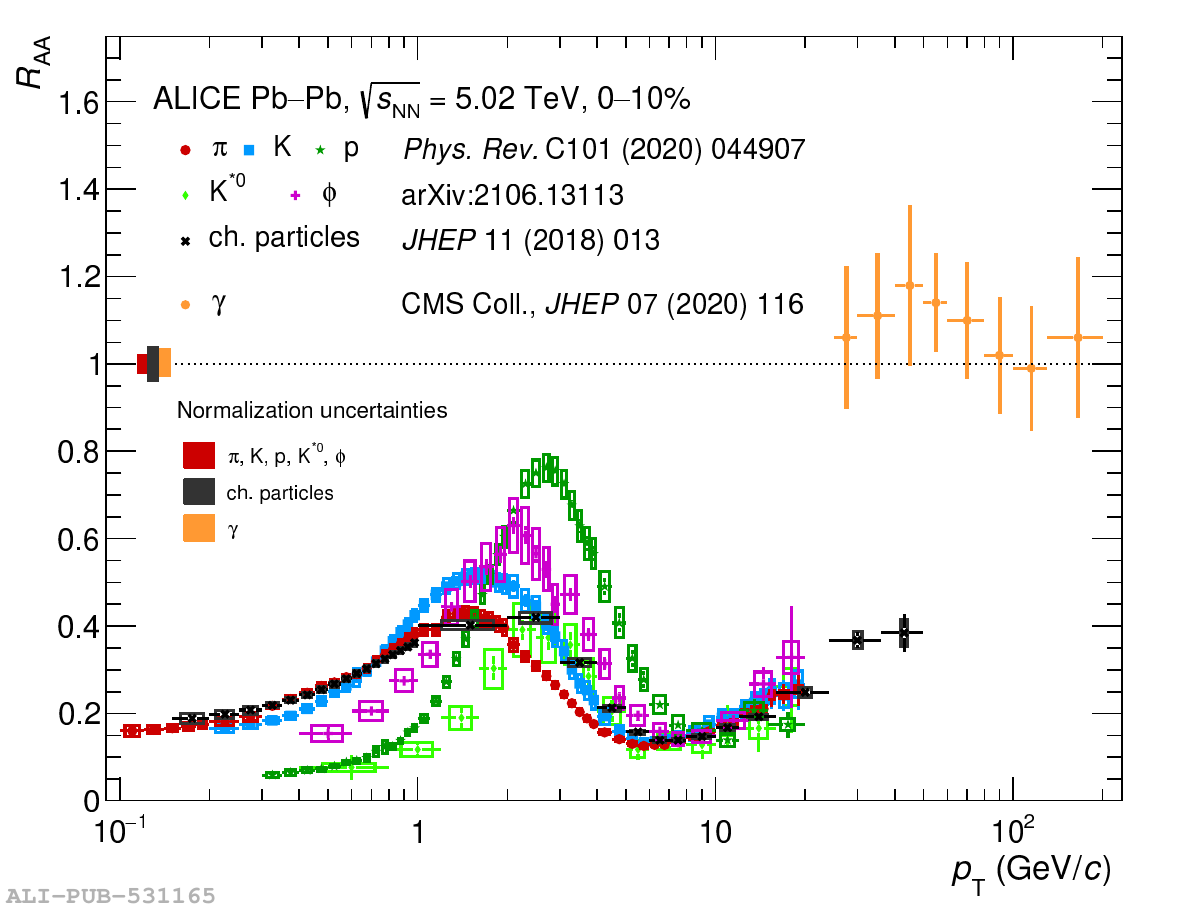}
\hfill
\includegraphics[width=0.49\textwidth]{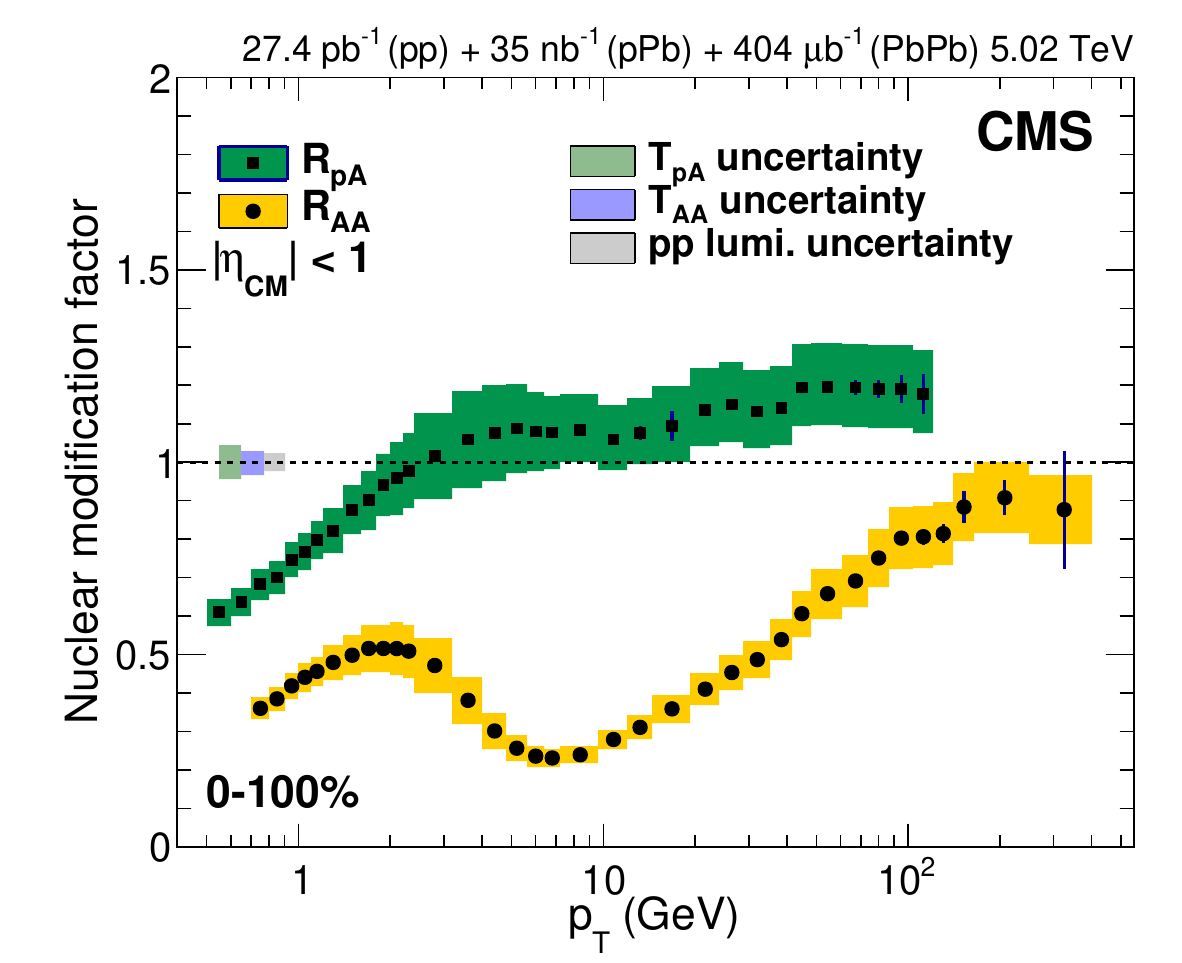}
\caption{\label{fig:raa} Left panel: \RAA in \PbPb collisions for charged particles as well as $\pi$, K, p and $\phi$ compared to $\gamma$ (Figure from Ref.~\citenum{ALICE:2022wpn}). Right panel: \RAA in \PbPb collisions for charged particles up to high \pT compared to the result in \pPb collisions (Figure from Ref.~\citenum{CMS:2016xef}).}
\end{figure}

\textbf{Parton energy loss}
In nucleus--nucleus collisions, jet quenching and its centrality dependence has been quantified through measurements of the nuclear modification factor $\RAA$ \eqref{RAA} 
~\cite{Aamodt:2010jd,ALICE:2012ab,Abelev:2012hxa,CMS:2012aa,Aad:2014bxa,Aad:2015wga,Adam:2015ewa,CMS:2016xef,ALICE:2018ekf,ALICE:2018hza,ALICE:2018vuu,CMS:2018yyx,ATLAS:2022kqu,ATLAS:2023iad} as well as correlation measurements\cite{Adam:2015doa,Adam:2016xbp,Aaboud:2017eww,CMS:2017ehl,Sirunyan:2017jic,Sirunyan:2018jqr,Sirunyan:2018qec,ATLAS:2022zbu,ATLAS:2023iad}. 
Fig.~\ref{fig:raa} (left panel) shows a strong suppression (\RAA as small as 0.13) in central \PbPb collisions for charged particles and various identified particles while the \RAA for $\gamma$ is consistent with unity.
Also, consistent with the qualitative expectations of parton energy loss, 
a large asymmetry in back-to-back jet \pT accompanied by slightly modified jet fragmentation functions inside small jet cone sizes ($\rm{R} = 0.4$) has been observed. The radiated energy appears mostly at large angles ($\rm{R} > 0.8$)\cite{Aad:2010bu,Chatrchyan:2011sx,Khachatryan:2016tfj,Sirunyan:2017bsd,Sirunyan:2018jju,CMS:2021nhn,ATLAS:2023xzy}. Furthermore, in semi-peripheral collisions, reduced but non-zero \vn coefficients have been found to persist up to the highest transverse momenta~\cite{CMS:2012tqw,ATLAS:2021ktw} and di-jet events have been found correlated with elliptic flow ~\cite{CMS:2022nsv}. These data are interpreted in terms of path-length dependent parton energy loss which in azimuthally asymmetric collision region leads to an aziumthal dependence of high-$p_T$ hadron yields. They illustrate the connection between jet quenching and flow emphasized in section~\ref{sec23}.

Early data on \RAA in \PbPb collisions 
show that a significant suppression $\RAA < 1$ persists even in the most peripheral event classes. By now, this phenomenon is understood as an intrinsic selection bias (events with less activity are biased towards larger nucleon--nucleon impact parameter which in turn have less high-$Q^2$ scatterings)~\cite{ALICE:2018ekf}. There is only a tight upper bound but no direct evidence for parton energy loss in the most peripheral \PbPb collisions~\cite{ALICE:2018ekf,ALICE:2022qxg}.

In \pPb collisions, measurements of inclusive high-\pT\ hadron and inclusive jet yields in minimum-bias \pPb collisions at the LHC are within the current accuracy of approximately 20\% consistent with $\RpPb = 1$\cite{ATLAS:2014cpa,Khachatryan:2015xaa,Adam:2016jfp,CMS:2016xef,Sirunyan:2016fcs,ALICE:2018vuu,ATLAS:2022kqu}, see Fig.~\ref{fig:raa} (right panel). As explained in section~\ref{sec23}, in small systems, the binning of \RAA and other measures of parton energy loss as a function of event activity (centrality) does not necessarily inform us about the system size dependence of jet quenching. In particular, inclusive jet yields \RpPb are strongly suppressed relative to unity in ``central'' \pPb collisions, and strongly enhanced in ``peripheral'' \pPb collisions~\cite{ATLAS:2014cpa}, but this is attributed to selection biases~\cite{Acharya:2017okq} and it does not inform us about the onset of jet quenching in ``sufficiently central" \pPb collisions. Also, signatures of jet quenching
are neither observed in inclusive hadron production nor in the dijet transverse momentum imbalance\cite{Chatrchyan:2014hqa,Khachatryan:2015xaa,ALICE:2018vuu}. 
No significant jet quenching effects have been found in \pPb collisions using correlations by either studying hadron production relative to the jet direction\cite{ATLAS:2022iyq} or by measuring the semi-inclusive yield of jets recoiling from a high-\pT trigger hadron\cite{Acharya:2017okq}. 
Ref.~\citenum{Acharya:2017okq} sets an upper limit of \unit[400]{\UMeV} (at 90\% CL) on medium-induced energy transport outside a jet cone with $R=0.4$. This is about a factor 20 smaller than the magnitude of energy transport from within the jet to outside the jet cone in \PbPb collisions~\cite{Adam:2015doa}. On the other hand, as mentioned, the presence of energy loss leads to sizable \vn coefficients at high-\pT\cite{CMS:2012tqw} in \PbPb collisions. Surprisingly, non-zero $\vtwo$ are measured above \unit[9]{\UGeVc} also in \pPb collisions despite the lack of other energy-loss signals\cite{ATLAS:2019vcm}. This indicates that our current understanding of the interplay between jet quenching and flow phenomena is incomplete. 

\textbf{\vn in presence of a hard probe.} The $\vtwo$ measured in events with a $Z$ candidate in \pp collisions are identical to the measurement in inclusive events\cite{ATLAS:2019wzn}. Similarly, selecting events with or without jets in pp collisions does not influence the measured $\vtwo$\cite{ATLAS:2023bmp}. Measurements in \pPb collisions in presence of a high-\pT jet are similar to the ones in inclusive events for \vtwo and \vthree\cite{ATLAS:2019vcm}. 

\textbf{Jet constituent \vn}
Azimuthal correlations have also been measured between jet constituents and the bulk. 
No significant correlation is found between constituents of high-\pT ($\pT > 15$ GeV) jets and the remaining hadrons in the event in pp collisions\cite{ATLAS:2023bmp}. 
On the contrary, attempts to isolate jet constituent \vtwo  in jet-like three-particle correlation in \pPb show significant non-zero values up to $\pT \sim \unit[8]{\UGeVc}$\cite{ALICE:2022cwa}.

\textbf{\vn in jets.} In an attempt to identify collective phenomena in the smallest systems, there has been a study of high-multiplicity high-energy ($\pT > \unit[550]{\UGeVc}$) jets with up to 100 charged constituents in the final state\cite{CMS:2023iam}. Two-particle correlations and \vtwo are extracted in a coordinate system defined relative to the jet direction. The measured \vone, \vtwo and \vthree show a distinct trend as a function of multiplicity which is reproduced by Monte Carlo simulations without collective phenomena except at multiplicities above 80 where the trend changes and departs from the Monte Carlo expectation. The authors interpret this result as possible collective phenomena occurring in the dense environment of the jet.

\begin{figure}[ht]
\includegraphics[width=0.467\textwidth]{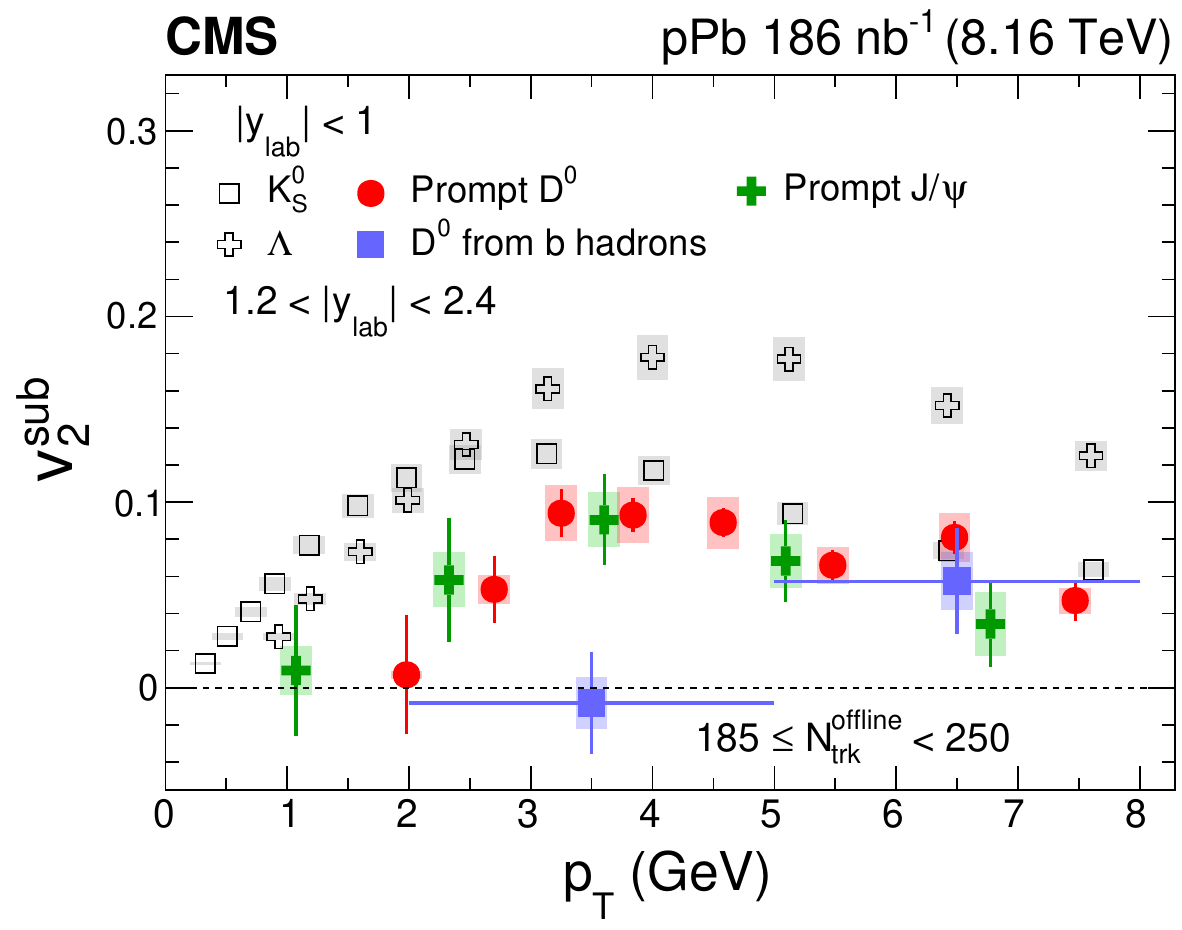}
\hfill
\includegraphics[width=0.513\textwidth]{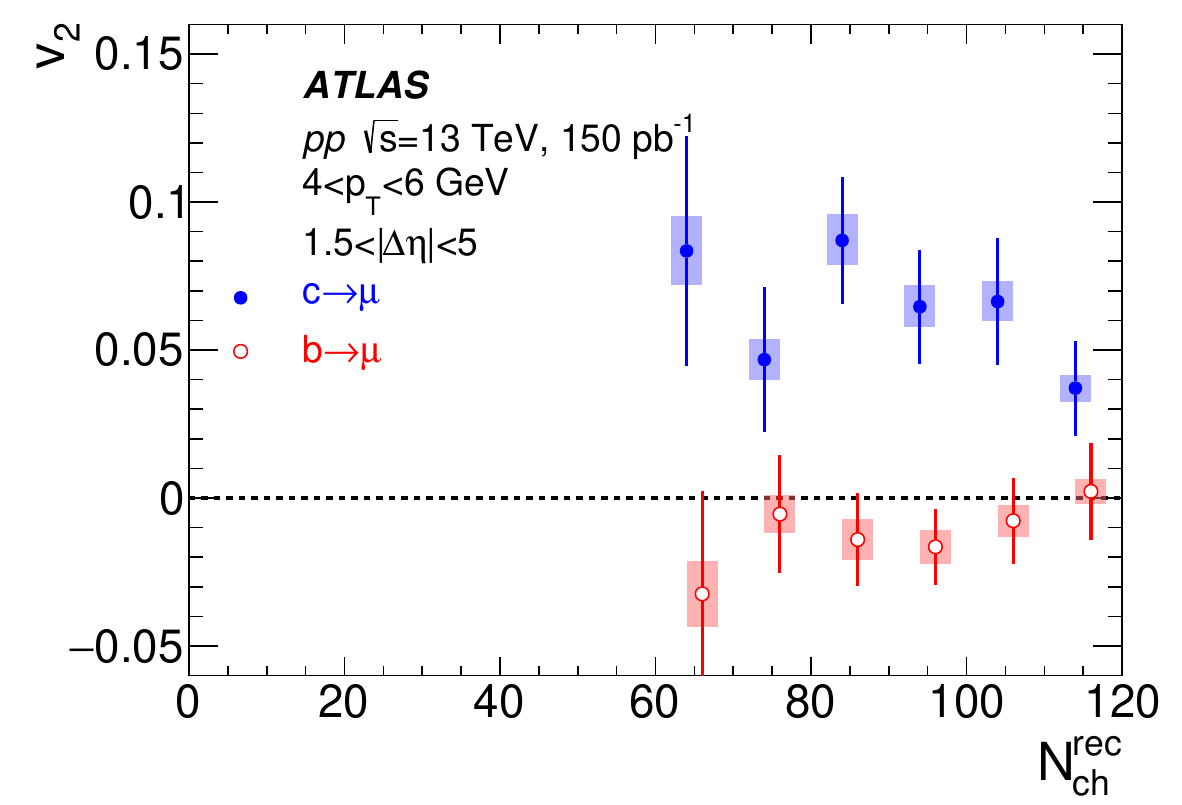}
\caption{\label{fig:hf_v2} Left panel: \vtwo coefficient of prompt $D^0$ mesons and J/$\psi$ as well as non-prompt $D^0$ compared to strange particles as a function of \pT in high-multiplicity \pPb collisions (Figure adapted from Ref.~\citenum{CMS:2020qul}). Right panel: \vtwo coefficient of muons from charm and beauty decays as a function of multiplicity in pp collisions (Figure from Ref.~\citenum{ATLAS:2019xqc}).}
\end{figure}

\textbf{Open heavy flavor and quarkonia}
Heavy flavor quarks (charm and beauty) are produced in the initial hard scattering and then interact and rescatter with the medium. Thus, also heavy-flavor particles exhibit finite anisotropies as shown with non-zero \vtwo measurements for $D$ mesons, J/$\psi$ and HF decay electrons and muons in both \PbPb and \pPb collisions\cite{ALICE:2013xna,Abelev:2013lca,Abelev:2014ipa,Adam:2015pga,ALICE:2016clc,Adam:2016ssk,Khachatryan:2016ypw,Acharya:2017qps,Acharya:2017tfn,Acharya:2017tgv,Sirunyan:2017plt,Acharya:2018dxy,CMS:2018loe,ALICE:2020pvw,ATLAS:2020yxw,CMS:2020qul,CMS:2021qqk,ALICE:2022ruh,CMS:2022vfn,CMS:2023mtk}. An example is shown in the left panel of Fig.~\ref{fig:hf_v2}. 
Non-zero \vtwo of particles from decays involving bottom quarks are measured in \PbPb and \pPb collisions but with large uncertainties\cite{CMS:2020qul} in \pPb collisions, see also the left panel of Fig.~\ref{fig:hf_v2}. 
\vtwo measurements for $\Upsilon$ are consistent with zero in \PbPb collisions\cite{ALICE:2019pox,CMS:2020efs} and \pPb collisions\cite{CMS:2023dse}.
In \pp collisions, significant \vtwo for charm are measured at high multiplicity\cite{ATLAS:2019xqc,CMS:2020qul} while the \vtwo for muons from b decays is consistent with zero\cite{ATLAS:2019xqc}, see the right panel of Fig.~\ref{fig:hf_v2}. 
Amongst measurements of collectivity seen in \PbPb collisions but not yet in smaller systems, there are data on non-zero \vtwo for $\psi(2S)$ as well as non-zero \vthree for $D$ mesons and J/$\psi$.\cite{ALICE:2020pvw,CMS:2022vfn,CMS:2023mtk}

The nuclear modification factor $R_{AA}$ for $J/\psi$ in large systems is enhanced at LHC with respect to RHIC energies\cite{Abelev:2012rv,CMS:2012bms,Adam:2015isa,Adam:2015rba,Adam:2016rdg,Khachatryan:2016ypw,CMS:2017uuv,ATLAS:2018hqe,STAR:2019fge,ALICE:2023gco,ALICE:2023hou}. This is qualitatively different from the $\sqrt{s}$-dependence of $R_{AA}^{J/\psi}$ from CERN SPS to RHIC energies. J/$\psi$ suppression has been expected due to the effect of the medium on the J/$\psi$ binding energy, while the enhancement at LHC energies can be incorporated as a J/$\psi$ regeneration component from deconfined charm quarks\cite{ALICE:2023gco}. $\Upsilon(1S)$, $\Upsilon(2S)$ and $\Upsilon(3S)$ production is equally found suppressed in \PbPb collisions with respect to the pp reference\cite{CMS:2012bms,Chatrchyan:2012lxa,ALICE:2018wzm,CMS:2018zza,ALICE:2020wwx,ATLAS:2022exb,CMS:2023lfu}.

In \pPb collisions, J/$\psi$ and $\Upsilon(1S)$ are suppressed relative to \pp collisions\cite{ALICE:2013snh,ALICE:2014ict,ALICE:2015sru,Aaij:2017cqq,Adamova:2017uhu,Sirunyan:2017mzd,ALICE:2018mml,ALICE:2018szk,ALICE:2019qie,ALICE:2022zig,CMS:2022wfi} which can be attributed to nuclear modification of the gluon PDF\cite{ALICE:2022wpn}. 
The production of the excited charmonium state, $\Psi(2S)$ as well as excited bottomonium states $\Upsilon(nS)$, $n \geq 2$ have been measured in both \PbPb and \pPb collisions\cite{Chatrchyan:2013nza,ALICE:2014ict,Abelev:2014zpa,Adam:2015isa,Adam:2015jsa,Adam:2016ohd,Khachatryan:2016xxp,Sirunyan:2016znt,Aaboud:2017cif,Sirunyan:2017lzi,ATLAS:2018hqe,ALICE:2022jeh,CMS:2022wfi,CMS:2023lfu} which shows a suppression w.r.t. the ground state.

\subsection{Towards even smaller systems...}

\begin{table}[t!]
\caption{Summary of observables measured in $\gamma$--p (realized in \pPb and ep colliders), $\gamma$--Pb, low multiplicity pp, \ee and ep collision systems. Low $\nch$ refers to values close to the minimum-bias multiplicity.}
\small
{
\begin{tabular}{|l|p{3cm}|p{2.8cm}|p{2cm}|}
\hline  
                            & \multicolumn{3}{c|}{Observable} \\
                            & \centering Near-side          & \centering Azimuthal & \centering Higher-order \tabularnewline 
System                      & \centering ridge yields       & \centering anisotropy (\vn) & \centering cumulants \tabularnewline \hline
$\gamma$–Pb                 & observed (template)           & up to \vthree (template)      & not studied \\
(in \PbPb UPC)              &                               &                               & \\ \hline
$\gamma$–p                  & not observed                  & up to \vthree $\approx$ MC    & not studied \\ 
(in \pPb UPC)               &                               & (template)                    & \\ \hline
$\gamma$–p                  & not studied                   & up to \vtwo $\approx$ MC      & n = 4 $\approx$ MC  \\ 
(in ep)                     &                               &                               & wrong sign \\ \hline
ep (DIS)                    & not studied                   & up to \vtwo $\approx$ MC      & not studied \\ \hline
pp (low \nch)               & observed                      & up to \vtwo (template)        & not studied \\ \hline
ee                          & limits + hint at $ 1.02\sigma $ & up to \vthree $\approx$ MC  & not studied \\ \hline
Refs.                       & \mbox{\citenum{Badea:2019vey,Belle:2022ars,Belle:2022fvl,CMS:2022doq,ALICE:2023ulm,Chen:2023njr}} 
                            & \mbox{\citenum{Aad:2015gqa,Khachatryan:2016txc,ZEUS:2019jya,ZEUS:2021qzg}} \mbox{\citenum{ATLAS:2021jhn,CMS:2022doq,Chen:2023njr}} 
                            & \mbox{\citenum{ZEUS:2021qzg}}  \\
\hline
\end{tabular} \label{table:smallsystems2}
}
\end{table}

Studies of the mentioned phenomena have been extended into even smaller systems, by studying low multiplicity pp collisions and by using ultra-peripheral (UPC) heavy-ion collisions where $\gamma$--p and $\gamma$--Pb processes occur. Furthermore, data of former \ee and ep experiments have been re-analyzed where the latter gives access to deep inelastic scattering (DIS) as well as resolved photoproduction resembling $\gamma$–p processes. While these studies only cover a fraction of the observables discussed in this section, they provide a number of interesting findings. Table~\ref{table:smallsystems2} summarizes the current status and a brief overview is given in the following.

\begin{figure}[ht]
\begin{minipage}[t]{.4\textwidth}
\includegraphics[width=\textwidth]{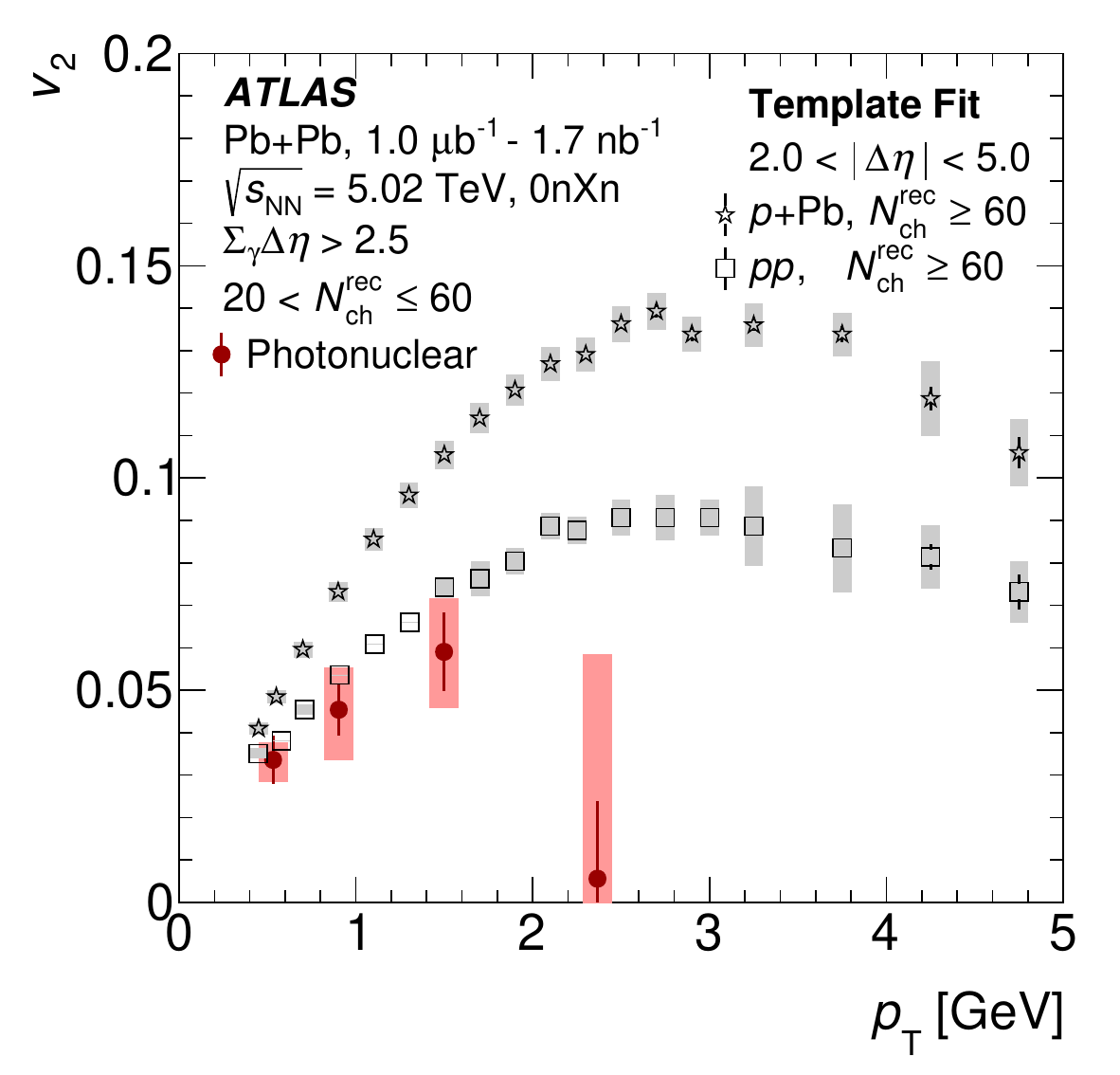}
\caption{\label{fig:upc} \vtwo coefficient measured in ultra-peripheral $\gamma$--Pb collisions compared to the measurement in \pPb and pp collisions (Figure adapted from Ref.~\citenum{ATLAS:2021jhn}).}
\end{minipage}
\hfill
\begin{minipage}[t]{.58\textwidth}
\includegraphics[width=\textwidth]{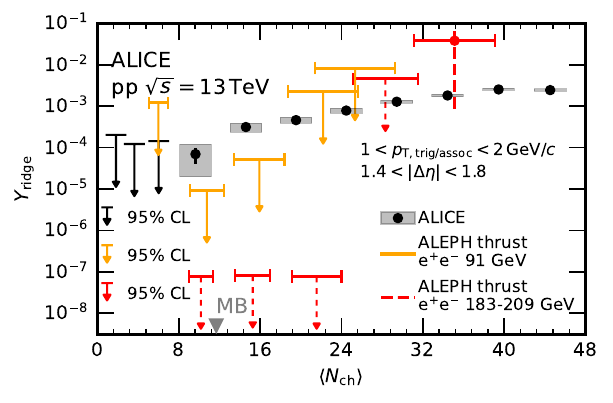}
\caption{\label{fig:lowmult} Near-side ridge yield measured in pp collisions compared to upper limits set in \ee collisions (Figure from Ref.~\citenum{ALICE:2023ulm}).}
\end{minipage}
\end{figure}

\textbf{Ultraperipheral collisions} 
The electromagnetic cloud traveling with Pb ions in a particle accelerator leads to collisions of photons with beam particles. This allows one to study $\gamma$--p ($\gamma$--Pb) collisions in a \pPb (\PbPb) collider configuration in so-called ultraperipheral collisions. Although the study of $\gamma$--p processes does not show any near-side ridge structure\cite{CMS:2022doq}, azimuthal Fourier coefficients up to third order have been measured. The positive \vtwo is reasonably described by simulations without involving collective phenomena\cite{CMS:2022doq}. The study of $\gamma$--Pb collisions shows a different picture: significant non-zero \vtwo and \vthree are measured, albeit with the template subtraction method. The \vn are lower than the equivalent measurement at the same multiplicity \pPb collisions but similar to the ones in pp collisions, see Fig.~\ref{fig:upc}. Contrary to the $\gamma$--p result, simulations without a collective component do not reproduce the result, even qualitatively\cite{ATLAS:2021jhn}.

\textbf{ep collisions}
Studies in ep collisions in the DIS region find no evidence of a \vtwo signal beyond the one expected from momentum conservation and are reproduced by MC generators without involving collective phenomena\cite{ZEUS:2019jya,ZEUS:2021qzg}. 
In the photoproduction region \vtwo is extracted with 2 and 4-particle correlations. No evidence of effects beyond momentum conservation are observed\cite{ZEUS:2021qzg}. The 4-particle signal at second order has the opposite sign\cite{ZEUS:2021qzg} to the one observed in high-multiplicity pp to \PbPb collisions.
The authors of Ref.~\citenum{ZEUS:2021qzg} report their \vtwo signal to be consistent with the one in $\gamma$--Pb collisions\cite{ATLAS:2021jhn} and attribute it to jet production unrelated to hydrodynamical behavior.

\textbf{Low-multiplicity pp collisions}
While high-multiplicity ridge yields in pp collisions have been the first observation of collective phenomena in pp collisions, see section~\ref{sect:longrangecorr}, studies at low multiplicity are intrinsically difficult as the signal scales with the square of the particle multiplicity and is dominated by jet fragmentation and resonance decays at low multiplicity. Recently, near-side ridge yields have been measured with great precision to multiplicities lower than the minimum-bias multiplicity\cite{ALICE:2023ulm}. Interestingly, significant long-range correlations emerge already at the minimum-bias multiplicity, see Fig.~\ref{fig:lowmult}. Studies beyond ridge yields, for instance of \vn coefficients are scarce at low multiplicity. Only a \vtwo measurement exists in pp collisions just above the minimum-bias multiplicity\cite{Aad:2015gqa,Khachatryan:2016txc} applying a template subtraction. Higher-order cumulants are only extracted at a multiplicity of 3--4 times the minimum-bias multiplicity at present.

\textbf{\ee collisions}
These collisions are free of hadronic effects in the initial state. As such, several mechanisms for collectivity invoked in other collision systems may be difficult to realize in \ee. This makes  it particularly interesting to check whether or not precursors of collective effects seen in other collision systems extend to \ee. 
These collisions have a randomly oriented event topology in the detector, therefore the mentioned studies are performed in the so-called thrust axis which is a proxy for the direction of the hard process occurring in e.g. $\ee \rightarrow q\bar{q}$.
Archived ALEPH data as well as data from Belle have been analyzed and limits on near-side ridge yields have been set at collisions energies of about \unit[10.5]{GeV}\cite{Belle:2022ars,Belle:2022fvl}, \unit[91]{GeV}\cite{Badea:2019vey} and at \unit[183--209]{GeV}\cite{Chen:2023njr}. 
In the highest multiplicity studied at \unit[183--209]{GeV} a non-zero near-side ridge yield has been found at $1.02\sigma$\cite{Chen:2023njr}, see rightmost ALEPH point in Fig.~\ref{fig:lowmult}.
A special case is the study of $\Upsilon(4S)$ on-resonance data. Here a non-zero near-side yield is found which is however attributed to the special topology in the decay of the $\Upsilon \rightarrow B\bar{B}$ system and reproduced by Monte Carlo simulations without collective phenomena. 
Furthermore, \vn coefficients are extracted and compared to MC simulations without collective phenomena which agree except at the largest studied multiplicities\cite{Chen:2023njr}. The statistical precision does not yet allow the conclusion that such phenomena are also found in \ee collisions.

This data allows for a clear comparison to previously mentioned low-multiplicity ridge yields measured in pp collisions. The latter are at least $5\sigma$ larger\cite{ALICE:2023ulm} than the ones in \ee collisions~\footnote{To be specific: The multiplicity range from 8 to 24 charged particles contributes most to this global assessment since ALEPH data allow one to establish the tightest upper bounds in this region. At high multiplicity, the upper bounds are much weaker. While this leaves more room for a potential signal, it must not be interpreted as a signal per se.
It needs to be mentioned that the comparison as a function of multiplicity of these different collision systems at very different center-of-mass energies is intrinsically difficult and requires modeling assumptions resulting in a 5--6.3$\sigma$ difference\cite{ALICE:2023ulm}.}, see Fig.~\ref{fig:lowmult}. This demonstrates that additional phenomena are observed in low-multiplicity pp collisions compared to \ee collisions.

\section{Theory overview}
\label{sec4}
In the phenomenological discussion of ultra-relativistic nucleus--nucleus collisions, it has been emphasized repeatedly that fluid dynamic models with close-to-minimal dissipative properties can account for a broad range of collective phenomena\cite{Heinz:2013th}. Data comparisons with 
iEBE-VISHNU\cite{Shen:2014vra},  
superSONIC\cite{Habich:2014jna}, 
MUSIC\cite{Schenke:2010nt}  
 and other fluid models\cite{Niemi:2015qia,Devetak:2019lsk} 
 support this conclusion.   However, the physics encoded in such phenomenologically successful ``fluid" models goes significantly beyond fluid dynamics\footnote{Here and in the following, we put the notion ``fluid" in quotation marks to highlight - where  required in content - that ``fluid" models invoke more than fluid dynamics.}. It includes suitably chosen initial conditions, hadronization prescriptions, phenomenological choices for kinetic transport in the pre-equilibrium dynamics, and hadronic rescattering. Also the modeling of the fluid dynamic phase itself includes more than fluid dynamics, as explained in section~\ref{sec411}. Early efforts to extend such fluid modeling to small collision systems are reviewed in section~\ref{sec43}. The current state of the art in model comparisons is set by Bayesian inference techniques~\cite{Bernhard:2018hnz}. 

The earliest Bayesian analyses of fluid models did not include any data from small systems and they were based on relatively few $p_T$-integrated measurements~\cite{Pratt:2015zsa,Bernhard:2016tnd,JETSCAPE:2020mzn,Parkkila:2021tqq,Parkkila:2021yha}. Qualitatively different initial conditions were also studied\cite{Heffernan:2023utr}. Early analyses that included some \pPb data remained restricted to $p_T$-integrated quantities~\cite{Moreland:2018gsh}. The most complete analysis as of today is given by Trajectum\cite{Nijs:2020ors,Nijs:2020roc}. It includes more than 500 data points on \pPb and \PbPb to overconstrain $O(20)$ model parameters, it extends to $p_T$-differential information and it has been used to illustrate the physics opportunities of future light ion beams by generating mock data for oxygen--oxygen collisions at the LHC~\cite{Nijs:2021clz}.

Remarkably, the inclusion of data on small collision systems has led so far to no significant additional constraints on the dynamical mechanisms and material properties invoked to describe successfully central nucleus--nucleus collisions. The reason for this is thought to lie in increased uncertainties associated to the initial conditions of small systems and in an increased sensitivity to final state effects, see section~\ref{sec432}. Within these complex multi-parameter models, our currently limited understanding of such uncertainties seems to offset the discriminatory power which a wider variation of system size could provide~\cite{Greif:2017bnr,Nijs:2020ors,Nijs:2020roc}. A similar statement applies to other complex simulation tools of ultra-relativistic heavy ion collisions such as A Multi Phase Transport (AMPT) model~\cite{Lin:2004en,Zhang:2019utb,Lin:2021mdn}, BAMPS~\cite{Xu:2007aa,Uphoff:2014cba} or EPOS~\cite{Pierog:2013ria,Werner:2023zvo}. We summarize in section~\ref{sec43} to what extent these codes have been compared to measures of collectivity in small systems, but we mention already here that more systematic studies are missing and that no significant further constraints on the dynamical mechanisms underlying collectivity have been reported. 

The situation is thus rather peculiar: On the one hand, experiments at the LHC have established the persistence of almost all soft \pT signatures of collectivity across all system sizes from central \PbPb down to the low multiplicity \pp collisions in which collectivity had not been expected. On the other hand, this wealth of qualitatively novel data does not seem to challenge any of the complex model frameworks used traditionally to simulate soft physics in heavy-ion collisions. The conclusion here is not that fluid-dynamic or transport models of ultra-relativistic heavy-ion collisions naturally explain collectivity in small systems. The conclusion is rather that the predictivity of these models for small systems is sufficiently weak to accommodate existing data. 

In this situation, a discussion of what we learn from small system collectivity profits from going beyond mere data-model comparisons. One should ask whether the experimentally established phenomena of collectivity are more generic than the concepts invoked for their explanation. To be specific: 
\begin{enumerate}[I]
\item The phenomenology of nucleus--nucleus collisions invokes the concept of rapid equilibration -- but can this equilibration be rapid enough to explain collectivity down to \pp? 
\item 
The phenomenology of nucleus--nucleus collisions invokes fluid dynamics -- but what is the smallest droplet to which fluid dynamics applies?  
\item 
The phenomenology of nucleus--nucleus collisions invokes close-to-minimal dissipative properties as the {\it sine qua non} for numerically large collective effects -- but can this be the main cause for collectivity in \pPb, in \pp or even in \ee? 
\end{enumerate}
The field of ultra-relativistic heavy-ion collisions knows of precedents where substantial insights were achieved by going beyond a mere model-data comparison. For instance, two decades ago, a calculation of the viscosity over entropy ratio in a maximally supersymmetric field theory had contributed to revolutionizing heavy-ion phenomenology not because it led itself to a direct data-model comparison, but because it indicated that a hitherto unimagined dynamics could be realized in quantum field theory. In the same spirit, the LHC discovery of collectivity in small systems has triggered since 2010 many studies in theoretically clean simplified set-ups to understand what is imaginable in terms of fast hydrodynamization, fast kinetic thermalization and fast chemical equilibration, and to what extent fluid dynamic or non-fluid dynamic degrees of freedom are at work in such fast processes. These lines of research are an intrinsic part of the LHC legacy of small system collectivity even if they address often ``only" the conceptual and not the phenomenological part of the problem. 
The present chapter summarizes the most pertinent findings of this broad theoretical research effort. Finally, the last subsection will discuss data-model comparisons. 

\subsection{Dynamical frameworks of collectivity}
\label{sec41}

Fluid dynamics, kinetic theory and the evolution of initial-state correlations (as well as various non-perturbative mechanisms) are the dynamical frameworks considered for the explanation of collective phenomena in ultra-relativistic nucleus--nucleus collisions. Here, we recall the assumptions on which they are based. This will enable a discussion about their applicability to collective effects in the smallest collision systems.

\subsubsection{Fluid dynamic descriptions of collectivity}
\label{sec411}
In nuclear physics, first fluid dynamic formulations of collectivity can be traced back to the liquid-drop model in the 1930s -- a very informative account of these earliest developments is given in Ref.~\citenum{Nagle:2018nvi}. In ultra-relativistic heavy ion collisions, it was Bjorken's work~\cite{Bjorken:1982qr} on longitudinally boost-invariant systems that initiated fluid dynamic formulations of the collective expansion. 

Fluid dynamics is formulated in terms of thermodynamic quantities, and it emerges in the long wavelength limit of any more complete dynamics. Different  derivations of relativistic fluid dynamics exist. 

\begin{enumerate}   
    \item {\it The thermodynamic (or entropy-wise) derivation }\\   Fluid dynamics can be derived directly from conservation laws in a gradient expansion. In the simplest case (expansion to zeroth order in gradient), the energy-momentum tensor $T_{\mu\nu} = (\varepsilon + p) u_\mu u_\nu - p\, g_{\mu\nu}$ depends only on energy density $\varepsilon$, pressure $p$ and  three independent components of the local flow field $u_\mu$. The collective dynamics of these five unknowns is fully determined in terms of four constraints from energy-momentum conservation $\nabla_\mu T^{\mu\nu} = 0$ supplemented by the equation of state $\varepsilon = \varepsilon(p)$. \\
    To first order in gradients, the tensor decomposition of $T_{\mu\nu}$ with respect to the flow field $u_\mu$ contains a shear viscous term 
    \begin{equation} \Pi_{\mu\nu}^{\rm constitutive} = \eta \left[ \Delta^{\mu\alpha} \nabla_{\alpha}u^\nu + \Delta^{\nu\alpha}\nabla_\alpha u^\mu - \tfrac{2}{3} \Delta^{\mu\nu} \nabla_\alpha u^\alpha \right]\, ,\label{constitutive}\end{equation}
    as well as a bulk viscous and a heat conducting one. 
    Here, $\eta$ denotes the shear viscosity and $\Delta^{\mu\nu}$ is the projector on the subspace orthogonal to $u^{\mu}$. Remarkably, requiring that entropy increases locally ($\nabla_\mu S^{\mu} \geq 0$) closes the dissipative fluid equations to first order in gradients~\cite{Rischke:1998fq}. In this sense, dissipative fluid dynamics results solely from the fundamental laws of thermodynamics supplemented by material properties ($\varepsilon = \varepsilon(p)$, $\eta$, ...) that are calculable in an equilibrium field theory.\\
    However, implementing the spatial gradients $\nabla_\mu u^\alpha$ in \eqref{constitutive} instantaneously would violate causality. Therefore, {\it causal} viscous fluid dynamics\footnote{Other causal formulations of fluid dynamics are possible~\cite{Hoult:2020eho,Bemfica:2020zjp} but these have not been employed in phenomenological studies so far.} specifies how the physical shear viscous tensor $\Pi_{\mu\nu}$ relaxes to its
    first-order constitutive form $\Pi_{\mu\nu}^{\rm constitutive} $.
This is done by ad hoc assumptions that lie outside the realm of fluid dynamics~\footnote{
As this relaxation dynamics must preserve transversality and orthogonality of $\Pi_{\mu\nu}$, its precise formulation is a bit technical and it will not be given here. For details, see e.g. Eq.~(3.12) of Ref.~\citenum{Baier:2007ix}.}. Israel-Stewart(IS)-type formulations of causal viscous fluid dynamics invoke a simple relaxation time $\tau_\Pi$; the temporal decay of a shear viscous excitation takes then the form
    \begin{equation} A_{\rm excitation\, decay}(t) \simeq c_{\rm hyd} \exp\left[-\frac{\eta}{\varepsilon + p} k^2\, t \right] + c_{\rm non-hyd} \exp\left[-\frac{t}{\tau_{\Pi}} \right]\, .\label{relax} \end{equation}
This time dependence reflects the precise statement that in IS fluid dynamics, the shear viscous channel of the retarded propagator of  $T_{\mu\nu}$ has poles at $-i\tfrac{\eta}{\varepsilon + p} k^2$ and $-i\tfrac{t}{\tau_{\Pi}}$, see e.g. Ref.\citenum{Kurkela:2019kip} for details.
Since hydrodynamics is a gradient expansion, the lifetime of fluid dynamic excitations must grow with the inverse of the squared wave number $k^2$. The $k$-independent decay with lifetime $\tau_\Pi$ is thus explicitly non-fluid dynamic. Israel-Stewart dynamics is fluid dynamics only for sufficiently large wavelengths $\lambda$
\begin{equation}
    \frac{1}{\tau_\Pi} > \frac{\eta}{\varepsilon + p} k^2 =
\frac{\eta}{s} \frac{(2\pi)^2}{T\, \lambda^2}\, .
\label{boundhydro}
\end{equation}
    \item {\it Derivation of relativistic fluid dynamics from transport theory}\\
      In weakly coupled theories, there is a scale separation between the typical size of the wave packet $1/T$ and the mean free path between individual scatterings. Therefore, for time separations much larger than $1/T$, when interference effects between subsequent scatterings can be neglected, the evolution is determined by Boltzmann transport in which collision kernels are given by in-medium scattering processes in the field theory. In QCD, this is the basis of AMY effective kinetic theory (EKT).\cite{Arnold:2002zm,Arnold:2003zc} \\
      In general, Boltzmann transport theory evolves one-point probability densities $f(x,p,t)$ in response to collision kernels.~\footnote{The evolution of  $f(x,p,t)$ is coupled to that of correlated $n$-point density functions in the so-called BBGKY-hierarchy, and Boltzmann transport results from the truncation of that hierarchy.} Fluid dynamics can be derived by restricting Boltzmann transport to the lowest momentum moments of $f(x,p,t)$, namely the local conserved currents $N_\mu = \int p_\mu f(x,p,t)$ and the energy-momentum tensor $T_{\mu\nu} = \int p_\mu p_\nu f(x,p,t)$. For relativistic dissipative fluid dynamics, this derivation was pioneered by Israel and Stewart~\cite{Israel:1979wp} (see also Refs.~\citenum{Muller:1967,Hiscock:1983zz,Hiscock:1985zz}) and it was adapted to heavy-ion collisions subsequently~\cite{Muronga:2001zk,Muronga:2003ta,Muronga:2003tb,Denicol:2012cn}. \\ 
      For fluid-dynamic excitations, the kinetic and entropy-wise derivations yield the same equations of motion. However, the spectrum of non-fluid dynamic excitations depends on how the collision kernel is modeled~\cite{Romatschke:2015gic,Kurkela:2017xis,Ochsenfeld:2023wxz,Du:2023bwi} and it differs from that of Israel-Stewart dynamics. In the simplest relaxation time approximation (RTA) of kinetic theory, the non-propagating non-fluid dynamic excitation of IS fluid dynamics in \eqref{relax} is replaced by propagating quasi-particle excitations with lifetime $\tau_\Pi = a\, \tfrac{\eta}{s\, T}$, $a = 5$. In close analogy to IS dynamics, this kinetic evolution will not be governed by fluid dynamic degrees of freedom on length scales $\lambda$ that are too short to satisfy \eqref{boundhydro}. 
    \item {\it Derivation of relativistic fluid dynamics from quantum field theory}\\
In general, dissipative transport coefficients can be calculated via the Green-Kubo formula as certain long-wavelength limits of correlation functions in the thermal field theory~\cite{Green:1954ubq,Kubo:1957mj}. For QCD, such transport coefficients have been calculated in finite temperature perturbation theory to (almost) next-to-leading order~\cite{Ghiglieri:2018dib}, and there are some exploratory studies in lattice QCD~\cite{Meyer:2007dy,Meyer:2007ic}. However, nothing is known rigorously about the spectrum of non-fluid excitations in QCD, except that they must exist, and that they may be more complicated than a few well-isolated poles~\cite{Moore:2018mma}. In this sense, the microscopic structure of the QGP is unknown to date.\\  
In the strong coupling limit of non-abelian quantum field theories with gravity dual, both fluid dynamics and the microscopic dynamics beyond fluid dynamics is known.
Dissipative fluid dynamics can be derived from the gradient expansion of gravitational perturbations in the dual classical supergravity theory~\cite{Baier:2007ix,Bhattacharyya:2007vjd}. String-theory inspired derivations have helped to identify the complete set of allowed second-order transport terms~\cite{Baier:2007ix}, including transport coefficients that can arise only in anomalous fluid dynamics~\cite{Erdmenger:2008rm} and that can lead to qualitatively novel phenomena such as the chiral magnetic effect~\cite{Kharzeev:2015znc}. These anomalous transport coefficients can also be identified in an entropy-wise derivation~\cite{Son:2009tf}. 
 \\
For ${\cal N}=4$ SYM theory in the limit of strong coupling $\lambda = g^2 N_c\to \infty$ and large number of colors $N_c\to \infty$, the relaxation time in \eqref{relax} 
is $\tau_\Pi = a\, \tfrac{\eta}{s\, T}$ with $a=4 - \log(4) \approx 2.61$ (see Ref.~\citenum{Casalderrey-Solana:2011dxg} for a more complete review of relaxation times). The excitation spectrum of non-fluid (so-called quasi-normal) modes is fully known in the strong-coupling limit~\cite{Son:2002sd,Starinets:2002br}. The question of how the non-fluid dynamic properties of these non-abelian plasmas change with coupling strength is subject of recent research\cite{Hartnoll:2005ju,Grozdanov:2016vgg,Casalderrey-Solana:2018rle,Grozdanov:2018gfx}. 
\end{enumerate}

In summary, these findings support the following qualitative statements: 
\begin{itemize}
\item 
The fluid dynamic evolution equations for relativistic collective dynamics are universal: they take the same form, irrespective of whether they are derived from conservation laws (entropy-wise derivation)\cite{Landau:1959}, from specific kinetic theories via Grad's 14 moment expansion~\cite{Grad:1949zza} 
or with string-theoretical techniques in strongly coupled QFTs~\cite{Baier:2007ix,Bhattacharyya:2007vjd}.
\item
Any dynamics limited to fluid dynamic excitations is acausal. Any causal relativistic collective dynamics  exhibits also non-fluid dynamic exitations. These are non-universal and they depend on details of the microscopic dynamics. As non-fluid dynamic exitations must exist,
the relevant question is not the qualitative one (to be a fluid or not to be a fluid?), but the quantitative one: to what extent is collectivity dominated by fluid-dynamic or non-fluid dynamic excitations? 
\item
Fluid-dynamic excitations dominate if they are more long-lived than the non-fluid ones. For this to be the case, the wavelengths $\lambda$ of excitations should 
satisfy~\footnote{Conditions for the applicability of fluid dynamics have been formulated alternatively in terms of Reynolds number and Knudsen 
number, as reviewed succinctly in Ref.~\citenum{Noronha:2024dtq}. Reynolds number is an intrinsically fluid dynamic concept used in the wider rheological practice to discriminate laminar from turbulent flow. Knudsen number measures system size in units of mean free path and thus requires knowledge about which degrees of freedom propagate with which cross section. Both concepts can be generalized\cite{Noronha:2024dtq}. Instead, we prefer here to formulate the applicability of fluid dynamics directly in terms of the decay time of non-fluid dynamic modes in \eqref{theBound}, as this brings to the forefront what is really experimentally at stage: to learn about the nature of the non-fluid degrees of freedom which are characteristic features of the QGP, to learn about their decay which governs hydrodynamization, and to be open to the logical possibility that over a significant period in the evolution, non-fluid degrees of freedom may be non-negligible and coexistent with the fluid dynamic ones.} 
\begin{equation}
    \lambda > \sqrt{
\frac{\eta}{s} \frac{(2\pi)^2}{T}\tau_\Pi}
= (2\pi) \frac{\eta}{s} \frac{\sqrt{a}}{T} \, .
\label{theBound}
\end{equation}
For an almost perfect plasma with $\tfrac{\eta}{s} = \tfrac{1}{4\pi}$ and varying $a$ from $a \approx 2.61$ (in ${\cal N}=4$ SYM) to $a=5$ (in kinetic theory), one finds $\lambda > (0.81 - 1.12) \tfrac{1}{T}$. In particular, $\lambda > 1\, {\rm fm}$ for $T = 200$ MeV. For a less perfect fluid, this bound is more stringent. 
\end{itemize}
According to these numerical estimates, the smallest \pp and \pPb collision systems have an initial transverse extension into which fluid dynamic excitations of wavelength \eqref{theBound} barely fit. 
For a predominantly fluid dynamic explanation of \pp and \pPb collisions, however, one would require that scales significantly smaller than the transverse system size can be resolved by excitations that propagate fluid dynamically. According to the estimate \eqref{theBound}, this must not be taken for granted. Apparently supportive of this parametric statement, there is some evidence that if ``fluid" modeling is applied to smaller and smaller systems, then the fluid dynamic aspects of that modeling become less and less relevant (see section~\ref{sec432} below).

\subsubsection{Kinetic theory and transport models}
\label{sec412}
The application of kinetic theory to ultra-relativistic heavy-ion collisions is as old as that of fluid dynamics. Around the same time at which Bjorken formulated boost-invariant fluid dynamics, Baym formulated boost-invariant kinetic theory to understand ``the approach to thermodynamic equilibrium" of ``the excitations present"~\cite{Baym:1984np}. As indicated by this wording, kinetic theory had been viewed from the very beginning as a tool for understanding how and on what time scale equilibrium is achieved (``thermalization") and how fluid dynamic behavior emerges (``hydrodynamization"). 
The potential relevance and the potential limitations of formulating small system collectivity in terms of kinetic transport theory may be summarized as follows:
\begin{itemize}
\item[{\bf Pro}] 
Unlike fluid dynamics, kinetic theory applies to arbitrarily small systems. It is a candidate dynamics that interpolates {\it smoothly} between free-streaming in sufficiently small and dilute systems (assumed to be realized in multi-purpose event generators for \pp collisions) and dissipative fluid dynamics in sufficiently large and dense systems (assumed to be realized in fluid simulations for \PbPb collisions). These qualitative features have motivated many recent studies of the onset of collective phenomena in small collision systems (see section~\ref{sec421}). 
\item[{\bf Con}]
The formulation of kinetic transport theory relies on a scale separation between the size $\sim 1/T$ of quantum mechanical wavepackets and the mean free path between individual scatterings. In the strong coupling limit of quantum field theories with gravity duals, this scale separation is not realized. In high temperature QCD, the scale separation is perturbatively realized and it allows one to derive an effective kinetic theory, called AMY EKT~\cite{Arnold:2002zm,Arnold:2003zc}. For phenomenologically relevant values of the running QCD coupling (say $\alpha_s (Q^2= 2 {\rm GeV}^2) \simeq 0.3$), a parametrically clean scale separation does not exist. In this case, the use of transport models may still promise qualitative insights but it pushes kinetic theory beyond its region of guaranteed applicability. 
\end{itemize}

In kinetic theory, collective dynamics arises from the microscopic dynamics implemented in the collision kernels of the transport equations. In the phenomenological practice, any formulation of this microscopic dynamics is necessarily incomplete. Without entering what would be an inevitably technical and model-dependent discussion of how different choices of collision kernels are motivated, we highlight here two qualitative insights that have emerged within the last decade:

\begin{enumerate}
    \item {\it Rapid kinetic hydrodynamization and thermalization}\\
    QCD effective kinetic theory, AMY EKT~\cite{Arnold:2002zm,Arnold:2003zc} propagates one-point probability densities in response to in-medium $2\to 2$ elastic and $1 \to 2$ LPM collision kernels.
    It implements the so-called BMSS bottom-up thermalization~\cite{Baier:2000sb} according to which a longitudinally Bjorken-expanding initially oversaturated system dilutes and becomes undersaturated while $1\to 2$ splittings build up a reservoir of soft scattering centers. Subsequent interactions with these soft scattering centers drive the system then to equilibrium. Numerical studies with phenomenologically realized values of $\alpha_s$ have demonstrated~\cite{Kurkela:2014tea,Kurkela:2015qoa} that the resulting evolution towards equilibrium can be sufficiently fast to hydrodynamize initial conditions in ultra-relativistic heavy ion collisions efficiently.   \\
    Beyond conceptual insights, this line of research has resulted by now in a simulation tool. A linearized version of AMY EKT underlies the KoMPoST code~\cite{Kurkela:2018vqr,Kurkela:2018wud} that sets the current state of the art of modeling the pre-equilibrium stage of ``fluid" models. The duration over which pre-equilibrium evolution deviates significantly from viscous fluid dynamics in AMY EKT is comparable to the transverse extent of a \pp collision (see discussion of Figure~\ref{fig12} below). This is another indication that if ``fluid" models are applied to small systems, the fluid part of their evolution plays an increasingly smaller role for smaller system size. It is in line with the conclusions summarized at the end of section~\ref{sec411}.\\
    Beyond the linearized treatement of KoMPoST, there is a first proof-of-principle study of a Lorentz invariant parton cascade ALPACA that solves the AMY EKT Boltzmann equations~\cite{Kurkela:2022qhn}.
    \item {\it Onset of collectivity in ``one-hit" kinetic theory}\\
    Evolving with $2\to 2$ Boltzmann transport an expanding mixture of several relativistic massive particle species, it was found early on that a single collision per particle in average can already lead to sizable elliptic flow, with mass ordering between the species~\cite{Borghini:2010hy}. Similar findings were reported for the phenomenologically more complete AMPT code\cite{He:2015hfa}, emphasizing that the contribution from hydrodynamic-type mechanisms is small, see section~\ref{sec43}.  \\
    In a sufficiently small and dilute system, the first correction to free-streaming is obtained by expanding the collective dynamics to first order in the collision term. This is an expansion to first order in opacity, where opacity measures a specific combination of transverse geometric extension and local transverse energy density~\cite{Kurkela:2018ygx,Romatschke:2018wgi}. In simple kinetic models (isotropization time approximation -- ITA), the one-hit approximation leads naturally and for generic reasons to a hierarchy of the linear response coefficients $v_n/\varepsilon_n$ in \eqref{trans} that is in line with experimental observations. Remarkably, also non-linear response coefficients result in one-hit kinematics as a consequence of simple scaling behaviors\cite{Borghini:2018xum}. In ITA kinetic theory, non-linear response coefficients like $v_4 / \varepsilon_2^2$ or $v_5 / (\varepsilon_2 \varepsilon_3)$ are found to have the phenomenologically required order of magnitude in one-hit dynamics\cite{Kurkela:2018ygx,Romatschke:2018wgi}. 
\end{enumerate}

In contrast to fluid dynamics, the region of validity of kinetic theory is not limited to quanta of low momentum. Kinetic theory is thus a framework in which bulk dynamics and the jet quenching of hard processes could emerge as different manifestations of the same underlying dynamics. This is particularly apparent in AMY EKT which is known to give rise to fast hydrodynamization while its $1 \to 2$ Landau-Pomeranchuk--Migdal collision kernel is known~\cite{Caron-Huot:2010qjx} to implement the BDMPS-Z formulation\cite{Baier:2000mf} of radiative parton energy loss that underlies many jet quenching formulations.

\subsubsection{Initial state quantum interference as a source of collectivity}
Both fluid dynamic and kinetic models attribute the observed phenomena of collectivity to final state partonic and hadronic scattering. In contrast, it has been suggested in the context of saturation physics that initial state correlations could be responsible for the flow-like phenomena observed in small systems. With the so-called IP-Glasma model\cite{Schenke:2012wb,Schenke:2012hg}, saturation physics has motivated an ab initio formulation of subnucleon fluctuations.
Models within this framework were shown to reproduce di-hadron correlation data in \pp\cite{Dusling:2012iga,Dusling:2013oia} and \pPb\cite{Dusling:2013oia} collisions, and in various small collision systems studied at RHIC 
\cite{Mace:2018vwq}. However, a reassessment of this saturation physics explanation in an independent model implementation~\cite{Nagle:2018ybc} called into question some of the essential elements reported in Ref.~\citenum{Mace:2018vwq}. In this context, the results of Ref.~\citenum{Mace:2018vwq} were referred to as ``counterintuitive". 

A short succinct review of the initial state perspective on collectivity in small collision systems can be found in Ref.~\citenum{Schlichting:2016sqo}. What is referred to as ``glasma graphs" in saturation physics can be viewed as the calculation of quantum interference effects in the superposition of emission amplitudes from different sources. It follows directly from Heisenberg's uncertainty principle that quantum interference can map spatially anisotropic distributions of these initial sources into momentum anisotropies in the final state. In this sense, non-vanishing elliptic flow from sources correlated in the initial state may be intuitively expected even though the size of their contribution may be difficult to calculate and even though current model implementations may be unsatisfactory.

The phenomenology of ultra-relativistic heavy ion collisions knows of collective effects that are of a pure quantum nature. In particular, HBT correlations are collective in the sense that they are correlations amongst all identical particles in an event and that they persist in the limit of a large number of sources; they also translate spatio-temporal anisotropies into momentum anisotropies\cite{Lisa:2005dd}. Explicit calculations in saturation physics\cite{Kovner:2010xk,Dusling:2013oia} have identified such HBT-like correlations. Different calculational set-ups indicate that momentum anisotropies $v_n$ resulting from quantum interference can persist in the limit of a large number of sources\cite{Gyulassy:2014cfa,McLerran:2014uka,Blok:2017pui} irrespective of whether the conditions for saturation physics are satisfied\cite{Gyulassy:2014cfa,Blok:2018xes}.
However, the size of possible effects depends on assumptions, e.g.,  about the orientation of color fields~\cite{Lappi:2015vta}. Also, saturation physics alone cannot be expected to explain all important manifestations of collectivity. In particular, the formalism tends to lead to relatively short-range correlations in rapidity\cite{Schenke:2022mjv} and in transverse momenta\cite{Lappi:2015vta}, it decreases with increasing system size\cite{Schenke:2015aqa,Mace:2018vwq} and it can be washed out by final-state interactions\cite{Greif:2017bnr,Schenke:2019pmk}. In addition, it does not address the observed mass ordering of collective flow and it does not explain naturally the system size dependence of flow from small to large systems. In summary, the observed collective phenomena are clearly more generic than their explanation within the framework of saturation physics although, from a theoretical viewpoint, quantum interference can be expected to contribute to momentum anisotropies at some (as yet not fully understood) scale.

\subsubsection{Strings, ropes, shoving, pomerons and all that}
The collective phenomena observed in small systems relate to physics at small momentum transfers which is nominally non-perturbative. Fluid dynamic descriptions aim at capturing this non-perturbative physics in terms of a few transport coefficients; saturation physics aims at capturing it by doing perturbation theory on top of a non-perturbative high-density background; kinetic theory views non-perturbative information as included in the collision kernels. All three frameworks have important limitations if applied to small systems. Also, they do not exhaust the range of non-perturbative concepts that can be motivated within QCD. This subsection reviews other non-perturbative approaches that have been explored in the context of small system collectivity. 

In QCD, partonic scatterings separate color charges.  
Separated color charges are connected via flux tubes that fragment eventually. The description of this non-perturbative dynamics is model-dependent, but it is an essential aspect of QCD. In the so-called Lund string, it is at the basis of a successful hadronization model~\cite{Andersson:1997xwk} that is employed in modern multi-purpose event generators~\cite{Sjostrand:2014zea}. It has been argued that this string fragmentation must be supplemented by thermal effects to account for the observation of heavy-ion like behavior in the hadrochemical composition of high-multiplicity pp collisions at the LHC~\cite{Fischer:2016zzs}. In inelastic hadronic or nuclear collisions, flux tubes stretch between partons at projectile and target rapidities.  With any interaction that shifts the distribution of such flux tubes in the transverse plane, these strings become candidates for long-range rapidity correlations~\cite{Bierlich:2016vgw,Bierlich:2017vhg}. Such a string shoving mechanism was shown to implement hadrochemical and kinematic signatures of collectivity in small systems~\cite{Bierlich:2018xfw}. 
A related mechanism has been suggested but not worked out in Ref.~\citenum{Shuryak:2010wp}. Even the founding father of fluid dynamic applications to ultra-relativistic heavy ion collisions has argued that while fluid dynamics may account for the observed ridge in nucleus--nucleus collisions, the apparently similar ridge effect in high-multiplicity pp collisions could be due to ``a quite different physical mechanism", namely the collision of aligned flux tubes~\cite{Bjorken:2013boa}.

In general, elevating strings and more complex, spatially extended colored structures (such as ropes, baryon junctions etc.) to dynamical degrees of freedom has a long history in the phenomenological modeling of nucleus--nucleus collisions. Early implementations include the dual parton model~\cite{Capella:1992yb}, RQMD~\cite{Sorge:1989vt},  URQMD~\cite{Bleicher:1999xi} and extensions of HiJING~\cite{Vance:1999pr}. 
Also AMPT enhances its initial energy density signficantly with a string melting mechanism~\cite{Molnar:2019yam} and it is thus more than a parton cascade. 
Another non-perturbative concept that emerges naturally in QCD is that of pomeron exchanges~\cite{Ryskin:2011qh}. 
Such mechanisms have been argued to give rise to fluid dynamic behavior~\cite{Werner:2010ny}.

\subsection{Hydrodynamization and Thermalization - insights from simple models}
\label{sec42}
In general, collectivity is thought to arise from far out-of-equilibrium initial conditions in a dynamics which thermalizes locally and which thus lends itself eventually to a fluid dynamic description. But in a smaller collision system, any collective dynamics is at work for a shorter time. In the following, we summarize efforts to understand in simplified model set-ups the central question I whether thermalization and hydrodynamization can occur sufficiently quickly for explaining collectivity in the smallest collision systems. 

\subsubsection{Studies in boost-invariant 1+1 dimensional systems}
\label{sec421}
\begin{figure}[t]
\includegraphics[width=0.49\textwidth,trim={0cm 0.5cm 0cm 0cm},clip]{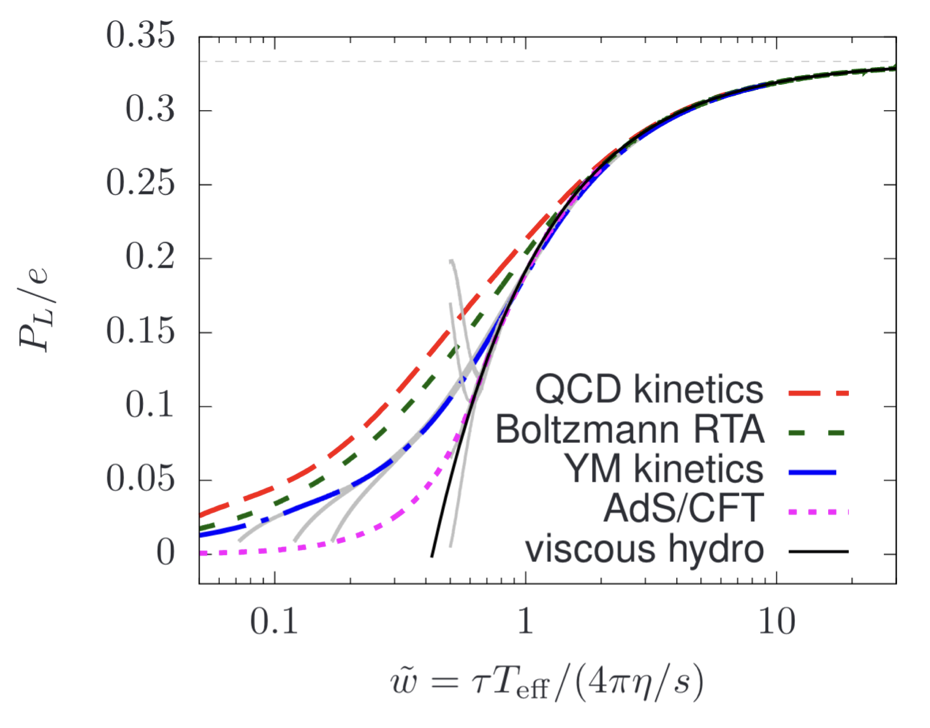}
\hfill
\includegraphics[width=0.49\textwidth,trim={0cm -1.4cm 0cm 0cm}]{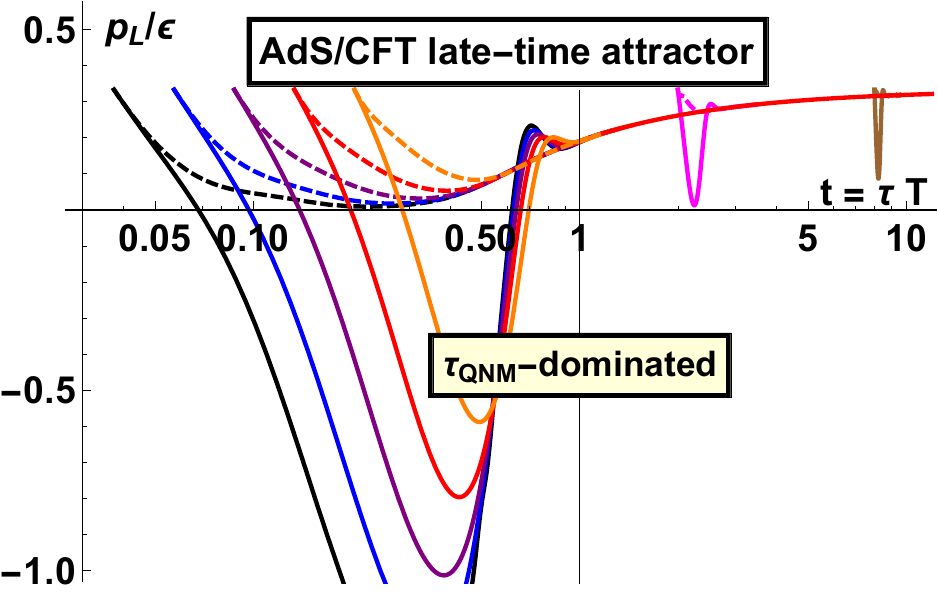}
\caption{
Evolution of the ratio of longitudinal pressure over energy density in different 1+1D models. Left panel: All microscopic theories (dashed curves) have the same late-time attractor which is 
relativistic viscous fluid dynamics. All microscopic theories hydrodynamize (i.e. approach this attractor) on the same time scale $\sim (4\pi) \tfrac{\eta}{sT}$, including variations of initial conditions within the same model (grey lines). Figure taken from Ref.~\citenum{Schlichting:2024uok}. Right panel: Not all microscopic theories have an early time attractor. For the strongly coupled QFT shown here, initial conditions set at five different times and two different choices all hydrodynamize at the time scale $\sim (4\pi) \tfrac{\eta}{sT}$ without displaying universal behavior at earlier times. Figure taken from Refs.~\citenum{Kurkela:2019set}. The hydrodynamization scale identified in these studies is consistent with the simple parametric estimate in \eqref{theBound}. 
}
\label{fig12}
\end{figure}

A dynamical system with boost-invariant initial conditions evolves in a boost-invariant way. Also, if initial conditions are translationally invariant in the transverse plane, this symmetry is preserved throughout the evolution. Invoking both symmetries, the resulting dynamics is essentially 1+1-dimensional. As first realized by Bjorken~\cite{Bjorken:1982qr} for hydrodynamics and by Baym~\cite{Baym:1984np} for kinetic theory, this offers the possibility of formulating technically simple, analytically tractable toy models that share important commonalities with physically realized almost boost-invariant collision systems. 

{\bf Timescale of hydrodynamization:}
Due to the symmetries in 1+1D systems, a very limited set of out-of-equilibrium conditions can be explored: Transverse gradients and density fluctuations are forbidden, conditions with initial eccentricity cannot be initialized and the dynamical mapping~\eqref{trans} from $\epsilon_n$ to $v_n$ cannot be explored. However, initial conditions that deviate {\it locally} from equilibrium can be imposed. In fluid dynamics~\cite{Romatschke:2017acs,Romatschke:2017vte,Strickland:2017kux,Kurkela:2019set} (in the absence of conserved currents), the only observables are the longitudinal pressure $p_L$ and the energy density $\varepsilon$; both depend only on time, and their ratio can be initialized arbitrarily far away from the equilibrium value  $p_L/\varepsilon = \tfrac{1}{3}$.
In Boltzmann kinetic theory~\cite{Blaizot:2017ucy,Romatschke:2017vte,Heller:2018qvh,Behtash:2019txb,Kurkela:2019set,Strickland:2019hff}, also higher moments of the one-particle distribution function $f$ can be studied. However, the timescales for their relaxation are rather trivially related and numerically comparable to those governing  $p_L$ and $\varepsilon$~\cite{Heller:2018qvh,Strickland:2018ayk,Kurkela:2019set}. In numerically tractable, strongly coupled field theories with gravity dual~\cite{Romatschke:2017vte,Spalinski:2017mel,Spalinski:2018mqg,Kurkela:2019set}, the evolution is not limited to a one-particle probability distribution and a larger variety of initial conditions may be initiated~\cite{Kurkela:2019set}.

Technically, the question of hydrodynamization amounts to asking on which timescale arbitrarily initialized components of the energy-momentum tensor relax to their constitutive expressions, (such as ~\eqref{constitutive}) in a fluid dynamic gradient expansion. In the simplest 1+1D models (which have only $\varepsilon$ and $p_L$ as observables), this reduces to the question on which time scale $p_L/\varepsilon$ reaches the viscous hydrodynamic correction (see black line in the left panel of Fig.~\ref{fig12}) to the equilibrium value $p_L/\varepsilon = \tfrac{1}{3}$. The answer is: in a Bjorken expanding system, hydrodynamization occurs on timescales $O\left( (4\pi) \tfrac{\eta}{sT}\right)$. This is so far all microscopic dynamics explored in Fig.~\ref{fig12}, irrespective of whether it is strongly coupled or weakly coupled.
This answer fully supports the parametric estimate \eqref{theBound} for the limited range of validity of fluid dynamics. In this sense, the problem of hydrodynamization is fully solved in 1+1D.  

{\bf Early-time and late-time attractors:}
In the recent literature on hydrodynamization, the more general concept of attractor solutions plays an important role, see Ref.~\citenum{Berges:2020fwq} for a broad review with relations to many neighboring fields. Attractors are the particular solutions of a dynamical system to which arbitrary initial conditions within the bassin of attraction relax at sufficiently late times. The late-time hydrodynamic attractor is clearly visible in Fig.~\ref{fig12} and it is reached for arbitrary initial conditions. Remarkably, in 1+1D  boost-invariant fluid dynamics and kinetic transport, attractor solutions are found not only at late times, but also at arbitrarily early times~\cite{Blaizot:2017ucy,Romatschke:2017vte,Heller:2018qvh,Behtash:2019txb,Kurkela:2019set,Strickland:2019hff}. However, the existence of an early-time attractor is not the tell-tale sign of a particularly efficient hydrodynamization. Rather, it is a consequence of the interplay between  local interaction and the rapid longitudinal expansion in boost-invariant systems~\cite{Kurkela:2019set}. In kinetic theory, this is a non-interacting free-streaming dynamics that drives the system to negligible $p_L$, in fluid dynamics, the early time attractor risks being unphysical since longitundinal gradients become arbitrarily large at arbitrarily early time and they thus lie outside a fluid dynamic description. In strongly coupled quantum field theories, no early time attractor exists and the system hydrodynamizes on the natural time scale $O(1/T)$ on which interactions drive the system to local equilibrium, see Fig.~\ref{fig12}. Related findings have been corroborated in QCD effective kinetic theory without~\cite{Kurkela:2015qoa,Kurkela:2018vqr} and with~\cite{Kurkela:2018xxd,Du:2020zqg} quark degrees of freedom. In summary, only the late-time attractor is independent of the microscopic dynamics, i.e. universal. It reflects the onset of fluid dynamics. In contrast, the early-time attractor is non-universal and it is not related to the onset of fluid dynamics. Fluid dynamics is not unreasonably effective. Rather, it sets in when standard textbook arguments expect it to set in, namely on scales on which fluid dynamic excitations live longer as non-fluid dynamic ones (see discussion leading to \eqref{theBound}).

The approach of particular initial conditions to known hydrodynamic late-time attractor solutions has also been explored to exhibit the asymptotic nature of the hydrodynamic gradient expansion, to develop resummation techniques~\cite{Denicol:2016bjh,Heller:2016rtz,Behtash:2017wqg,Kurkela:2017xis,Denicol:2018pak,Behtash:2019txb,Kurkela:2019set} and it has been used to illustrate the role of non-hydrodynamic poles in the relaxation to the attractor~\cite{Spalinski:2018mqg}.  

\subsubsection{Studies in boost-invariant higher-dimensional systems}

The fact that arbitrary initial conditions relax to a universal late-time fluid-dynamic attractor solution is equivalent to saying that initial conditions are forgotten. In the boost-invariant 1+1-dimensional systems discussed so far, essentially all measurable quantities forget initial conditions on a hydrodynamization time scale which is significantly shorter than the thermalization time scale at which the system approaches local equilibrium (here $p_L/\varepsilon = 1/3$), see Figure~\ref{fig12}. Having identified this universal hydrodynamization time scale in multiple models based on qualitatively different physics assumptions is arguably one of the main insights of these 1+1-dimensional studies. However, while thermalization is about forgetting, hydrodynamization is not. A fluid dynamic evolution that implements the mapping \eqref{trans} from initial spatial eccentricities $\epsilon_n$ to final momentum anisotropies $v_n$ is a dynamics that remembers the initial conditions $\epsilon_n$. The fact that the evolution of 1+1D models forgets any initial condition at late times is not a general physics feature but it points to an intrinsic limitation of 1+1D studies: these models simply do not allow for the description of measurable quantitites that are not forgotten in the evolution. Higher dimensional systems need to be studied to this end. 

While most studies of hydrodynamization and thermalization are carried out in 1+1D models because of their technical simplicity, there have been studies in higher dimensions ~\cite{Behtash:2019txb,Brewer:2019oha,Kurkela:2019kip,Kurkela:2019set,Kurkela:2020wwb,Ambrus:2021fej,Chattopadhyay:2021ive,Ambrus:2022koq,Borghini:2022iym}. Early numerical studies of radially symmetric colliding show waves mimicking small systems in strongly coupled field theory corroborated the importance of the time and length scale $O(1/T)$ for driving small systems toward fluid dynamic evolution\cite{Chesler:2015bba,Chesler:2015wra,Chesler:2016ceu}. More generally, once radial symmetry is not imposed, one can study how efficiently initial conditions $\epsilon_n$ are remembered, and to what extent the response depends on the assumed dynamics. For instance, evolving a system with initial spatial eccentricity $\epsilon_2$, first with kinetic theory up to a switching time $\tau_s$ and then with the corresponding Israel-Stewart dynamics to later times informs us about the sensitivity of collective response to the nature of non-fluid dynamic degrees of freedom. In such model studies, one third to one half of the elliptic flow signal seen in fully hydrodynamized systems can be built up in small collision systems that extend over only one mean free path and that do not hydrodynamize\cite{Kurkela:2018qeb}.  Along a different line of investigation, there has been a significant computational effort to quantify in simple conformal kinetic theories the full opacity dependence of all relevant linear and non-linear response coefficients 
    in the mapping $\lbrace \varepsilon_n\rbrace \longrightarrow \lbrace v_n\rbrace $
    \cite{Kurkela:2019kip,Kurkela:2020wwb,Ambrus:2021fej,Ambrus:2022koq,Borghini:2022iym}. This opacity dependence interpolates smoothly between free-streaming and almost perfect fluid dynamics as a function of opacity. Despite their simplicity, these models have led to statements about the (in)applicability of hydrodynamics in \pp, \pPb and light nucleus collisions\cite{Ambrus:2022qya}.
    
\subsection{Model studies and data comparisons}
\label{sec43}

Up to this point, our discussion has focused mainly on the description of basic physics concepts that may underlie the formulation of collectivity in small systems (section~\ref{sec41}), and on the illustration of these concepts in highly simplified model studies whose conceptual transparency offers unique opportunities for understanding how and how efficiently collectivity may be built up in small systems. However, such simple models do not reflect the full complexity of ultra-relativistic hadronic and nuclear collisions. Here, we finally summarize data comparisons with phenomenologically more complete model frameworks. 

\subsubsection{Theory expectations prior to LHC data}
Prior to the experimental observations of collectivity in small systems, different works had contemplated the possibility that azimuthal asymmetries may be seen in \pp collisions at LHC as a consequence of various nuclear-like 
effects~\cite{Bautista:2009my,Cunqueiro:2009zem}, as a consequence of fluid dynamic behavior~\cite{Luzum:2009sb,Ortona:2009yc,Prasad:2009bx,Bozek:2010pb} or more generally as a consequence of a collective dynamical response to initial spatial eccentricities in the \pp overlap~\cite{Casalderrey-Solana:2009rtc,dEnterria:2010xip}. 
The very idea that \pp collisions may develop fluid dynamic behavior is not new and had been considered early on, see e.g. Ref.~\citenum{VonGersdorff:1986tqh}. However, predictions of the size of $v_2$ in \pp varied from zero~\cite{Luzum:2009sb} to values of $O(0.1)$, depending on model assumptions, and mechanisms ranged from collective explosion~\cite{Shuryak:2010wp} to initial state effects in saturation physics~\cite{Dumitru:2010iy}. In short, these early works formulated interesting expectations, but none of them constitutes a well-motivated {\it prediction} that  can be regarded as having received detailed experimental support. 

\subsubsection{Small system collectivity in fluid models}
\label{sec432}
Following the phenomenological success in ultra-relativistic nucleus--nucleus collisions, several works aimed at including small-system collectivity within the same ``fluid" modeling in small and large systems. This working hypothesis ``one fluid to rule them all" has been supported by a viscous fluid dynamic description of 
high-multiplicity \pp, \pPb and \PbPb collisions  at $\sqrt{s}=5.02$ TeV~\cite{Romatschke:2015gxa,Weller:2017tsr}. However, it also became clear that ``one fluid might not rule them all"~\cite{Zhou:2020pai}. While the wordings cited here sound like contradicting each other, the scientific facts reported in these studies are consistent with each other. The main message is simply that ``fluid" models are more than fluid models: they include additional physics. Additional physics found numerically relevant for small collision systems includes the physics of hadronization (recombination, see section~\ref{sec22} for details)
\cite{Zhao:2020wcd}, the physics of hadronic\cite{Romatschke:2015dha,Zhou:2015iba} non-fluid contributions to flow, and the physics of initial state fluctuations\cite{Welsh:2016siu,Greif:2017bnr,Zhao:2017rgg,Zhao:2020pty}. These additional physical mechanisms appear to become numerically more relevant for smaller collision systems. In this sense, fluid dynamics does not become abruptly invalid below a certain system size, but it becomes gradually less and less relevant for the description of the system. The studies of ``fluid" models cited above compare to a small subset of the measures of collectivity in Table~\ref{table:smallsystems} only (mainly measurements of $v_2(p_T)$ and $v_3(p_T)$). Where more detailed sets of data were included in the analysis, issues of fine-tuning and of a possibly limited range of validity of hydrodynamics became more important~\cite{Zhou:2020pai,Wu:2023vqj}. 

The EPOS event generator employs a core-corona model. The initial conditions are generated in a model of pomeron-type multiple scattering\cite{Pierog:2013ria}. The corona of hadronic and nuclear collisions is modeled in a microscopic-dynamic picture that is assumed to be free of collective dynamics. The dynamics in the core (which becomes more important in large systems) is described by viscous fluid dynamics in recent versions of EPOS\cite{Werner:2023zvo}. In this sense, EPOS is a particular variant of a viscous ``fluid" model. 
Hadronic rescattering is simulated with UrQMD, but the fluid dynamic phase seems essential for the dynamical translation \eqref{trans} from spatial to momentum anisotropies. Different versions of EPOS reproduce the ridge in pp collisions\cite{Werner:2010ss}, anisotropic flow phenomena in \pPb collisions\cite{Werner:2013ipa}, and radial flow phenomena in \pp and \pPb collisions including their mass hierarchy\cite{Werner:2013tya}. 
But EPOS also predicts a characteristic nuclear modification of heavy-flavor hadrons\cite{Vogel:2010et} 
and a characteristic enhancement of thermal photon production\cite{Liu:2011dk} in \pp collisions which have not been observed, yet. 

\subsubsection{Small system collectivity in transport models}
 \label{sec433}
In ultra-relativistic heavy ion collisions, the modeling of the bulk of the collective evolution with partonic transport codes has a long history~\cite{Geiger:1991nj,Peter:1994yq,Gyulassy:1997ib,Zhang:1997ej,Zhang:1998tj,Borchers:2000wf,Molnar:2000jh,Bass:2002fh,Lin:2004en,Xu:2007aa}. For all these simulation packages,  the code {\it is} the model in the sense that the code is more than a tool for solving an easily stated set of transport equations of motion. The interpretational question of why a transport code (that implement much more physics than transport physics) does or does not agree with data is as old as the comparison of transport codes to data and it can be difficult to address. Amongst the above-mentioned partonic transport codes, mainly A Multi Phase Transport (AMPT) model~\cite{Lin:2004en,Zhang:2019utb,Lin:2021mdn} and BAMPS~\cite{Xu:2007aa,Uphoff:2014cba} have been compared recently to data from the LHC. 

AMPT successfully reproduces elliptic and triangular flow in \pPb, peripheral \PbPb\cite{Bzdak:2014dia,Ma:2014pva,Bozek:2015swa} and \HeAu collisions\cite{OrjuelaKoop:2015jss}. It has also been shown to satisfactorily reproduce mass ordering of $v_2$ and $v_3$ in settings in which coalescence was enabled~\cite{Li:2016ubw}. The question for why AMPT is successful has focused in particular on the problem that the AMPT code performs very differently from Molnar's Parton Cascade MPC\cite{Molnar:2000jh} although it uses nominally similarly small partonic cross sections. This links the discussion of AMPT to the time-honored problem of understanding the so-called ``opacity puzzle"~\cite{Molnar:2000jh} at the beginning of the RHIC heavy-ion program. As Molnar had realized at the time, to account for the size of the elliptic flow at RHIC, MPC had to be initialized either with unphysically high parton densities per unit rapidity, or it had to be evolved with unphysically large partonic cross sections~\cite{Molnar:2001ux}. There is a detailed analysis in Ref.~\citenum{Molnar:2019yam} of which model-dependent choices in AMPT may underlie its  phenomenological success where other transport codes fail.
    
We mention in passing the role of hadronic transport in heavy-ion phenomenology. In simulations of nucleus--nucleus collisions that involve fluid dynamic modeling, the system is hadronized along Cooper-Frye~\cite{Cooper:1974mv} freeze-out hypersurfaces. The resulting hadron gas is sufficiently dilute to be evolved with Boltzmann transport codes. Interactions in this hadronic phase affect kinetic and hadrochemical distributions. RQMD~\cite{Sorge:1989vt}, URQMD~\cite{Bleicher:1999xi} and most recently SMASH~\cite{Petersen:2018jag} are the main hadronic transport codes used to this end. 

\subsection{Model comparisons to hadrochemical measures of collectivity}

In principle, any phenomenologically complete model of hadronic collisions should describe both hadrochemical and kinematic characteristics of multi-particle distributions. In practice, however, several microscopic mechanisms that affect kinematic contributions at best mildly are invoked to account for the observed hadrochemical distributions. These include statistical hadronization, recombination, string breaking, color reconnection,  rope formation etc. Here, we comment on model comparisons invoking such hadronization models.  

It is one profound consequence of the smooth evolution of strange particle production with event multiplicity that the hadronization mechanisms used in multi-purpose event generators need to be supplemented with qualitatively novel physics~\cite{Fischer:2016zzs}. Hadronization mechanisms where individual color degrees of freedom hadronize without seeing the nuclear environment cannot account for the observed multiplicity dependence. For instance, string breaking tuned to \ee collisions cannot account for strangeness enhancement. Several mechanisms have been considered in the literature. These include color reconnections beyond leading color\cite{Christiansen:2015yqa} as implemented in PYTHIA, color ropes where several strings hadronize together\cite{Bierlich:2015rha} or the explicit addition of collective phenomena as for instance in core/corona models like EPOS\cite{Werner:2007bf}.
Statistical models using a Grand Canonical ensemble including a strangeness undersaturation factor $\gamma_{S}$ (which is close to 1 in large systems and smaller than 1 in small collision systems) describe particle ratios in all systems within an accuracy of about 20\%\cite{ALICE:2022wpn}.

While the above-mentioned models consider mechanisms of strangeness enhancement that are operational at the time of hadronization, there is also the question whether (some of) the observed strangeness enhancement could be generated dynamically by sufficiently efficient partonic interaction rates prior to hadronization, i.e., by turning an overdense gluon-rich system into a strangeness rich one via $gg \to s\bar{s}$. This motivates efforts to model the microscopic partonic (and hadronic) processes that can contribute to strangeness enhancement. Fermion production has been studied using perturbative reaction rates~\cite{Rafelski:1982pu,Koch:1986ud}, classical real-time lattice simulations of nonequilibrium production~\cite{Tanji:2017xiw}, rate equations~\cite{Biro:1993qt,Elliott:1999uz} and various partonic Boltzmann transport models~\cite{Geiger:1991nj,Borchers:2000wf,Xu:2004mz,Blaizot:2014jna,Ruggieri:2015tsa}.Inspired by the signs of collective behavior in the hadrochemical composition of small collision systems, there have been first studies of chemical equilibration in QCD effective kinetic theory~\cite{Kurkela:2018xxd}, estimating that chemical equilibration can be reached for final state multiplicities $dN_{\rm ch}/d\eta \gsim 100$.    Measurements of the change of hadrochemical composition as a function of system size inform this discussion.

\section{Concluding discussion and outlook}

The experimental program reviewed here was not thought of or proposed prior to the start of LHC operation, but it emerged in response to a surprising discovery. Over the last decade, this led to a thorough and complete experimental exploration of the system-size dependence of all soft multi-particle phenomena including the traditional hallmarks of collectivity in flow observables, hadrochemical abundances and the study of heavy quarks. In line with the quote of T.D. Lee with which we have started this review, one may note that qualitatively novel collective phenomena had been thought to arise as a function of both, an increasing volume of the collision system and an increasing initial local energy density (which increases with $\sqrt{s}$). What we know now is that almost all collective phenomena observed in large \AOnA collisions can be identified also in small \pp and \pPb collisions. Collective phenomena are numerically larger in larger collision systems but they remain sizable across all hadronic collision systems and they evolve smoothly with the spatial extension of the colliding system, with event multiplicity and with $\sqrt{s}$ without any experimental indication for a finite minimal scale required for collectivity. 

Future experimental small-system runs at the LHC are foreseen to include proton--ion collisions, oxygen--oxygen collisions and possibly other light-ion collisions with increased integrated luminosity. In addition to increased experimental accuracy, this will substantially extend our experimental knowledge of collectivity into the heavy-quark sector in small systems and possibly identify the onset of jet quenching as a function of system size.

\begin{figure}[ht]
\includegraphics[width=\textwidth,trim={5 0 0 0},clip]{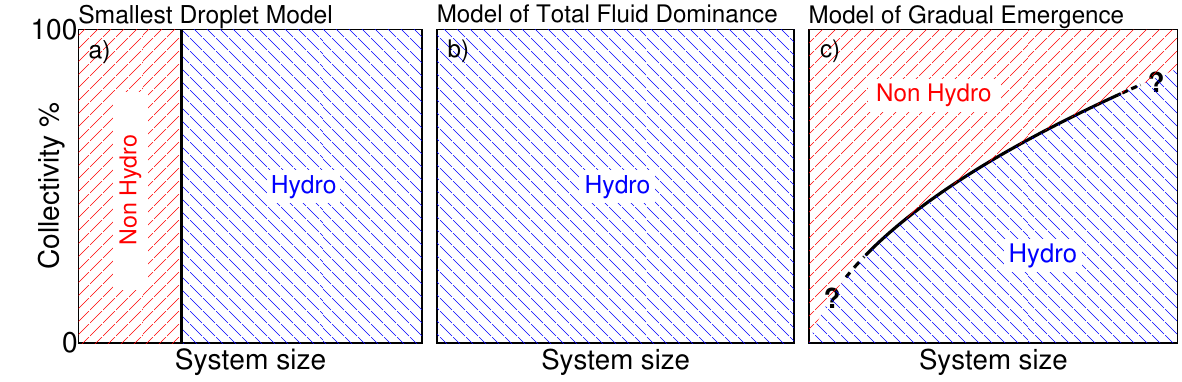}
\caption{\label{fig:summary} Three simple sketches of the degree to which hydrodynamic and non-hydrodynamic mechanisms may contribute to the observed collectivity as a function of system size. There is strong support for the idea that the range of experimentally accessible system sizes allows one to vary significantly the relative importance of fluid- and non-fluid degrees of freedom in collective phenomena. }
\label{BigPictureFig}
\end{figure}

One decade into this program, what is the big picture that has emerged? Fig.~\ref{BigPictureFig} sketches three different ways to think about the onset of collectivity as a function of system size. According to the first picture (a), a mesoscopic system displays fluid dynamic behavior only above a finite minimal size. In this “minimal droplet” model, the HEP phenomenology of \pp collisions (as embodied in standard multi-purpose event generators) and the HIP phenomenology of heavy ion collisions (based on the notion of an almost perfect fluid) could have lived side-by-side within limited, mutually exclusive regions of validity. This picture is falsified. Almost all collective phenomena observed in large \AOnA collisions have been identified by now also in small \pp and \pPb collisions. After a decade of small system physics, we know that the dependence of collective phenomena on system size does not show any clear onset of qualitatively novel physics above a minimal characteristic scale. Also, we do not have any evidence that the physical mechanisms responsible for that collectivity change abruptly or qualitatively as a function of system size. Second, there is the simple view (b) that if one observes collectivity in the smallest systems, then even the smallest systems flow. However, the identification of such observed “flow” with a signature of fluid dynamics is largely semantic and not tenable. A closer look at model studies reveals that there is no model in which fluid dynamics dominates collectivity for systems of all size. Rather, fluid dynamics is the universal long wavelength limit of any more complete and more microscopic description of collectivity. As the system size is reduced, this long wavelength limit becomes gradually less important and the non-fluid dynamic degrees of freedom play an increasingly important role in building up collectivity. The relative contribution of fluid dynamics to collectivity changes gradually with system size. The observed commonalities of collective phenomena in small and large systems thus give support to the idea that essentially the same microscopic  mechanisms are at work in both \pp and \AOnA, though their relative importance changes gradually with system size. This idea of a gradual emergence of fluid dynamics with system size is captured in the third caricature (c) of Fig.~\ref{BigPictureFig}. It is the experimentally and theoretically supported one.

How can this big picture guide future studies? What is surprising in hindsight is not so much that the picture of gradual emergence of fluid dynamics is the supported one. The fortunate surprise is rather that the transition from a collective dynamics dominated by non-fluid degrees of freedom to a collective dynamics dominated by fluid ones seems to occur exactly within the experimentally accessible range of system size. The picture of gradual emergence of fluid dynamics thus indicates that system size can be used as a tool to vary the relative importance of fluid- and non-fluid dynamic degrees of freedom in controlled experimentation. It also raises the question to what extent the impact of non-fluid dynamic degrees of freedom in large collision systems, even if subleading, could be quantified better. Small systems provide a natural inroad to understanding the QGP beyond its fluid dynamic manifestations. 

\section*{Acknowledgments}

We thank 
Katarina Krizkova Gajdosova,
Aleksi Kurkela,
Andreas Morsch,
Krishna Rajagopal and
Soeren Schlichting
for constructive input and criticism at various stages of this work.

\clearpage

\bibliographystyle{unsrtnat}

\bibliography{biblio}

\end{document}